\documentclass[english,11pt]{elsarticle} 

\usepackage[T1]{fontenc}
\usepackage{float}
\usepackage{amsmath}
\usepackage{amsthm}
\usepackage{amssymb}
\usepackage{amsfonts}
\usepackage{bbm}
\usepackage{graphicx}
\usepackage{epstopdf}
\usepackage[pdftex,colorlinks,linkcolor={blue},citecolor={blue}]{hyperref}

 \usepackage{pgfplots}
  \pgfplotsset{compat=newest}
  \usetikzlibrary{plotmarks}
  \usetikzlibrary{arrows.meta}
  \usepgfplotslibrary{patchplots}
  \usepackage{grffile}
  \usepackage{amsmath}

\usepackage{epsfig}
\usepackage{subfigure}
\usepackage{amssymb,amsmath,amsfonts,layout,graphicx}
\usepackage{makeidx}
\usepackage{babel}
\usepackage{tikz}
\usepackage{sublabel}
\usepackage{cases}
\usepackage{algorithm}
\usepackage{algorithmic}

\usepackage
[
        letterpaper,
        left=1.91cm,
        right=1.91cm,
        top=1.91cm,
        bottom=2cm,
]
{geometry}

\newtheorem{assumption}{Assumption}
\newtheorem{proposition}{Proposition}
\newtheorem{theorem}{Theorem}

\newtheorem{remark}{Remark}

\newcommand{\x}{{\bf x}}

\begin{document}
\begin{frontmatter}
 \title{\textbf{Impact of Congestion Charge and Minimum Wage on TNCs: \\ A Case Study for San Francisco}}

\author[1staddress]{Sen Li}
\ead{cesli@ust.hk}
\author[2ndaddress]{Kameshwar Poolla}
\ead{poolla@berkeley.edu}
\author[2ndaddress]{Pravin Varaiya}
\ead{varaiya@berkeley.edu}

\address[1staddress]{Department of Civil and Environmental Engineering, The Hong Kong University of Science and Technology}
\address[2ndaddress]{Department of Electrical Engineering and Computer Science, The University of California, Berkeley}

\begin{abstract}
This paper describes the  impact on transportation network companies (TNCs) of the imposition of a congestion charge and  a driver minimum wage.  The impact is assessed using a market equilibrium model to calculate the  changes in the number of passenger trips and trip fare, number of drivers employed, the TNC platform profit, the number of   TNC vehicles, and city revenue.  Two  types of charges are considered: (a)  a charge  per TNC trip  similar to an excise tax, and (b) a charge per vehicle operating hour (whether or not it has a passenger) similar to a road tax. Both charges  reduce the number of TNC trips, but this reduction is limited by the wage floor, and  the number of TNC vehicles reduced is not significant. The time-based charge is preferable to the trip-based charge since,  by penalizing idle vehicle time, the former increases  vehicle occupancy. In a case study for San Francisco, the  time-based charge  is found to be Pareto superior to the trip-based charge as it yields  higher passenger surplus, higher platform profits, and higher tax revenue for the city. 
\end{abstract}


\end{frontmatter}

\section{Introduction}
{ 
Transportation network companies (TNCs) like Uber, Lyft and Didi, have dramatically changed  urban transportation. While the emergence of TNC significantly benefits passengers and drivers, it also brings negative externalities that have to be addressed by regulatory intervention. In recent years, this concern has prompted several cities to take actions to regulate TNCs \cite{NYC2019surcharge, ban2018gan, SFSPUR, seattleregulation2020}. Despite numerous works on the operation and management strategies of TNC platforms, only a handful of works have considered the mathematical model for policy analysis on the ride-hailing market.  This paper aims to formulate an economic equilibrium model to evaluate the impacts of various regulations on the TNC economy.

{\bf Background and Motivation} \\
TNCs are disrupting the urban transportation systems. On the one hand, they offer on-demand ride services at prices that many riders can afford.  On the other hand, they create numerous job opportunities for drivers working as independent contractors.  These favorable demand and supply factors  led to the TNC's explosive growth.} However, the resulting growth has raised two public concerns in large metropolitan areas. The first  is due to increased traffic congestion. In New York City, Uber, Lyft, Juno and Via together dispatch nearly 600,000 rides per day, involving about 80,000 vehicles.  Schaller \cite{schaller2017empty} estimates that from 2013 to 2017 TNC trips in NYC increased by 15\%, traffic speed dropped by 15\%, VMT increased by 36\%, and the number of TNC vehicles increased by 59\%.  He suggested regulation to reduce TNC vehicles deadhead time (when  vehicles are carrying no passengers) in order to limit  congestion. Two reports \cite{castiglione2016tncs,castiglione2018tncs} by the San Francisco County Transportation Authority identified TNC impact  on traffic congestion and estimated that TNCs account for approximately 50 percent of the increase in congestion in San Francisco between 2010 and 2016.  More recently, Uber and Lyft  commissioned Fehr \& Peer to estimate the TNC share of VMT in six US Metropolitan Regions,  Boston, Chicago, Los Angeles, Seattle, San Francisco and Washington. Their report \cite{balding2019}  concludes that Uber and Lyft have a nontrivial  impact in core urban areas such as San Francisco County, where they account for 12.8\% of total VMT.  

The second concern is provoked by the very low earnings of TNC drivers. The success of the on-demand ride-hailing business relies on short passenger waiting times that require a large pool of available but idle TNC drivers. This pushes down driver wages. Parrott and Reich \cite{parrott2018earning} revealed that the majority of  for-hire vehicle drivers in NYC work full-time. They found that the median driver earnings  declined almost $\$$3 per hour from $\$$25.67 in September 2016 to $\$$22.90 in October 2017, and  that 85 percent of drivers made less than the minimum wage after deducting  vehicle expenses. A follow-up study \cite{parrott2020minimum} examined the payments of drivers working for TNCs in Seattle and discovered that their average net earning is \$9.73/hour (after expenses), well below the \$16.39/hour minimum wage.  Further, more than four-fifths of full time drivers purchased their vehicle primarily or partly to provide TNC services, and nearly three-fourths rely on TNC driving as their sole source of income. These drivers are hired as independent contractors, who can not unionize to negotiate for  labor rights such as minimum wage, overtime compensation, and paid time-off.

These concerns have prompted  cities to regulate TNCs. To address congestion,  New York City Taxi and Limousine Commission (NYCTLC)  introduced a  \$2.75 charge on all for-hire vehicle trips that pass through the ``congestion zone'' of the city \cite{NYC2019surcharge}. The congestion zone is  the area south of 96th Street in Manhattan, and the charge is assessed on each trip  that starts from, ends in, or passes through the congestion area.  To protect TNC drivers, NYCTLC imposed a minimum per-trip wage for drivers  amounting to a wage floor of \$25.76/hour or \$17.22/hour after vehicle expenses \cite{ban2018gan}. This is equivalent to the \$15/hour minimum wage after deducting a paid-time off supplement of \$2.22/hour. 
In addition to NYC, similar  regulations are being considered by other U.S. cities.  In November 2019 Chicago  approved a congestion tax on ride-hailing services for weekday single-passenger trips (and lowered the tax on shared trips) in the downtown area to raise \$40 million per year \cite{chicago_surcharge}. Also in November 2019 San Francisco passed a special 3.25\% excise tax on TNC rides to raise \$30-\$35 million per year for congestion mitigation projects 
\cite{SFSPUR}. At around the same time, the Seattle City Council unanimously approved the ``Fare Share'' plan, which provides TNC driver protections including  a fair wage after expenses and a first-in-the-nation Driver Resolution Center to offer support services for drivers to fight against unwarranted deactivations \cite{seattleregulation2020}. In September 2019
California passed bill AB5  \cite{vox_gig} which classifies hundreds of thousands of independent contractors (gig workers) including TNC drivers as employees to protect them with minimum wage and other employee benefits. These actions imply a changing regulatory environment to address TNC-provoked concerns in large cities. 

{ {\bf Research Problem and Contribution} }\\
This paper presents a study calculating the  impact  on TNCs of the joint imposition of a congestion charge and a driver minimum wage. The  impact is formulated within a framework comprised of a queuing theoretic model of the  arrivals of passengers and drivers, a general equilibrium model that predicts  market prices, passenger
demand and driver supply, and a profit maximizing  model of the TNC platform decisions. This framework enables the assessment of the  impact in terms of changes in  ride prices, passenger waiting time, driver wage, numbers of passengers and drivers, vehicle occupancy rate, platform rent, and city tax revenue. The key conclusions of this study are:
\begin{itemize}
    \item  The congestion charge does not significantly affect TNC ridership. It does not directly curb  traffic congestion by reducing the number of TNC vehicles on the road. This is because the impact of the surcharge is mitigated by the wage floor on TNC drivers.
    \item The time-based congestion charge is preferable to the trip-based charge because the former penalizes idle vehicle hours, thereby increasing  vehicle occupancy (we  use the terms congestion charge and tax interchangeably.) Furthermore, the increased occupancy generates a surplus that offers a Pareto improvement in a certain regime, bringing  higher consumer surplus, higher platform profit and higher tax revenue for the city.  
    \item The case study for San Francisco employs a model whose parameters are calibrated to match  reported San Francisco TNC data, and the model is used to predict the likely effect of regulatory policies on the San Francisco TNC market. 
    \item Through numerical simulation, we show that the tax burden mainly falls on the ride-hailing platform as opposed to passengers and drivers. Under a trip-based tax of \$2/trip (with average trip fare of \$11.6),  passenger travel cost increases by 0.6\%,  driver wage remains unchanged, while the platform profit is reduced by 59.5\%. Under a time-based tax in the regime of practical interest, both passengers and drivers are unaffected, while the platform assumes all of the tax burden.
\end{itemize}

\section{Related Works}
There is an extensive literature on app-based ride-hailing platforms. Many studies investigated the platform pricing strategy under various interacting factors. Zha et al \cite{zha2016economic}
developed an aggregate model to capture the interactions among passengers, drivers and the platform, and found that the first-best solution is not sustainable when the matching function exhibits increasing returns to scale and the cost function of the platform is subject to economies of scale. 
Bai et al \cite{Bai2018coordinating} considered an on-demand service platform using earning-sensitive independent providers with heterogeneous reservation price, and concluded that it is optimal to charge a higher price when demand increases, and that the platform should offer a higher payout ratio as demand increases, capacity decreases, or customers become more sensitive to waiting time. Taylor \cite{taylor2018demand} examined how delay sensitivity and agent independence affect the platform's optimal price and wage and identified the complexity caused by uncertainty in  customer valuation. Hu and Zhou \cite{hu2019price} studied the commission setting of the ride-sourcing platform and showed that an optimal fixed-commission contract can achieve at least 75\% of the optimal profit when there is no pre-committed relationship between  price and wage. 

Platform pricing has also been studied with temporal and spatial considerations. From the temporal aspect, Cachon et al \cite{cachon2017role} showed that surge pricing can significantly increase  platform profit relative to contracts that have a fixed price or fixed wage, and that all stakeholders can benefit from the use of surge pricing on a platform with driver self-scheduling capacity. Castillo et al \cite{castillo2017surge} showed that surge pricing can avoid cases where vehicles are sent on a wild goose chase to pick up distant customers, wasting driver time and reducing earnings. Zha et al \cite{zha2017surge} investigated the impact of surge pricing using a bi-level programming framework, and showed that compared to static pricing, the platform and drivers are found  generally to enjoy higher revenue while customers may be made worse off during highly surged periods. 
Banerjee et al \cite{banerjee2015pricing} developed a queuing theoretic model to study the optimal (profit-maximizing) pricing of ride-sharing platforms. They show that the performance of a dynamic price (in terms of revenue and throughput) does not exceed that of a static price, but it is more robust to fluctuations of model parameters. From the spatial aspect, Bimpikis et al \cite{bimpikis2019spatial}  considered the price discrimination of a ride-sourcing platform over a transportation network and established that profits and consumer surplus at the equilibrium corresponding to the platform's optimal pricing are maximized when the demand pattern is ``balanced'' across the network's locations. Guda and Subramanian \cite{guda2019your} studied the spatial pricing of a ride-sourcing platform over a transportation network and showed that surge pricing can be useful even in zones where supply exceeds demand. Zha et al \cite{zha2018geometric} developed a model to investigate the effects of spatial pricing on ride-sourcing markets and found that the platform may resort to relatively higher price to avoid an inefficient supply state if spatial price differentiation is not allowed.
In addition to platform pricing, studies also touch upon driver supply \cite{Hall_Kreuger}, \cite{gurvich2019operations}, platform operations \cite{yang2020optimizing}, \cite{vazifeh2018addressing},  platform competition \cite{nikzad2017thickness}, \cite{bernstein2019competition}, and regulations \cite{li2019regulating}, \cite{benjaafar2018labor}, \cite{yu2019balancing}, \cite{vignon2020regulating}. Please see  \cite{wang2019ridesourcing} for a comprehensive literature review. 

Road pricing has attracted substantial research attention for decades. The idea was initially proposed by Pigou \cite{pigou2017economics}, which inspired several seminal works including Vickery \cite{vickrey1955some}, Walters \cite{walters1961theory} and Beckmann \cite{beckmann1967optimal}. Since then, various taxing schemes have been proposed in the literature, including charge based on cordon-crossing, distance traveled, time spent traveling, or time spent in congestion \cite{may2000effects}. For instance, Zhang and Yang \cite{zhang2004optimal} investigate the cordon-based second-best congestion pricing problem on road networks that jointly consider toll levels and toll locations. Yang et al \cite{yang2010road} study road pricing for effective congestion control without knowing the link travel time and travel demand. Liu and Li \cite{liu2017pricing} derive a time-varying toll combined with a flat ride-sharing price to nudge morning travelers to depart in off-peak hours. Despite this large literature in transportation  economics, the research on congestion charges for TNCs is relatively scarce. A TNC congestion charge  is distinctive since it involves decisions  of the profit-maximizing platform and the passengers and drivers in the two-sided ride-hailing market.  Li et al \cite{li2019regulating} proposed a market equilibrium model to evaluate the impact of various regulatory policies and analyzed the incidence of a TNC tax on passengers, drivers, and the TNC platform.  Schaller \cite{schaller2018making}  conducted an in-depth analysis of how to apply pricing to new mobility services, and recommended that a surcharge on taxi/for-hire trips in central Manhattan be applied as an hourly charge. Recent work of Vignon and Yin \cite{vignon2020regulating} investigated the performance of various regulation policies  on ride-sourcing platforms with congestion externality and product differentiation taken into account. They compared a uniform toll that treats all vehicles identically with a differentiated toll that treats idle vehicles, solo rides and pooled rides differently, and showed that a differentiated toll offers little advantage over a uniform one.  

Only a handful of studies considered wage regulation of TNCs. Gurvich \cite{gurvich2016operations}
 studied the platform's profit maximizing wage level for self-scheduling drivers, and showed that under a minimum wage, the platform limits agent flexibility by restricting the number of agents that can work during some time intervals. Parrott and Reich \cite{parrott2018earning} utilized administrative data of New York City and showed by simulation that the proposed minimum wage standard in New York City will increase driver wage by 22.5 percent while hurting  passengers by slightly increasing  ride fare and  waiting time.  Li et al. \cite{li2019regulating} and Benjaafar et al. \cite{benjaafar2018labor} developed market equilibrium models to show that wage regulations on TNC will benefit both passengers and drivers, because  wage regulation curbs  TNC labor market power \cite{li2019regulating}. 
Zhang and Nie \cite{zhang2019pool} proposed a market equilibrium model for ride-sourcing platforms that offers a mix of solo and pooled rides. They showed that a wage floor on TNC drivers will force the platform to hire more drivers, which will reduce the appeal of collective modes and the supply efficiency and  is likely to worsen traffic congestion.

 This paper differs from the aforementioned works in that we explore the {\em joint} impact of congestion charge and driver minimum wage on the TNC market. We are the first to point out that distinct regulatory policies on TNCs do interfere with each other when they are jointly implemented, which may produce surprising market outcomes that deviate from the expectation of the policy maker. We are also the first to establish models that compare the trip-based congestion charge and the time-based congestion charge and identify the superiority of time-based congestion in certain regimes of practical interest. These results will provide valuable insights for city planners who are considering  implementing (e.g., San Francisco), or have already implemented (e.g., NYC and Seattle) a congestion charge and a minimum wage to address  TNC externalities.


\section{Market Equilibrium Model} 
\label{lowerlevel}

We consider a transportation system comprised of  a city council, a TNC platform, and a group of passengers and drivers. The city council approves legislation (e.g., cap on the total number of vehicles, minimum wage for TNC drivers, congestion charge on TNC trips) to regulate the operations of the TNC platform. The platform sets fares and wages and hires drivers to maximize its profit under these regulations. The pricing decisions  affect the choices of  passengers and drivers, and these choices collectively  determine the platform's profit. We will describe a market equilibrium model to capture the decisions of passengers, drivers, and the TNC platform. The model will be used to investigate how TNC market outcomes are affected by  regulation.

\subsection{Matching passengers and drivers}

The TNC platform matches randomly arriving passengers to idle TNC  drivers. Upon arrival, each passenger  joins a queue and waits until she or he is matched to an idle driver\footnote{For simplicity, we do not consider the case of multiple passengers sharing the same vehicle.}. This matching is modeled  as a continuous-time queuing process, in which each passenger defines a ``job'' and each driver is a ``server''. The server is ``idle'' if the vehicle is not occupied, and it is ``busy'' if a passenger is on board or if the vehicle is dispatched and on its way to pick up a passenger. Assume that passenger arrivals form a Poisson process with rate $\lambda  > 0$, and denote $N$ as the total number of TNC drivers.  This matching process forms a M/G/N queue, and the expected number of idle servers (vehicles) is ${N_I} = N - \lambda /\mu $,  with $\mu $ being the service rate ($1/\mu$ is the amount of time a passenger occupies a vehicle on average).  We assume that $N > \lambda /\mu $. { Given the model parameters, the average waiting time for the M/G/N queue can be derived approximately in an analytical form. We comment that this is the ride confirmation time, which represents the time elapsed after the ride is requested and before the ride is confirmed. It differs from the pickup time (from ride confirmation to pickup), which will be treated below. }

\subsection{Passenger incentives}
The total travel cost of the TNC passenger consists of the waiting time for pickup, the travel time during the trip, and the monetary payment for the ride service. We refer to this total travel cost as the ``generalized cost'' and define it as the weighted sum of waiting time,  travel time, and trip fare. It may differ for distinct passengers due to the randomness in trip length, trip duration, and the matching process of the TNC platform. Since we primarily focus on  aggregate market outcomes, we define the average generalized cost as:
\begin{equation}
\label{cost_definition}
	c = \alpha {t_w}  + \beta t_0+ {p_f},
	\end{equation}
where ${t_w}$ is the average waiting time,  $t_0$ is the average trip duration (in minutes),  and $p_f $ is the average price of a TNC ride.  The parameters $\alpha$ and $\beta$ specify the passenger trade-off between time and money. Note that $\alpha$ is generally larger than $\beta$ since empirical study suggests that the value of time while waiting is larger than the value of time while traveling in the vehicle. 
It is important to emphasize that we do not need to assume that all passengers have the same travel cost. The heterogeneity in passengers is irrelevant as  we focus on  the aggregate market outcome, which typically depends on the average cost $c$. A widely-studied example is the logit choice model, where the total number of agents choosing a particular mode only depends on the average cost of each mode. In this spirit, we define a demand function that determines the arrival rate of TNC passengers as a function of the average generalized cost:
\begin{equation}
\label{demand_function}
	\lambda  = {\lambda _0}{F_p}(c),
	\end{equation}
where \({\lambda _0}\) is the arrival rate of potential passengers (total travel demand in the city), and \({F_p}( \cdot )\) is the proportion of potential passengers who choose a TNC ride. We assume that \({F_p}( \cdot )\) is a strictly decreasing and continuously differentiable function so that a higher  TNC travel cost $c$ will lead to fewer TNC passengers. The logit model is a special case of (\ref{demand_function}).

The passenger waiting time $t_w$ intimately interacts with other endogenous decision variables $\lambda$ and $N$. To delineate this relation, we divide a TNC ride into three time periods: (1) from ride being requested to the ride being confirmed, (2) from the ride being confirmed to passenger pickup, (3) from passenger pickup to drop-off. Let ${t_m}$, ${t_p}$, and  ${t_0}$ represent the length of these three periods, respectively, then we have ${t_w} = {t_m} + {t_p}$, and  ${t_0}$ as the average trip distance $L$ divided by traffic speed $v$, i.e., ${t_0} = L/v$.   Since the platform immediately matches each newly arrived passenger to the nearest idle vehicle, ${t_m}$ is the average waiting time in the queue, and $t_p$ depends on the traffic speed $v$ and the  distance of the passenger to  the nearest idle vehicle, which further depends on the number of idle vehicles ${N_I}$. Therefore, we  write $t_p$  as a function of $N_I$ and $v$, i.e., $t_p(N_I, v)$. The following assumption is imposed on $t_p(\cdot)$:

\begin{assumption}
\({t_p}({N_I},v)\) is twice differentiable with respect to \({N_I}\) and \(v.\) It is decreasing and strictly convex with respect to \({N_I},\) and it is decreasing with respect to traffic speed \(v.\)
\label{assumption1}
\end{assumption}
	
Assumption \ref{assumption1} requires that  the pickup time decreases with respect to the number of idle vehicles and the traffic speed. We suppose traffic speed $v(N)$ is a function of the total number $N$ of vehicles  and impose the following assumption on $v(\cdot)$:

\begin{assumption}
$v(N)$ is decreasing and continuously differentiable with respect to $N$.
\label{assumption2}
\end{assumption}

{ Using data of San Francisco and New York City for the M/G/N queue, we find that the ride confirmation time $t_m$ is very short, i.e., less than 1 seconds. This is negligible compared to the pickup time \({t_p}\), which is typically around 3-5 minutes.} Therefore we ignore \({t_m}\) and express the total waiting time \({t_w}\) as\footnote{The waiting time can be significantly larger in rush hours. In this case, one can add $t_m$ to $t_w$ as the waiting time in the queue. We believe that this will not affect our conclusion, but we neglect this term in this paper for analytic tractability.}
\begin{equation}
{t_w} = {t_p}({N_I},v).
\end{equation}
The number of idle vehicles $N_I$ depends on  $\lambda$ and $N$, whereas the average traffic speed $v $ depends on $N$.

\subsection{ Driver incentives}

In the TNC market, drivers can decide whether to remain subscribed to the  TNC platform depending on the long-term average earnings offered by the platform. The average hourly wage of drivers depends on the ride fare of the TNC trip, the commission rate set by the platform, and the occupancy rate of the vehicles. It can be described as:
\begin{equation}
\label{driver_wage_def}
	w = \frac{{\lambda {p_d}}}{N},
\end{equation}
where $p_d$ is the average per-trip payment to drivers. The driver payment \({p_d}\) differs from the passenger trip fare $p_f$.  The difference $p_f-p_d$ is kept by the platform as profit. Therefore, the commission rate of the platform (typically 25\%-40\%) can be written as $(p_f-p_d)/p_f$. The average hourly wage (\ref{driver_wage_def}) is just the total platform payment to all drivers $\lambda {p_d}$ divided by the total number of drivers $N$.  Each driver may have an hourly earning that differs from the earning of others  due to the randomness in work schedule, driver location, and repositioning strategy. However, as we primarily focus on the aggregated market outcome, the heteregeneity in driver earnings is irrelevant as far as the aggregate market outcome (e.g.,total number of TNC passengers or drivers) only depends on the average hourly earning over all TNC drivers. Note that this is the case for the well-established logit choice model. More generally, we define a supply function that determines the total number of TNC drivers as a function of the average hourly wage:  
\begin{equation}
\label{supply_function}
	N = {N_0}{F_d}(w),
\end{equation}
where $N_0$ is the number of potential drivers (all drivers seeking a job), and $F_d(w)$ is a strictly increasing and continuously differentiable function that gives the proportion of drivers willing to join TNC. Note that the logit model is a special case of (\ref{supply_function})

\subsection{ Platform decisions in absence of regulation}
The TNC platform determines the ride prices and the driver payment to gauge passengers and drivers  to maximize its profit. In each time period, the platform revenue is the total ride fares received from passengers, i.e.,  $\lambda p_f$, and the platform cost is the total payment made to the drivers, i.e., $\lambda p_d$. The profit of the platform can be thus written as the difference between the revenue and the cost
\begin{equation}
\label{optimalpricing}
 \hspace{-1.5cm} \mathop {\max }\limits_{{p_f}, {p_d}} \quad \lambda ({p_f} - {p_d})
\end{equation}
\begin{subnumcases}{\label{constraint_optimapricing_TNC}}
\lambda  = {\lambda _0}{F_p}\left(\alpha {t_p} + \beta t_0+ {p_f} \right) \label{demand_constraint}\\
N = {N_0}{F_d}\left(\frac{{\lambda {p_d}}}{N}\right) \label{supply_constraint} 
\end{subnumcases}
where (\ref{demand_constraint}) is the demand function and (\ref{supply_constraint}) is the supply function. Note that $t_p$ depends on $\lambda$ and $N$, and $t_0$ depends on the traffic speed which is a function of $N$. The overall problem not only involves $p_f$ and $p_d$ as decision variables, but also involves $N$, $\lambda$, $t_p$, $v$ and $t_0$ as endogenous variables. The optimal solution to (\ref{optimalpricing}) represents the platform's profit-maximizing pricing decision in absence of the regulatory intervention. 

{ The profit maximization problem (\ref{optimalpricing}) is a constrained optimization which can be solved by various gradient-based algorithms \cite{bertsekas1997nonlinear}. However, since the problem is non-concave with respect to $p_d$ and $p_f$, it is difficult to assert whether the obtained solution is globally optimal. To address this concern, we apply a change of variable and treat $\lambda$ and $N$ as the new decision variables. More specifically, given $\lambda$ and $N$, we can use (\ref{demand_constraint})-(\ref{supply_constraint}) to uniquely determine $p_f$ and $p_d$ as follows: 
\begin{subnumcases}{\label{changeofvariable}}
p_f= F_p^{-1} \left(\dfrac{\lambda}{\lambda_0}\right) - \alpha t_p(N_I, v)-\beta p_0  \label{changeofvariable1}\\
p_d= \dfrac{N}{\lambda} F_d^{-1}\left( \dfrac{N}{N_0}\right)  \label{changeofvariable2} 
\end{subnumcases}
where (\ref{changeofvariable1}) is derived from  (\ref{demand_constraint}), and (\ref{changeofvariable2}) is derived from (\ref{supply_constraint}). Note that the right-hand sides of (\ref{changeofvariable1}) and (\ref{changeofvariable2}) are both functions of $\lambda$ and $N$. By plugging (\ref{changeofvariable1}) and (\ref{changeofvariable2}) into (\ref{optimalpricing}), we can transform the profit maximization problem (\ref{optimalpricing}) into the following unconstrained optimization:
\begin{equation}
\label{optimalpricing_transformed}
 \hspace{-1.5cm} \mathop {\max }\limits_{\lambda, N} \quad \lambda \left(F_p^{-1} \left(\dfrac{\lambda}{\lambda_0}\right)- \alpha t_p(N_I, v)-\beta p_0  \right)- N F_d^{-1}\left( \dfrac{N}{N_0}\right) 
\end{equation}
where $\lambda$ and $N$ are decision variables. Clearly,  (\ref{optimalpricing_transformed}) is equivalent to (\ref{optimalpricing}). We note that although (\ref{optimalpricing_transformed}) is non-concave with respect to $\lambda$ and $N$, under certain mild conditions, it is concave with respect to $\lambda$ for fixed $N$. We formally summarize this result as the following proposition:
\begin{proposition}
\label{prop_concave}
Assume the demand function $F_p(\cdot)$ is a logit model represented as:
\begin{equation}
\label{logit_demand_prop}
	\lambda =\lambda_0 \frac{e^{-\epsilon c}}{e^{-\epsilon c}+e^{-\epsilon c_0}},
\end{equation}
where $\epsilon>0$ and $c_0$ are parameters. Further assume that given $v$, the waiting time function $t_p(N_I, v)$ is convex with respect to $N_I$, then we have the following results: \\
(1) the profit maximization problem (\ref{optimalpricing_transformed}) is concave with respect to $\lambda$ under a fixed $N$, \\
(2) Given $N$, there exists a unique $\lambda$ that maximizes the platform profit (\ref{optimalpricing_transformed})
\end{proposition}

The proof can be found in Appendix A. Proposition \ref{prop_concave} suggests that for any fixed $N$, we can efficiently derive the unique optimal $\lambda$ that maximizes the profit by solving a concave program. This result is based on a  few mild assumptions: (a) the logit model (\ref{logit_demand_prop}) is used for studying customer discrete choice, (b) the convexity of $t_p(\cdot)$ simply requires that the marginal benefit of adding extra idle vehicles in reducing passenger waiting time decreases with respect to $N_I$, which is consistent with intuition. Based on this result, we can obtain the optimal combination of $(\lambda, N)$ by enumerating over $N$. This provides the globally optimal solution to (\ref{optimalpricing_transformed}).
}
\begin{remark}
Many works study the spatial and temporal aspects of the TNC market. These aspects are neglected in our model since we primarily focus on the  evaluation of regulatory policies (e.g., minimum wage) that are imposed on a uniform basis regardless of the time of the day or the location of the driver. This makes it legitimate to consider the impact of these policies at the aggregate scale, which suffices to provide valuable insights for city planners to assess their policies. 
A spatial-temporal analysis is necessary if policy makers further consider fine-tuning these policies so that they differentiate trips at different time instances or different locations. This is left for future work. 
\end{remark}

\subsection{Modeling regulation policies}
Regulation policies, such as congestion charge and driver minimum wage, modify the incentives of passengers and drivers and affect the pricing decision of the TNC platform. To capture this effect, we  formulate the platform pricing problem under the  minimum wage, trip-based congestion charge, and  time-based congestion charge. 

{\bf Minimum wage:} To capture the impact of a driver wage floor $w_0$, we impose the constraint that requires the driver hourly earning to be greater than  $w_0$. The optimal pricing problem under minimum wage regulation can be formulated as:
\begin{equation}
\label{optimalpricing_wage}
 \hspace{-1.5cm} \mathop {\max }\limits_{{p_f}, {p_d}, N} \quad \lambda ({p_f} - {p_d})
\end{equation}
\begin{subnumcases}{\label{constraint_optimapricing_TNC_wage}}
\lambda  = {\lambda _0}{F_p}\left(\alpha {t_p} + \beta t_0+ {p_f} \right) \label{demand_constraint_wage}\\
N \leq  {N_0}{F_d}\left(\frac{{\lambda {p_d}}}{N}\right) \label{supply_constraint_wage}  \\
\frac{{\lambda {p_d}}}{N} \ge {w_0} \label{min_wage_wage}
\end{subnumcases}
where constraint (\ref{min_wage_wage}) captures the wage floor on TNC driver earnings. Note that we relax the equality constraint (\ref{supply_constraint}) to inequality constraint (\ref{supply_constraint_wage}). This permits the TNC platform to hire a subset of drivers who are willing to work for TNC in case the minimum wage is set so high that it is unprofitable for the platform to hire all the willing drivers in the market.  

{
\begin{remark}
Note that the minimum wage constraint (\ref{min_wage_wage}) places a lower bound on the {\em average} driver wage $w$. Since the hourly wage may differ from one driver to another, when (\ref{min_wage_wage}) is satisfied, it does not necessarily mean that all drivers  earn at least the minimum wage. Instead, it only indicates that drivers can earn more than the minimum wage on average. We emphasize that this formulation is consistent with the practice: the minimum wage for TNC drivers in New York City and Seattle are both implemented on a platform-wide average basis, instead of an individual driver basis \cite{ban2018gan}, \cite{ban2018sea}.   
\end{remark}}

{\bf Trip-based congestion charge:} Many existing congestion charge schemes are trip-based (e.g., New York City, Seattle, Chicago). The trip-based congestion charge assesses an extra fee of $p_t$ on each TNC trip in the congestion area.  When a congestion charge $p_t$ and a minimum wage $w_0$ are imposed concurrently,  the optimal pricing problem can be formulated as follows:
\begin{equation}
\label{optimalpricing_trip}
 \hspace{-2cm} \mathop {\max }\limits_{{p_f}, {p_d}, N} \quad \lambda ({p_f} - {p_d})
\end{equation}
\begin{subnumcases}{\label{constraint_optimapricing_trip}}
\lambda  = {\lambda _0}{F_p}\left(\alpha {t_p} + \beta t_0+ {p_f}+ p_t \right) \label{demand_constraint_wage}\\
N \le {N_0}{F_d}\left(\frac{{\lambda {p_d}}}{N}\right) \label{supply_constraint_trip} \\
\frac{{\lambda {p_d}}}{N} \ge {w_0}
\label{min_wage_const}
\end{subnumcases}
where the per-trip congestion charge $p_t$ is incorporated into the passenger travel cost within the demand function (\ref{demand_constraint_wage}). Another way to formulate the congestion charge is by adding it to the cost of the platform. This is easier to implement as it only requires the platform to transfer the accumulated congestion charge of all trips within certain period to the city. In this case, the optimal pricing problem can be written as:
\begin{equation}
\label{optimalpricing_trip2}
 \hspace{-1cm} \mathop {\max }\limits_{{p_f}, {p_d}, N} \quad \lambda ({p_f} - {p_d})- \lambda p_t
\end{equation}
\begin{subnumcases}{\label{constraint_optimapricing2}}
\lambda  = {\lambda_0}{F_p}\left(\alpha {t_p} + \beta t_0+ {p_f}\right) \label{demand_constraint_wage2}\\
N \le {N_0}{F_d}\left(\frac{{\lambda {p_d}}}{N}\right) \label{supply_constraint_trip2} \\
\frac{{\lambda {p_d}}}{N} \ge {w_0}
\label{min_wage_const2}
\end{subnumcases}
where $p_t$ is incorporated into the profit of the platform instead of the travel cost of the passengers. Economists find that whether a tax is levied on the buyer or seller of the good does not matter because they always share the tax burden based on their elasticities \cite[Chap. 16]{varian2014intermediate}. This principle also applies here:
\begin{proposition}
\label{prop1}
Let $(p_f^*,p_d^*,N^*,\lambda^*)$ and $(p_f^{**},p_d^{**},N^{**},\lambda^{**})$ denote the optimal solutions to (\ref{optimalpricing_trip}) and (\ref{optimalpricing_trip2}), respectively, then we have $p_f^*+p_t=p_f^{**}, p_d^*=p_d^{**}, \lambda^*=\lambda^{**},$ and $N^*=N^{**}$. 
\end{proposition}
Proposition \ref{prop1} states that the two formulations of trip-based congestion charge, i.e., (\ref{optimalpricing_trip}) and (\ref{optimalpricing_trip2}),  lead to the same market outcome. The proof is omitted since it can be simply derived by a change of variable.


{\bf Time-based congestion charge:} Distinct from the trip-based congestion charge, the time-based charge is levied on TNC vehicles based on vehicle hours instead of trip volumes. The key difference between the two congestion charge schemes is that time-based congestion charge not only penalizes TNC trips, but also penalizes idle TNC hours and thus incentivizes the platform to increase vehicle utilization. When the time-based congestion charge $p_h$ and a minimum wage $w_0$ are concurrently levied on TNC drivers, we have the following formulation: 
\begin{equation}
\label{optimalpricing_time}
 \hspace{-2.5cm} \mathop {\max }\limits_{{p_f}, {p_d}, N} \quad \lambda ({p_f} - {p_d})
\end{equation}
\begin{subnumcases}{\label{constraint_optimapricing_time}}
\lambda  = {\lambda _0}{F_p}\left(\alpha {t_p} + \beta t_0+ {p_f}+ p_t \right) \label{demand_constraint_wage_time}\\
N \le {N_0}{F_d}\left(\frac{{\lambda {p_d}}}{N}-p_h\right) \label{supply_constraint_time} \\
\frac{{\lambda {p_d}}}{N} -p_h\ge {w_0}
\label{min_wage_const_time}
\end{subnumcases}

When the time-based congestion charge is levied on the TNC platform, we have the following formulation:
\begin{equation}
\label{optimalpricing_time2}
 \hspace{-0.5cm} \mathop {\max }\limits_{{p_f}, {p_d}, N} \quad \lambda ({p_f} - {p_d}) -Np_h
\end{equation}
\begin{subnumcases}{\label{constraint_optimapricing_time2}}
\lambda  = {\lambda _0}{F_p}\left(\alpha {t_p} + \beta t_0+ {p_f} \right) \label{demand_constraint_wage_time2}\\
N \le {N_0}{F_d}\left(\frac{{\lambda {p_d}}}{N}\right) \label{supply_constraint_time2} \\
\frac{{\lambda {p_d}}}{N} \ge {w_0}
\label{min_wage_const_time2}
\end{subnumcases}

Similar to the trip-based congestion charge, these two forms of formulations are equivalent.
\begin{proposition}
\label{prop2}
Let $(\bar{p}_d,\bar{p}_d,\bar{N},\bar{\lambda})$ and $(\tilde{p}_f,\tilde{p}_d,\tilde{N},\tilde{\lambda})$ denote the optimal solutions to (\ref{optimalpricing_time}) and (\ref{optimalpricing_time2}), respectively, then we have $\bar{p}_f=\tilde{p}_f, \dfrac{\bar{\lambda}\bar{p}_d}{\bar{N}}-p_h=\dfrac{\tilde{\lambda}\tilde{p}_d}{\tilde{N}}, \bar{\lambda}=\tilde{\lambda},$ and $\bar{N}=\tilde{N}$. 
\end{proposition}
Proposition \ref{prop2} states that the two formulations of time-based congestion charge, i.e., (\ref{optimalpricing_time}) and (\ref{optimalpricing_time2}),  lead to the same market outcome. The proof is omitted since it is similar to  that of Proposition \ref{prop1}.
 
\section{Profit maximization under trip-based congestion charge}
This section analyzes the joint impact of a trip-based congestion charge and a minimum wage for TNC drivers. We consider a platform that determines the ride fare $p_f$ and the per-trip driver payment $p_d$ to maximize its profit $\lambda(p_f-p_d)$ under the trip-based congestion charge $p_t$ and a minimum wage $w_0$. The optimal pricing problem can be formulated as (\ref{optimalpricing_trip}) or (\ref{optimalpricing_trip2}). For sake of exposition, we will start with a realistic numerical example for San Francisco. The numerical example will be complemented by a theoretical analysis presented later that shows the insights derived from the numerical example can be generalized.

\subsection{Numerical example}
\label{parameter_section}

We investigate the impact of the proposed regulations via a case study for San Francisco (followed by theoretical analysis in the next subsection). Assume that passengers choose their transport mode based on the total travel cost. We use a logit model so the demand function for TNC rides is 
\begin{equation}
\label{logit_demand}
	\lambda =\lambda_0 \frac{e^{-\epsilon c}}{e^{-\epsilon c}+e^{-\epsilon c_0}},
\end{equation}
where $c$ is the total travel cost of a TNC trip, and $\epsilon>0$ and $c_0$ are parameters. Similarly, drivers choose to work for the TNC  depending on its wage. Under a logit model, the supply function is 
\begin{equation}
\label{logit_supply}
	N =N_0 \frac{e^{\sigma w}}{e^{\sigma w}+e^{\sigma w_0}},
\end{equation}
where $\sigma$ is a parameter. We note that  (\ref{logit_demand}) is a special case of the general demand function (\ref{demand_function}), and  (\ref{logit_supply}) is a special case of the general supply function (\ref{supply_function}).

Passenger pickup time $t_p$ follows the ``square root law'' established in \cite{arnott1996taxi} and \cite{li2019regulating}:
\begin{equation}
	{t_p}({N_I},v) = \frac{M}{{v\sqrt {N - \lambda /\mu } }},
\label{pickuptime_func}
\end{equation}
where the constant $M$ depends on the travel times in  the city. The square root law establishes that the average pickup time is inversely proportional to the square root of the number  of idle vehicles in the city, $(N - \lambda /\mu)$. The intuition behind (\ref{pickuptime_func}) is straightforward.  Suppose all idle vehicles are uniformly distributed throughout the city, then the distance between any two nearby idle vehicles is inversely proportional to the square root of the total number of idle vehicles. This distance is proportional to that between the passenger and the closest idle vehicle, which determines the pickup time.  A  justification of the square root law can be found in \cite{li2019regulating}.

The average traffic speed $v$ is a function of the total traffic. Using  Greenshield model \cite{greenshields1953study} gives  the linear speed-density relation\footnote{Since TNC vehicles only account for a small percentage of the overall traffic, the Greenshield model can be regarded as a linear approximation in a small neighborhood of a nonlinear speed-density function.},
\begin{equation}
\label{greens_model}
v = {v_0} - \kappa (N+N_b),
\end{equation}
where $N_b$ is the background traffic\footnote{TNC trips may substitute taxis or private vehicles. This may introduce coupling between the TNC demand and the background traffic $N_b$. For simplicity, we neglect this substitution effect and assume $N_b$ is exogenous. We leave it for future work to investigate how the coupling between $\lambda$ and $N_b$ affect the conclusion of this paper. }, $N$ is the number of TNC vehicles, and $v_0$ and $\kappa$ are  model parameters. Assuming that $N_b$ is constant,  (\ref{greens_model}) is equivalent to
\begin{equation}
v = {v_f} - \kappa N.
\label{greenshiledmodel}
\end{equation}
In summary, the model parameters are
\[{\Theta}=\{\lambda_0, N_0,  M, L, v_f, \kappa, \alpha,  \epsilon, c_0, \sigma, w_0\}.
\]
In the numerical study we set the  parameters values so that the optimal solution to (\ref{optimalpricing}) matches the real data of San Francisco city. The values of these model parameters are summarized below:
\[ \lambda_0=1049/\text{min}, \, N_0=10000,  \, M = 41.18, \quad L=2.6 \text{ mile}, \quad {v_f} = 15 \text{ mph},\quad \kappa  = 0.0003,\]
	\[\alpha=2.33,  \,  \epsilon=0.33, \, c_0=31.2, \, \sigma=0.089, \, w_0=\$31.04/\text{hour}.\]	
For the data source (from San Francisco) and justification of these parameter values, please refer to Appendix B. 

We solve the profit maximizing problem (\ref{optimalpricing_trip}) for different values of congestion charge $p_t$ under a fixed wage floor $w_0$, and plot all the variables as a function of $p_t$. The minimum wage of TNC drivers in San Francisco is set in a way similar  to that in NYC. Under current NYC regulations, the TNC driver minimum wage is $\$25.76$/hour, which is equivalent to the $\$15$/hour  minimum wage of NYC after deducting  vehicle expenses such as insurance, maintenance and taxes. Since the hourly minimum wage of San Francisco is $\$0.59$ higher than in NYC, we set $w_0=\$25.76+\$0.59=\$26.35$/hour to compensate for this difference. 

\begin{figure*}[bt]
\begin{minipage}[b]{0.32\linewidth}
\centering
%
%
\definecolor{mycolor1}{rgb}{0.00000,0.44700,0.74100}%
\begin{tikzpicture}

\pgfplotsset{every axis y label/.style={
at={(-0.47,0.5)},
xshift=30pt,
rotate=90}}

\begin{axis}[%
width=1.794in,
height=1.03in,
at={(1.358in,0.0in)},
scale only axis,
xmin=0,
xmax=3,
xtick={0,1,2,3},
xticklabels={0,1,2,3},
xlabel style={font=\color{white!15!black}},
xlabel={per-trip tax},
ymin=3200,
ymax=4100,
ytick={3200,3600,4000},
yticklabels={{3.2K},{3.6K},{4K}},
ylabel style={font=\color{white!15!black}},
ylabel={Number of drivers},
axis background/.style={fill=white},
legend style={legend cell align=left, align=left, draw=white!15!black}
]
\addplot [color=black, line width=1.0pt]
  table[row sep=crcr]{%
0	3968.14712146414\\
0.0303030303030303	3968.14712146414\\
0.0606060606060606	3968.14712146414\\
0.0909090909090909	3968.14712146414\\
0.121212121212121	3968.14712146414\\
0.151515151515152	3968.14712146414\\
0.181818181818182	3968.14712146414\\
0.212121212121212	3968.14712146414\\
0.242424242424242	3968.14712146414\\
0.272727272727273	3968.14712146414\\
0.303030303030303	3968.14712146414\\
0.333333333333333	3968.14712146414\\
0.363636363636364	3968.14712146414\\
0.393939393939394	3968.14712146414\\
0.424242424242424	3968.14712146414\\
0.454545454545455	3968.14712146414\\
0.484848484848485	3968.14712146414\\
0.515151515151515	3968.14712146414\\
0.545454545454545	3968.14712146414\\
0.575757575757576	3968.14712146414\\
0.606060606060606	3968.14712146414\\
0.636363636363636	3968.14712146414\\
0.666666666666667	3968.14712146414\\
0.696969696969697	3968.14712146414\\
0.727272727272727	3968.14712146414\\
0.757575757575758	3968.14712146414\\
0.787878787878788	3968.14712146414\\
0.818181818181818	3968.14712146414\\
0.848484848484849	3968.14712146414\\
0.878787878787879	3968.14712146414\\
0.909090909090909	3968.14712146414\\
0.939393939393939	3968.14712146414\\
0.96969696969697	3968.14712146414\\
1	3968.14712146414\\
1.03030303030303	3968.14712146414\\
1.06060606060606	3968.14712146414\\
1.09090909090909	3968.14712146414\\
1.12121212121212	3968.14712146414\\
1.15151515151515	3968.14712146414\\
1.18181818181818	3968.14712146414\\
1.21212121212121	3968.14712146414\\
1.24242424242424	3968.14712146414\\
1.27272727272727	3968.14712146414\\
1.3030303030303	3968.14712146414\\
1.33333333333333	3968.14712146414\\
1.36363636363636	3968.14712146414\\
1.39393939393939	3968.14712146414\\
1.42424242424242	3968.14712146414\\
1.45454545454545	3968.14712146414\\
1.48484848484848	3968.14712146414\\
1.51515151515152	3968.14712146414\\
1.54545454545455	3968.14712146414\\
1.57575757575758	3968.14712146414\\
1.60606060606061	3968.14712146414\\
1.63636363636364	3968.14712146414\\
1.66666666666667	3968.14712146414\\
1.6969696969697	3968.14712146414\\
1.72727272727273	3968.14712146414\\
1.75757575757576	3968.14712146414\\
1.78787878787879	3968.14712146414\\
1.81818181818182	3968.14712146414\\
1.84848484848485	3968.14712146414\\
1.87878787878788	3968.14712146414\\
1.90909090909091	3968.14712146414\\
1.93939393939394	3968.14712146414\\
1.96969696969697	3968.14712146414\\
2	3968.14712146414\\
2.03030303030303	3968.14712146414\\
2.06060606060606	3968.14712146414\\
2.09090909090909	3968.14712146414\\
2.12121212121212	3962.5332677229\\
2.15151515151515	3944.49494467638\\
2.18181818181818	3926.40837145426\\
2.21212121212121	3908.27355554703\\
2.24242424242424	3890.08920145607\\
2.27272727272727	3871.8552668891\\
2.3030303030303	3853.57070110621\\
2.33333333333333	3835.23522860657\\
2.36363636363636	3816.84796024135\\
2.39393939393939	3798.4084773862\\
2.42424242424242	3779.91611290089\\
2.45454545454545	3761.36998722856\\
2.48484848484848	3742.76918172402\\
2.51515151515152	3724.11346244639\\
2.54545454545455	3705.40162183681\\
2.57575757575758	3686.63299036701\\
2.60606060606061	3667.80707697129\\
2.63636363636364	3648.92243919869\\
2.66666666666667	3629.97871103748\\
2.6969696969697	3610.97444509346\\
2.72727272727273	3591.90928976276\\
2.75757575757576	3572.78210810502\\
2.78787878787879	3553.59186553545\\
2.81818181818182	3534.33735429618\\
2.84848484848485	3515.01775231061\\
2.87878787878788	3495.63190359766\\
2.90909090909091	3476.17871459978\\
2.93939393939394	3456.65653501054\\
2.96969696969697	3437.06489359696\\
3	3417.40194940702\\
};
\end{axis}
\end{tikzpicture}%
\vspace*{-0.3in}
\caption{Number of drivers under different trip-based congestion charge. }
\label{figure1_trip}
\end{minipage}
\begin{minipage}[b]{0.005\linewidth}
\hfill
\end{minipage}
\begin{minipage}[b]{0.32\linewidth}
\centering
%
%
\definecolor{mycolor1}{rgb}{0.00000,0.44700,0.74100}%
\begin{tikzpicture}

\pgfplotsset{every axis y label/.style={
at={(-0.45,0.5)},
xshift=32pt,
rotate=90}}

\begin{axis}[%
width=1.794in,
height=1.03in,
at={(1.358in,0.0in)},
scale only axis,
xmin=0,
xmax=3,
xtick={0,1,2,3},
xticklabels={0,1,2,3},
xlabel style={font=\color{white!15!black}},
xlabel={per-trip tax},
ymin=150,
ymax=220,
ylabel style={font=\color{white!15!black}},
ylabel={Passenger arrival},
axis background/.style={fill=white},
legend style={legend cell align=left, align=left, draw=white!15!black}
]
\addplot [color=black, line width=1.0pt]
  table[row sep=crcr]{%
0	208.456289196894\\
0.0303030303030303	208.289112671055\\
0.0606060606060606	208.121523331934\\
0.0909090909090909	207.953509171651\\
0.121212121212121	207.785094022954\\
0.151515151515152	207.616276720102\\
0.181818181818182	207.447038495743\\
0.212121212121212	207.277391134145\\
0.242424242424242	207.107342445751\\
0.272727272727273	206.936858894481\\
0.303030303030303	206.765958158003\\
0.333333333333333	206.594642290514\\
0.363636363636364	206.422912163495\\
0.393939393939394	206.250767863404\\
0.424242424242424	206.078190582081\\
0.454545454545455	205.905183403046\\
0.484848484848485	205.731763957387\\
0.515151515151515	205.55791564656\\
0.545454545454545	205.38363939665\\
0.575757575757576	205.208946096188\\
0.606060606060606	205.033806407476\\
0.636363636363636	204.858240904308\\
0.666666666666667	204.682248766482\\
0.696969696969697	204.505817234518\\
0.727272727272727	204.328933742089\\
0.757575757575758	204.151629741407\\
0.787878787878788	203.973905205787\\
0.818181818181818	203.795713408607\\
0.848484848484849	203.617103345923\\
0.878787878787879	203.438038937015\\
0.909090909090909	203.25852818558\\
0.939393939393939	203.078583947501\\
0.96969696969697	202.898187774106\\
1	202.717349611199\\
1.03030303030303	202.53604246707\\
1.06060606060606	202.354306271866\\
1.09090909090909	202.172109208596\\
1.12121212121212	201.989478637702\\
1.15151515151515	201.806373071061\\
1.18181818181818	201.622823799756\\
1.21212121212121	201.438812547961\\
1.24242424242424	201.254349279923\\
1.27272727272727	201.069416838842\\
1.3030303030303	200.884015501089\\
1.33333333333333	200.698166774303\\
1.36363636363636	200.511861309716\\
1.39393939393939	200.325082903638\\
1.42424242424242	200.137839617865\\
1.45454545454545	199.950116361796\\
1.48484848484848	199.761949048336\\
1.51515151515152	199.573285108601\\
1.54545454545455	199.384164294319\\
1.57575757575758	199.194566774232\\
1.60606060606061	199.004493870116\\
1.63636363636364	198.813947865372\\
1.66666666666667	198.622910944653\\
1.6969696969697	198.431422738262\\
1.72727272727273	198.239426240081\\
1.75757575757576	198.046962031154\\
1.78787878787879	197.854011413978\\
1.81818181818182	197.660577687036\\
1.84848484848485	197.466657479488\\
1.87878787878788	197.272239814682\\
1.90909090909091	197.077360383644\\
1.93939393939394	196.881964460052\\
1.96969696969697	196.686081041383\\
2	196.489712609193\\
2.03030303030303	196.292851070209\\
2.06060606060606	196.095478934545\\
2.09090909090909	195.897634649617\\
2.12121212121212	195.432800140791\\
2.15151515151515	194.378083706048\\
2.18181818181818	193.321115253612\\
2.21212121212121	192.261912909174\\
2.24242424242424	191.20040509905\\
2.27272727272727	190.136596146736\\
2.3030303030303	189.070445874682\\
2.33333333333333	188.001944026521\\
2.36363636363636	186.931045684034\\
2.39393939393939	185.857745933967\\
2.42424242424242	184.782012641274\\
2.45454545454545	183.703802771451\\
2.48484848484848	182.623083151755\\
2.51515151515152	181.539847315898\\
2.54545454545455	180.454040316934\\
2.57575757575758	179.365634311814\\
2.60606060606061	178.274613861095\\
2.63636363636364	177.180914287211\\
2.66666666666667	176.084516398521\\
2.6969696969697	174.985363121421\\
2.72727272727273	173.883441353693\\
2.75757575757576	172.778698930786\\
2.78787878787879	171.671095748932\\
2.81818181818182	170.560577535687\\
2.84848484848485	169.447109071496\\
2.87878787878788	168.330643789579\\
2.90909090909091	167.211133085796\\
2.93939393939394	166.088497733281\\
2.96969696969697	164.962728678435\\
3	163.833739824088\\
};

\end{axis}
\end{tikzpicture}%
\vspace*{-0.3in}
\caption{Passenger arrivals  under different trip-based congestion charge.} 
\label{figure2_trip}
\end{minipage}
\begin{minipage}[b]{0.005\linewidth}
\hfill 
\end{minipage}
\begin{minipage}[b]{0.32\linewidth}
\centering
%
%
\definecolor{mycolor1}{rgb}{0.00000,0.44700,0.74100}%
\begin{tikzpicture}

\pgfplotsset{every axis y label/.style={
at={(-0.44,0.5)},
xshift=32pt,
rotate=90}}

\begin{axis}[%
width=1.794in,
height=1.03in,
at={(1.358in,0.0in)},
scale only axis,
xmin=0,
xmax=3,
xtick={0,1,2,3},
xticklabels={0,1,2,3},
xlabel style={font=\color{white!15!black}},
xlabel={per-trip tax},
ymin=0.52,
ymax=0.60,
ylabel style={font=\color{white!15!black}},
ylabel={occupancy},
axis background/.style={fill=white},
legend style={legend cell align=left, align=left, draw=white!15!black}
]
\addplot [color=black, line width=1.0pt]
  table[row sep=crcr]{%
0	0.597762201599234\\
0.0303030303030303	0.597282811850303\\
0.0606060606060606	0.596802238331974\\
0.0909090909090909	0.596320446610851\\
0.121212121212121	0.595837505028923\\
0.151515151515152	0.595353410243346\\
0.181818181818182	0.594868108437499\\
0.212121212121212	0.594381633403547\\
0.242424242424242	0.593894007538456\\
0.272727272727273	0.593405134675296\\
0.303030303030303	0.592915065505949\\
0.333333333333333	0.592423805920959\\
0.363636363636364	0.591931358419358\\
0.393939393939394	0.591437723249071\\
0.424242424242424	0.590942846476454\\
0.454545454545455	0.590446736943704\\
0.484848484848485	0.589949445209132\\
0.515151515151515	0.589450923675313\\
0.545454545454545	0.588951174997862\\
0.575757575757576	0.588450230400359\\
0.606060606060606	0.587948005755014\\
0.636363636363636	0.587444560058542\\
0.666666666666667	0.586939890958961\\
0.696969696969697	0.586433961867616\\
0.727272727272727	0.585926736749736\\
0.757575757575758	0.585418305796617\\
0.787878787878788	0.584908668931739\\
0.818181818181818	0.584397692163421\\
0.848484848484849	0.583885515991097\\
0.878787878787879	0.583372036950915\\
0.909090909090909	0.582857277994003\\
0.939393939393939	0.58234127598546\\
0.96969696969697	0.581823978022496\\
1	0.581305412625363\\
1.03030303030303	0.580785502393545\\
1.06060606060606	0.580264361829379\\
1.09090909090909	0.579741899695545\\
1.12121212121212	0.579218194449915\\
1.15151515151515	0.578693127122554\\
1.18181818181818	0.578166787442711\\
1.21212121212121	0.577639123003246\\
1.24242424242424	0.577110162377297\\
1.27272727272727	0.576579856366604\\
1.3030303030303	0.576048205763682\\
1.33333333333333	0.575515272243103\\
1.36363636363636	0.574981028996664\\
1.39393939393939	0.574445429559207\\
1.42424242424242	0.573908497048464\\
1.45454545454545	0.573370188191141\\
1.48484848484848	0.572830605969872\\
1.51515151515152	0.572289599639898\\
1.54545454545455	0.571747283191933\\
1.57575757575758	0.57120359975853\\
1.60606060606061	0.57065855312997\\
1.63636363636364	0.570112149854056\\
1.66666666666667	0.569564338844107\\
1.6969696969697	0.569015233742631\\
1.72727272727273	0.568464671080817\\
1.75757575757576	0.567912767227494\\
1.78787878787879	0.567359468566333\\
1.81818181818182	0.566804784555954\\
1.84848484848485	0.56624870553025\\
1.87878787878788	0.565691200012944\\
1.90909090909091	0.565132370350419\\
1.93939393939394	0.564572059611319\\
1.96969696969697	0.564010350947831\\
2	0.563447251475981\\
2.03030303030303	0.562882737987773\\
2.06060606060606	0.562316760329803\\
2.09090909090909	0.561749428752579\\
2.12121212121212	0.561141512204177\\
2.15151515151515	0.560444208576862\\
2.18181818181818	0.559742851350545\\
2.21212121212121	0.559037425819922\\
2.24242424242424	0.558327825813069\\
2.27272727272727	0.557613999999983\\
2.3030303030303	0.556895898121656\\
2.33333333333333	0.556173453866022\\
2.36363636363636	0.555446578107108\\
2.39393939393939	0.554715235833543\\
2.42424242424242	0.553979342632672\\
2.45454545454545	0.553238807765524\\
2.48484848484848	0.552493573240331\\
2.51515151515152	0.551743567180067\\
2.54545454545455	0.550988700841582\\
2.57575757575758	0.550228892216719\\
2.60606060606061	0.549464070606931\\
2.63636363636364	0.548694142462944\\
2.66666666666667	0.547919003270441\\
2.6969696969697	0.547138578291814\\
2.72727272727273	0.546352774700609\\
2.75757575757576	0.545561484318982\\
2.78787878787879	0.544764619531068\\
2.81818181818182	0.543962069304597\\
2.84848484848485	0.543153725199225\\
2.87878787878788	0.542339486795899\\
2.90909090909091	0.541519234776503\\
2.93939393939394	0.540692826962566\\
2.96969696969697	0.539860174902659\\
3	0.539021133748207\\
};

\end{axis}
\end{tikzpicture}%
\vspace*{-0.3in}
\caption{Occupancy rate under different  trip-based congestion charge.}
\label{figure3_trip}
\end{minipage}
\begin{minipage}[b]{0.32\linewidth}
\centering
%
%
\definecolor{mycolor1}{rgb}{0.00000,0.44700,0.74100}%
\definecolor{mycolor2}{rgb}{0.85000,0.32500,0.09800}%
\begin{tikzpicture}

\pgfplotsset{every axis y label/.style={
at={(-0.42,0.5)},
xshift=32pt,
rotate=90}}

\begin{axis}[%
width=1.794in,
height=1.03in,
at={(1.358in,0.0in)},
scale only axis,
xmin=0,
xmax=3,
xtick={0,1,2,3},
xticklabels={0,1,2,3},
xlabel style={font=\color{white!15!black}},
xlabel={per-trip tax},
ymin=4,
ymax=14,
ylabel style={font=\color{white!15!black}},
ylabel={price/payment},
axis background/.style={fill=white},
legend style={at={(0.65,0.0)}, anchor=south west, legend cell align=left, align=left, draw=white!15!black}
]
\addplot [color=black, line width=1.0pt]
  table[row sep=crcr]{%
0	11.6282174100135\\
0.0303030303030303	11.6375382566739\\
0.0606060606060606	11.646872749036\\
0.0909090909090909	11.6562215224015\\
0.121212121212121	11.66558321836\\
0.151515151515152	11.6749578715563\\
0.181818181818182	11.6843464925866\\
0.212121212121212	11.6937483957887\\
0.242424242424242	11.7031631182581\\
0.272727272727273	11.7125924847753\\
0.303030303030303	11.7220354845881\\
0.333333333333333	11.7314919730709\\
0.363636363636364	11.7409618714867\\
0.393939393939394	11.7504451447636\\
0.424242424242424	11.7599427974303\\
0.454545454545455	11.7694546271539\\
0.484848484848485	11.7789796339134\\
0.515151515151515	11.7885186990463\\
0.545454545454545	11.7980717403788\\
0.575757575757576	11.8076381309175\\
0.606060606060606	11.8172194458493\\
0.636363636363636	11.8268145278226\\
0.666666666666667	11.8364233917109\\
0.696969696969697	11.8460467033699\\
0.727272727272727	11.855685115647\\
0.757575757575758	11.8653368839353\\
0.787878787878788	11.8750019817269\\
0.818181818181818	11.8846829186697\\
0.848484848484849	11.8943770018654\\
0.878787878787879	11.9040861599154\\
0.909090909090909	11.9138099271861\\
0.939393939393939	11.923547577592\\
0.96969696969697	11.9333000796086\\
1	11.9430668652828\\
1.03030303030303	11.9528493610592\\
1.06060606060606	11.9626453824736\\
1.09090909090909	11.9724566154568\\
1.12121212121212	11.9822815575125\\
1.15151515151515	11.9921224105985\\
1.18181818181818	12.0019774625206\\
1.21212121212121	12.0118476653818\\
1.24242424242424	12.0217324551972\\
1.27272727272727	12.0316327215709\\
1.3030303030303	12.0415484189174\\
1.33333333333333	12.0514783681784\\
1.36363636363636	12.061423041336\\
1.39393939393939	12.0713832741061\\
1.42424242424242	12.0813586070503\\
1.45454545454545	12.0913498143566\\
1.48484848484848	12.1013549567967\\
1.51515151515152	12.1113768007371\\
1.54545454545455	12.1214132050906\\
1.57575757575758	12.1314651946632\\
1.60606060606061	12.1415326706636\\
1.63636363636364	12.1516154840371\\
1.66666666666667	12.1617145485204\\
1.6969696969697	12.1718277421028\\
1.72727272727273	12.1819580475214\\
1.75757575757576	12.1921032947185\\
1.78787878787879	12.2022644424042\\
1.81818181818182	12.2124412891648\\
1.84848484848485	12.2226339849732\\
1.87878787878788	12.2328430786684\\
1.90909090909091	12.2430666701677\\
1.93939393939394	12.2533076034905\\
1.96969696969697	12.2635643302835\\
2	12.2738366947568\\
2.03030303030303	12.2841250943647\\
2.06060606060606	12.2944304157039\\
2.09090909090909	12.3047506257847\\
2.12121212121212	12.3163786923096\\
2.15151515151515	12.3308617723836\\
2.18181818181818	12.3453465208373\\
2.21212121212121	12.359831824301\\
2.24242424242424	12.3743183148737\\
2.27272727272727	12.3888054753346\\
2.3030303030303	12.4032928171334\\
2.33333333333333	12.4177800371664\\
2.36363636363636	12.4322672514539\\
2.39393939393939	12.4467535157665\\
2.42424242424242	12.4612387547428\\
2.45454545454545	12.4757229715953\\
2.48484848484848	12.4902055252307\\
2.51515151515152	12.5046859395198\\
2.54545454545455	12.5191640398296\\
2.57575757575758	12.5336394458354\\
2.60606060606061	12.5481115114573\\
2.63636363636364	12.562579975219\\
2.66666666666667	12.5770446764671\\
2.6969696969697	12.5915048777697\\
2.72727272727273	12.6059600780723\\
2.75757575757576	12.6204099989822\\
2.78787878787879	12.6348539125674\\
2.81818181818182	12.6492914349662\\
2.84848484848485	12.6637220552638\\
2.87878787878788	12.6781450328597\\
2.90909090909091	12.6925598808141\\
2.93939393939394	12.7069664175224\\
2.96969696969697	12.7213634252656\\
3	12.7357505650028\\
};
\addlegendentry{$p_f$}

\addplot [color=mycolor1, dashed, line width=1.0pt]
  table[row sep=crcr]{%
0	8.35992020624679\\
0.0303030303030303	8.36663002606621\\
0.0606060606060606	8.37336723408913\\
0.0909090909090909	8.38013242054925\\
0.121212121212121	8.38692473283874\\
0.151515151515152	8.39374432345557\\
0.181818181818182	8.40059205864294\\
0.212121212121212	8.40746757107011\\
0.242424242424242	8.41437065242054\\
0.272727272727273	8.42130277557243\\
0.303030303030303	8.42826333551793\\
0.333333333333333	8.43525236112259\\
0.363636363636364	8.4422699297841\\
0.393939393939394	8.44931615154267\\
0.424242424242424	8.45639191247764\\
0.454545454545455	8.46349720475543\\
0.484848484848485	8.47063142149158\\
0.515151515151515	8.47779536338911\\
0.545454545454545	8.48498911254934\\
0.575757575757576	8.49221233931724\\
0.606060606060606	8.49946637928091\\
0.636363636363636	8.50675050456165\\
0.666666666666667	8.51406487215471\\
0.696969696969697	8.52141013758017\\
0.727272727272727	8.52878695278665\\
0.757575757575758	8.53619413366298\\
0.787878787878788	8.54363180632428\\
0.818181818181818	8.55110205719734\\
0.848484848484849	8.55860296379779\\
0.878787878787879	8.5661361723796\\
0.909090909090909	8.57370148122508\\
0.939393939393939	8.58129848210308\\
0.96969696969697	8.58892808898087\\
1	8.59659001816418\\
1.03030303030303	8.60428555307469\\
1.06060606060606	8.6120131381589\\
1.09090909090909	8.61977426559014\\
1.12121212121212	8.62756791061424\\
1.15151515151515	8.63539598703701\\
1.18181818181818	8.64325730259137\\
1.21212121212121	8.65115278497492\\
1.24242424242424	8.6590821535611\\
1.27272727272727	8.66704629462918\\
1.3030303030303	8.67504534807991\\
1.33333333333333	8.68307853621844\\
1.36363636363636	8.69114641295234\\
1.39393939393939	8.69924983390428\\
1.42424242424242	8.70738860529191\\
1.45454545454545	8.7155635409738\\
1.48484848484848	8.72377323348333\\
1.51515151515152	8.73202013600181\\
1.54545454545455	8.74030267320475\\
1.57575757575758	8.74862187456896\\
1.60606060606061	8.75697784650896\\
1.63636363636364	8.76537065375515\\
1.66666666666667	8.77380124925233\\
1.6969696969697	8.78226805073603\\
1.72727272727273	8.79077374883963\\
1.75757575757576	8.79931671914361\\
1.78787878787879	8.80789796336279\\
1.81818181818182	8.81651750980708\\
1.84848484848485	8.82517568495005\\
1.87878787878788	8.83387315829847\\
1.90909090909091	8.84260851060679\\
1.93939393939394	8.85138437619525\\
1.96969696969697	8.86019963868044\\
2	8.86905436948968\\
2.03030303030303	8.87794911875331\\
2.06060606060606	8.88688486672366\\
2.09090909090909	8.89586006126768\\
2.12121212121212	8.90440358777023\\
2.15151515151515	8.91196509147983\\
2.18181818181818	8.91960339770204\\
2.21212121212121	8.92731921699906\\
2.24242424242424	8.93511447716041\\
2.27272727272727	8.94299049123192\\
2.3030303030303	8.95094837133274\\
2.33333333333333	8.9589896527459\\
2.36363636363636	8.96711613493011\\
2.39393939393939	8.97532885309378\\
2.42424242424242	8.98362960687613\\
2.45454545454545	8.99202027650088\\
2.48484848484848	9.00050222169665\\
2.51515151515152	9.00907717932082\\
2.54545454545455	9.01774699012173\\
2.57575757575758	9.02651351143581\\
2.60606060606061	9.03537847079536\\
2.63636363636364	9.044343805284\\
2.66666666666667	9.05341186836279\\
2.6969696969697	9.06258433380978\\
2.72727272727273	9.07186341306353\\
2.75757575757576	9.08125144391403\\
2.78787878787879	9.0907504695105\\
2.81818181818182	9.10036291614322\\
2.84848484848485	9.11009127281761\\
2.87878787878788	9.11993785822898\\
2.90909090909091	9.12990535172737\\
2.93939393939394	9.13999674276024\\
2.96969696969697	9.15021438194312\\
3	9.160561337321\\
};
\addlegendentry{$p_d$}

\end{axis}
\end{tikzpicture}%
\vspace*{-0.3in}
\caption{Per-trip ride price and driver payment under different trip-based congestion charge.} 
\label{figure4_trip}
\end{minipage}
\begin{minipage}[b]{0.005\linewidth}
\hfill
\end{minipage}
\begin{minipage}[b]{0.32\linewidth}
\centering
%
%
\definecolor{mycolor1}{rgb}{0.00000,0.44700,0.74100}%
\begin{tikzpicture}

\pgfplotsset{every axis y label/.style={
at={(-0.42,0.5)},
xshift=32pt,
rotate=90}}

\begin{axis}[%
width=1.794in,
height=1.03in,
at={(1.358in,0.0in)},
scale only axis,
xmin=0,
xmax=3,
xtick={0,1,2,3},
xticklabels={0,1,2,3},
xlabel style={font=\color{white!15!black}},
xlabel={per-trip tax},
ymin=4.2,
ymax=4.7,
ylabel style={font=\color{white!15!black}},
ylabel={pickup time},
axis background/.style={fill=white},
legend style={legend cell align=left, align=left, draw=white!15!black}
]
\addplot [color=black, line width=1.0pt]
  table[row sep=crcr]{%
0	4.5113226177619\\
0.0303030303030303	4.508636705822\\
0.0606060606060606	4.50594897088109\\
0.0909090909090909	4.50325924207303\\
0.121212121212121	4.50056792207342\\
0.151515151515152	4.49787501235266\\
0.181818181818182	4.49518023443258\\
0.212121212121212	4.49248379708062\\
0.242424242424242	4.48978584465945\\
0.272727272727273	4.48708586635222\\
0.303030303030303	4.48438416399627\\
0.333333333333333	4.48168079102637\\
0.363636363636364	4.47897578196405\\
0.393939393939394	4.47626915883364\\
0.424242424242424	4.47356064731129\\
0.454545454545455	4.470850317793\\
0.484848484848485	4.46813846753639\\
0.515151515151515	4.46542485791879\\
0.545454545454545	4.46270952493447\\
0.575757575757576	4.45999265920068\\
0.606060606060606	4.45727382613659\\
0.636363636363636	4.45455336711559\\
0.666666666666667	4.45183129050797\\
0.696969696969697	4.44910742055095\\
0.727272727272727	4.44638158583286\\
0.757575757575758	4.44365429276609\\
0.787878787878788	4.44092556125877\\
0.818181818181818	4.43819469611709\\
0.848484848484849	4.43546247029369\\
0.878787878787879	4.43272835330399\\
0.909090909090909	4.42999248992262\\
0.939393939393939	4.42725509769379\\
0.96969696969697	4.42451591774503\\
1	4.4217751231611\\
1.03030303030303	4.4190323275667\\
1.06060606060606	4.41628815711498\\
1.09090909090909	4.41354215305559\\
1.12121212121212	4.41079474954899\\
1.15151515151515	4.40804534478471\\
1.18181818181818	4.405294431284\\
1.21212121212121	4.4025417571912\\
1.24242424242424	4.3997874938905\\
1.27272727272727	4.39703140776381\\
1.3030303030303	4.39427352611842\\
1.33333333333333	4.39151419108294\\
1.36363636363636	4.38875328546715\\
1.39393939393939	4.38599059175441\\
1.42424242424242	4.38322625206437\\
1.45454545454545	4.38046006670197\\
1.48484848484848	4.37769258700496\\
1.51515151515152	4.37492306257858\\
1.54545454545455	4.3721521001894\\
1.57575757575758	4.36937943158733\\
1.60606060606061	4.36660509911158\\
1.63636363636364	4.36382915887214\\
1.66666666666667	4.36105137467907\\
1.6969696969697	4.35827234514252\\
1.72727272727273	4.35549126595164\\
1.75757575757576	4.35270874853756\\
1.78787878787879	4.34992454520666\\
1.81818181818182	4.34713872673787\\
1.84848484848485	4.34435126769059\\
1.87878787878788	4.34156203390439\\
1.90909090909091	4.33877155994032\\
1.93939393939394	4.33597908539685\\
1.96969696969697	4.33318504889571\\
2	4.33038950865653\\
2.03030303030303	4.32759237243591\\
2.06060606060606	4.32479341567535\\
2.09090909090909	4.3219932066038\\
2.12121212121212	4.32152618124085\\
2.15151515151515	4.32625170386345\\
2.18181818181818	4.33103590818659\\
2.21212121212121	4.3358797719104\\
2.24242424242424	4.34078444677008\\
2.27272727272727	4.34575080761424\\
2.3030303030303	4.35078023050977\\
2.33333333333333	4.35587368246876\\
2.36363636363636	4.36103233794517\\
2.39393939393939	4.36625744743878\\
2.42424242424242	4.37155016987785\\
2.45454545454545	4.37691177178285\\
2.48484848484848	4.3823437384861\\
2.51515151515152	4.38784719015231\\
2.54545454545455	4.39342368069486\\
2.57575757575758	4.39907457501159\\
2.60606060606061	4.40480124412952\\
2.63636363636364	4.41060547055077\\
2.66666666666667	4.41648848707385\\
2.6969696969697	4.42245227059409\\
2.72727272727273	4.42849819815224\\
2.75757575757576	4.43462802658879\\
2.78787878787879	4.4408436161793\\
2.81818181818182	4.44714686253146\\
2.84848484848485	4.45353952901893\\
2.87878787878788	4.46002365646573\\
2.90909090909091	4.46660122490694\\
2.93939393939394	4.47327447595706\\
2.96969696969697	4.48004533772495\\
3	4.48691628984236\\
};

\end{axis}
\end{tikzpicture}%
\vspace*{-0.3in}
\caption{Passenger pickup time in minutes under different trip-based congestion charge.}
\label{figure5_trip}
\end{minipage}
\begin{minipage}[b]{0.005\linewidth}
\hfill
\end{minipage}
\begin{minipage}[b]{0.32\linewidth}
\centering
%
%
\definecolor{mycolor1}{rgb}{0.00000,0.44700,0.74100}%
\begin{tikzpicture}

\pgfplotsset{every axis y label/.style={
at={(-0.45,0.5)},
xshift=32pt,
rotate=90}}

\begin{axis}[%
width=1.794in,
height=1.03in,
at={(1.358in,0.0in)},
scale only axis,
xmin=0,
xmax=3,
xtick={0,1,2,3},
xticklabels={0,1,2,3},
xlabel style={font=\color{white!15!black}},
xlabel={per-trip tax},
ymin=35,
ymax=37,
ylabel style={font=\color{white!15!black}},
ylabel={passenger cost},
axis background/.style={fill=white},
legend style={legend cell align=left, align=left, draw=white!15!black}
]
\addplot [color=black, line width=1.0pt]
  table[row sep=crcr]{%
0	35.4300486745168\\
0.0303030303030303	35.4331023933174\\
0.0606060606060606	35.4361655041508\\
0.0909090909090909	35.4392382436307\\
0.121212121212121	35.4423201929234\\
0.151515151515152	35.4454113901047\\
0.181818181818182	35.4485121959882\\
0.212121212121212	35.4516224120356\\
0.242424242424242	35.454741912189\\
0.272727272727273	35.4578713293226\\
0.303030303030303	35.4610103569715\\
0.333333333333333	35.4641589751913\\
0.363636363636364	35.467317185795\\
0.393939393939394	35.4704850051009\\
0.424242424242424	35.4736627975489\\
0.454545454545455	35.476850525063\\
0.484848484848485	35.4800478812239\\
0.515151515151515	35.4832551905823\\
0.545454545454545	35.4864724549514\\
0.575757575757576	35.4896994921113\\
0.606060606060606	35.4929368632268\\
0.636363636363636	35.4961842074844\\
0.666666666666667	35.4994415592883\\
0.696969696969697	35.5027091743809\\
0.727272727272727	35.5059873056492\\
0.757575757575758	35.509275390115\\
0.787878787878788	35.5125734477228\\
0.818181818181818	35.5158823660017\\
0.848484848484849	35.5192012556095\\
0.878787878787879	35.5225308073502\\
0.909090909090909	35.5258708933978\\
0.939393939393939	35.5292212952697\\
0.96969696969697	35.5325823774058\\
1	35.5359539757175\\
1.03030303030303	35.539336615107\\
1.06060606060606	35.542729572134\\
1.09090909090909	35.5461334623121\\
1.12121212121212	35.5495477961856\\
1.15151515151515	35.5529733714883\\
1.18181818181818	35.5564096252421\\
1.21212121212121	35.5598569218868\\
1.24242424242424	35.5633150973338\\
1.27272727272727	35.5667844960786\\
1.3030303030303	35.5702651362525\\
1.33333333333333	35.5737566370974\\
1.36363636363636	35.5772591971515\\
1.39393939393939	35.5807731445919\\
1.42424242424242	35.5842983515927\\
1.45454545454545	35.5878351263867\\
1.48484848484848	35.5913828162004\\
1.51515151515152	35.5949424364793\\
1.54545454545455	35.5985132619246\\
1.57575757575758	35.6020956914258\\
1.60606060606061	35.6056897249828\\
1.63636363636364	35.6092953444643\\
1.66666666666667	35.612912912497\\
1.6969696969697	35.6165417038275\\
1.72727272727273	35.6201828244674\\
1.75757575757576	35.6238355310317\\
1.78787878787879	35.6275002042786\\
1.81818181818182	35.6311768079454\\
1.84848484848485	35.6348654326434\\
1.87878787878788	35.6385663141708\\
1.90909090909091	35.642278799754\\
1.93939393939394	35.6460039591419\\
1.96969696969697	35.6497412674324\\
2	35.6534907046809\\
2.03030303030303	35.6572524531074\\
2.06060606060606	35.6610268753386\\
2.09090909090909	35.6648132642524\\
2.12121212121212	35.6737209817289\\
2.15151515151515	35.6939935056539\\
2.18181818181818	35.714394761212\\
2.21212121212121	35.7349259467941\\
2.24242424242424	35.7555900351236\\
2.27272727272727	35.7763885672733\\
2.3030303030303	35.7973239910645\\
2.33333333333333	35.818398210197\\
2.36363636363636	35.8396138547119\\
2.39393939393939	35.8609728083644\\
2.42424242424242	35.8824775389202\\
2.45454545454545	35.9041307832594\\
2.48484848484848	35.9259351341179\\
2.51515151515152	35.9478926948213\\
2.54545454545455	35.9700066029071\\
2.57575757575758	35.9922795047344\\
2.60606060606061	36.0147138470367\\
2.63636363636364	36.037313149197\\
2.66666666666667	36.0600800583001\\
2.6969696969697	36.0830180719715\\
2.72727272727273	36.1061298391089\\
2.75757575757576	36.1294188936006\\
2.78787878787879	36.1528885914665\\
2.81818181818182	36.1765426669008\\
2.84848484848485	36.2003845295468\\
2.87878787878788	36.2244179125562\\
2.90909090909091	36.2486466798076\\
2.93939393939394	36.2730754537948\\
2.96969696969697	36.2977074257238\\
3	36.3225475643969\\
};

\end{axis}
\end{tikzpicture}%
\vspace*{-0.3in}
\caption{Passenger travel cost in \$ per trip under different trip-based congestion charge.}
\label{figure6_trip}
\end{minipage}

\begin{minipage}[b]{0.32\linewidth}
\centering
%
%
\definecolor{mycolor1}{rgb}{0.00000,0.44700,0.74100}%
\begin{tikzpicture}

\pgfplotsset{every axis y label/.style={
at={(-0.42,0.5)},
xshift=32pt,
rotate=90}}

\begin{axis}[%
width=1.794in,
height=1.03in,
at={(1.358in,0.0in)},
scale only axis,
xmin=0,
xmax=3,
xtick={0,1,2,3},
xticklabels={0,1,2,3},
xlabel style={font=\color{white!15!black}},
xlabel={per-trip tax},
ymin=26.349999999997,
ymax=26.350000000003,
ylabel style={font=\color{white!15!black}},
ylabel={driver revenue},
axis background/.style={fill=white},
legend style={legend cell align=left, align=left, draw=white!15!black}
]
\addplot [color=black, line width=1.0pt]
  table[row sep=crcr]{%
0	26.35\\
0.0505050505050505	26.35\\
0.101010101010101	26.35\\
0.151515151515152	26.35\\
0.202020202020202	26.35\\
0.252525252525253	26.35\\
0.303030303030303	26.35\\
0.353535353535354	26.35\\
0.404040404040404	26.35\\
0.454545454545455	26.35\\
0.505050505050505	26.35\\
0.555555555555556	26.35\\
0.606060606060606	26.35\\
0.656565656565657	26.35\\
0.707070707070707	26.35\\
0.757575757575758	26.35\\
0.808080808080808	26.35\\
0.858585858585859	26.35\\
0.909090909090909	26.35\\
0.95959595959596	26.35\\
1.01010101010101	26.35\\
1.06060606060606	26.35\\
1.11111111111111	26.35\\
1.16161616161616	26.35\\
1.21212121212121	26.35\\
1.26262626262626	26.35\\
1.31313131313131	26.35\\
1.36363636363636	26.35\\
1.41414141414141	26.35\\
1.46464646464646	26.35\\
1.51515151515152	26.35\\
1.56565656565657	26.35\\
1.61616161616162	26.35\\
1.66666666666667	26.35\\
1.71717171717172	26.35\\
1.76767676767677	26.35\\
1.81818181818182	26.35\\
1.86868686868687	26.35\\
1.91919191919192	26.35\\
1.96969696969697	26.35\\
2.02020202020202	26.35\\
2.07070707070707	26.35\\
2.12121212121212	26.35\\
2.17171717171717	26.35\\
2.22222222222222	26.35\\
2.27272727272727	26.35\\
2.32323232323232	26.35\\
2.37373737373737	26.35\\
2.42424242424242	26.35\\
2.47474747474747	26.35\\
2.52525252525253	26.35\\
2.57575757575758	26.35\\
2.62626262626263	26.35\\
2.67676767676768	26.35\\
2.72727272727273	26.35\\
2.77777777777778	26.35\\
2.82828282828283	26.35\\
2.87878787878788	26.35\\
2.92929292929293	26.35\\
2.97979797979798	26.35\\
3.03030303030303	26.35\\
3.08080808080808	26.35\\
3.13131313131313	26.35\\
3.18181818181818	26.35\\
3.23232323232323	26.35\\
3.28282828282828	26.35\\
3.33333333333333	26.35\\
3.38383838383838	26.35\\
3.43434343434343	26.35\\
3.48484848484848	26.35\\
3.53535353535354	26.35\\
3.58585858585859	26.35\\
3.63636363636364	26.35\\
3.68686868686869	26.35\\
3.73737373737374	26.35\\
3.78787878787879	26.35\\
3.83838383838384	26.35\\
3.88888888888889	26.35\\
3.93939393939394	26.35\\
3.98989898989899	26.35\\
4.04040404040404	26.35\\
4.09090909090909	26.35\\
4.14141414141414	26.35\\
4.19191919191919	26.35\\
4.24242424242424	26.35\\
4.29292929292929	26.35\\
4.34343434343434	26.35\\
4.39393939393939	26.35\\
4.44444444444444	26.35\\
4.49494949494949	26.35\\
4.54545454545455	26.35\\
4.5959595959596	26.35\\
4.64646464646465	26.35\\
4.6969696969697	26.35\\
4.74747474747475	26.35\\
4.7979797979798	26.35\\
4.84848484848485	26.35\\
4.8989898989899	26.35\\
4.94949494949495	26.35\\
5	26.35\\
};

\end{axis}
\end{tikzpicture}%
\vspace*{-0.3in}
\caption{Per-hour driver wage under different trip-based congestion charge.}
\label{figure7_trip}
\end{minipage}
\begin{minipage}[b]{0.005\linewidth}
\hfill
\end{minipage}
\begin{minipage}[b]{0.32\linewidth}
\centering
%
%
\definecolor{mycolor1}{rgb}{0.00000,0.44700,0.74100}%
\begin{tikzpicture}

\begin{axis}[%
width=1.794in,
height=1.03in,
at={(1.358in,0.0in)},
scale only axis,
xmin=0,
xmax=3,
xtick={0,1,2,3},
xticklabels={0,1,2,3},
xlabel style={font=\color{white!15!black}},
xlabel={per-trip tax},
ymin=5000,
ymax=45000,
ylabel style={font=\color{white!15!black}},
ylabel={platform revenue},
axis background/.style={fill=white},
legend style={legend cell align=left, align=left, draw=white!15!black}
]
\addplot [color=black, line width=1.0pt]
  table[row sep=crcr]{%
0	40877.8264253881\\
0.0303030303030303	40498.9669013202\\
0.0606060606060606	40120.4117249855\\
0.0909090909090909	39742.1616371414\\
0.121212121212121	39364.2173808241\\
0.151515151515152	38986.5797013587\\
0.181818181818182	38609.2493463613\\
0.212121212121212	38232.2270657481\\
0.242424242424242	37855.5136117395\\
0.272727272727273	37479.1097388677\\
0.303030303030303	37103.0162039799\\
0.333333333333333	36727.2337662471\\
0.363636363636364	36351.7631871691\\
0.393939393939394	35976.605230579\\
0.424242424242424	35601.760662651\\
0.454545454545455	35227.2302519045\\
0.484848484848485	34853.0147692124\\
0.515151515151515	34479.1149878038\\
0.545454545454545	34105.5316832716\\
0.575757575757576	33732.2656335783\\
0.606060606060606	33359.3176190616\\
0.636363636363636	32986.6884224391\\
0.666666666666667	32614.3788288149\\
0.696969696969697	32242.3896256859\\
0.727272727272727	31870.7216029458\\
0.757575757575758	31499.3755528921\\
0.787878787878788	31128.3522702312\\
0.818181818181818	30757.6525520831\\
0.848484848484849	30387.2771979884\\
0.878787878787879	30017.227009913\\
0.909090909090909	29647.5027922537\\
0.939393939393939	29278.1053518433\\
0.96969696969697	28909.0354979568\\
1	28540.2940423157\\
1.03030303030303	28171.8817990948\\
1.06060606060606	27803.7995849261\\
1.09090909090909	27436.0482189035\\
1.12121212121212	27068.6285225915\\
1.15151515151515	26701.5413200257\\
1.18181818181818	26334.7874377214\\
1.21212121212121	25968.3677046773\\
1.24242424242424	25602.2829523808\\
1.27272727272727	25236.5340148137\\
1.3030303030303	24871.1217284552\\
1.33333333333333	24506.0469322899\\
1.36363636363636	24141.3104678097\\
1.39393939393939	23776.9131790202\\
1.42424242424242	23412.8559124457\\
1.45454545454545	23049.1395171336\\
1.48484848484848	22685.7648446584\\
1.51515151515152	22322.7327491286\\
1.54545454545455	21960.0440871877\\
1.57575757575758	21597.699718022\\
1.60606060606061	21235.7005033641\\
1.63636363636364	20874.0473074968\\
1.66666666666667	20512.7409972582\\
1.6969696969697	20151.7824420458\\
1.72727272727273	19791.1725138213\\
1.75757575757576	19430.9120871128\\
1.78787878787879	19071.0020390221\\
1.81818181818182	18711.4432492266\\
1.84848484848485	18352.2365999841\\
1.87878787878788	17993.3829761365\\
1.90909090909091	17634.8832651149\\
1.93939393939394	17276.7383569417\\
1.96969696969697	16918.9491442347\\
2	16561.5165222127\\
2.03030303030303	16204.4413886969\\
2.06060606060606	15847.7246441156\\
2.09090909090909	15491.3671915082\\
2.12121212121212	15135.445448984\\
2.15151515151515	14781.0715804206\\
2.18181818181818	14428.6174168321\\
2.21212121212121	14078.0870481204\\
2.24242424242424	13729.4846060128\\
2.27272727272727	13382.814265349\\
2.3030303030303	13038.0802454206\\
2.33333333333333	12695.2868113649\\
2.36363636363636	12354.4382756184\\
2.39393939393939	12015.5389994299\\
2.42424242424242	11678.593394438\\
2.45454545454545	11343.6059243175\\
2.48484848484848	11010.5811064952\\
2.51515151515152	10679.5235139412\\
2.54545454545455	10350.4377770407\\
2.57575757575758	10023.3285855467\\
2.60606060606061	9698.20069062258\\
2.63636363636364	9375.05890697717\\
2.66666666666667	9053.90811509731\\
2.6969696969697	8734.75326358504\\
2.72727272727273	8417.5993716045\\
2.75757575757576	8102.45153144492\\
2.78787878787879	7789.31491120695\\
2.81818181818182	7478.19475761962\\
2.84848484848485	7169.09639899628\\
2.87878787878788	6862.02524833731\\
2.90909090909091	6556.986806589\\
2.93939393939394	6253.98666607005\\
2.96969696969697	5953.03051407354\\
3	5654.12413665825\\
};

\end{axis}
\end{tikzpicture}%
\vspace*{-0.3in}
\caption{Per-hour platform profit under different trip-based congestion charge.}
\label{figure8_trip}
\end{minipage}
\begin{minipage}[b]{0.005\linewidth}
\hfill
\end{minipage}
\begin{minipage}[b]{0.32\linewidth}
\centering
%
%
\definecolor{mycolor1}{rgb}{0.00000,0.44700,0.74100}%
\begin{tikzpicture}

\begin{axis}[%
width=1.794in,
height=1.03in,
at={(1.358in,0.0in)},
scale only axis,
xmin=0,
xmax=3,
xtick={0,1,2,3},
xticklabels={0,1,2,3},
xlabel style={font=\color{white!15!black}},
xlabel={per-trip tax},
ymin=0,
ymax=32000,
ylabel style={font=\color{white!15!black}},
ylabel={tax revenue},
axis background/.style={fill=white},
legend style={legend cell align=left, align=left, draw=white!15!black}
]
\addplot [color=black, line width=1.0pt]
  table[row sep=crcr]{%
0	0\\
0.0303030303030303	378.707477583736\\
0.0606060606060606	756.805539388852\\
0.0909090909090909	1134.291868209\\
0.121212121212121	1511.16432016694\\
0.151515151515152	1887.42069745547\\
0.181818181818182	2263.05860177174\\
0.212121212121212	2638.07588716184\\
0.242424242424242	3012.47043557456\\
0.272727272727273	3386.23950918242\\
0.303030303030303	3759.38105741824\\
0.333333333333333	4131.89284581028\\
0.363636363636364	4503.7726290217\\
0.393939393939394	4875.01814949863\\
0.424242424242424	5245.62666936206\\
0.454545454545455	5615.59591099217\\
0.484848484848485	5984.92404239672\\
0.515151515151515	6353.60830180276\\
0.545454545454545	6721.646380254\\
0.575757575757576	7089.0363196865\\
0.606060606060606	7455.77477845368\\
0.636363636363636	7821.8601072554\\
0.666666666666667	8187.28995065928\\
0.696969696969697	8552.06144798892\\
0.727272727272727	8916.17165420025\\
0.757575757575758	9279.61953370031\\
0.787878787878788	9642.4027915463\\
0.818181818181818	10004.5168400589\\
0.848484848484849	10365.9616248833\\
0.878787878787879	10726.7329621335\\
0.909090909090909	11086.8288101225\\
0.939393939393939	11446.2474588592\\
0.96969696969697	11804.9854704934\\
1	12163.040976672\\
1.03030303030303	12520.4098979644\\
1.06060606060606	12877.0922173005\\
1.09090909090909	13233.0835118354\\
1.12121212121212	13588.3831083545\\
1.15151515151515	13942.9857758188\\
1.18181818181818	14296.8911421645\\
1.21212121212121	14650.0954580335\\
1.24242424242424	15002.5969463215\\
1.27272727272727	15354.3918313298\\
1.3030303030303	15705.4775755397\\
1.33333333333333	16055.8533419442\\
1.36363636363636	16405.5159253404\\
1.39393939393939	16754.4614792134\\
1.42424242424242	17102.6881127994\\
1.45454545454545	17450.1919733931\\
1.48484848484848	17796.9736424881\\
1.51515151515152	18143.0259189637\\
1.54545454545455	18488.3497800187\\
1.57575757575758	18832.9408586547\\
1.60606060606061	19176.7966820294\\
1.63636363636364	19519.9148813274\\
1.66666666666667	19862.2910944653\\
1.6969696969697	20203.9266788049\\
1.72727272727273	20544.8132648811\\
1.75757575757576	20884.952359649\\
1.78787878787879	21224.3394062267\\
1.81818181818182	21562.972111313\\
1.84848484848485	21900.8474659068\\
1.87878787878788	22237.9615791096\\
1.90909090909091	22574.3158257629\\
1.93939393939394	22909.9013189879\\
1.96969696969697	23244.7186685271\\
2	23578.7655131032\\
2.03030303030303	23912.03822128\\
2.06060606060606	24244.5319409983\\
2.09090909090909	24576.2487105883\\
2.12121212121212	24873.2654724643\\
2.15151515151515	25092.4435329626\\
2.18181818181818	25307.4914513819\\
2.21212121212121	25518.3993497632\\
2.24242424242424	25725.1454133268\\
2.27272727272727	25927.7176563731\\
2.3030303030303	26126.0979754106\\
2.33333333333333	26320.272163713\\
2.36363636363636	26510.2210242812\\
2.39393939393939	26695.9307796062\\
2.42424242424242	26877.3836569126\\
2.45454545454545	27054.5600445229\\
2.48484848484848	27227.4414880799\\
2.51515151515152	27396.0133222173\\
2.54545454545455	27560.2534302226\\
2.57575757575758	27720.1434845531\\
2.60606060606061	27875.6668946439\\
2.63636363636364	28026.7991690679\\
2.66666666666667	28173.5226237634\\
2.6969696969697	28315.8133051028\\
2.72727272727273	28453.6540396952\\
2.75757575757576	28587.021095821\\
2.78787878787879	28715.8923798213\\
2.81818181818182	28840.2431105799\\
2.84848484848485	28960.0513685829\\
2.87878787878788	29075.2930182\\
2.90909090909091	29185.9432295207\\
2.93939393939394	29291.971418415\\
2.96969696969697	29393.3589281575\\
3	29490.0731683358\\
};

\end{axis}
\end{tikzpicture}%
\vspace*{-0.3in}
\caption{Per-hour tax revenue  under different trip-based congestion charge.}
\label{figure9_trip}
\end{minipage}
\end{figure*}

\subsection{Analysis}
\label{trip_analysis_sec}
Figure \ref{figure1_trip}- Figure \ref{figure3_trip} show the number of drivers, passenger arrival rate, and the occupancy rate of  TNC vehicles  as a function of the congestion charge $p_t$  when the minimum wage is set at $w_0=\$26.35$/hour. Figure \ref{figure4_trip} shows the per-trip ride fare $p_f$ and the driver payment $p_d$. Figure \ref{figure5_trip}-Figure \ref{figure6_trip} show the passenger pickup time and travel cost. Figure \ref{figure7_trip} shows the driver wage (which equals the minimum wage). Figure \ref{figure8_trip} and Figure \ref{figure9_trip} show the platform profit and city's tax revenue under different values of ${p_t}$, respectively.

Clearly, the optimal solution as a function of $p_t$ has two distinct regimes: 
\begin{itemize}
    \item when ${p_t} \leq \$2.1$/trip,  the number of drivers remains constant, while the number of passengers reduces; vehicle occupancy drops,  passenger pickup time decreases,  ride fare increases, and the passenger total travel cost increases. At the same time, driver wage remains constant and equals the minimum wage,  platform profit reduces, and the tax revenue increases. 
    \item when ${p_t} > \$2.1$/trip, both the passenger arrival rate and number of TNC drivers reduce sharply;  vehicle occupancy reduces,  ride fare and pickup time increase, and the total travel cost increases. The driver wage remains constant and equals the minimum wage, while the platform revenue declines, and the tax revenue increases. 
\end{itemize}

This is a surprising result: the number of drivers is unaffected by the congestion charge $p_t$ when $p_t\leq \$2.1$/trip. It is in  contrast with the case when there is only a congestion charge and no minimum wage (see \cite{li2019regulating}).  Therefore, this set of result indicates that the effect of a congestion charge on congestion relief is mitigated by the wage floor on TNC drivers. In certain regimes, the congestion charge cannot directly curb traffic congestion by reducing the number of TNC vehicles.  

The  reason behind this surprising result is rooted in the platform's   power in the labor market. The platform is a monopoly in the labor market and sets driver wages. When there is no regulation (i.e., ${p_t} = 0$ and ${w_0} =0$), the platform hires fewer drivers to maximize its profit compared to a competitive labor market where the TNC faces the competitive driver wage. In a certain regime, the minimum wage squeezes the platform's market power and induces it to hire more drivers \cite{li2019regulating}. This indicates that the marginal profit of hiring additional drivers under the minimum wage regulation is positive. When the  congestion charge is insignificant, this marginal profit reduces but remains positive, and thus the platform still hires all drivers available in the labor market. The number of drivers is upper bounded by $N\leq N_0F_d(w_0)$. Therefore, in the first regime, $N$ remains constant and satisfies $N=N_0F_d(w_0)$. If the congestion charge is further increased, the marginal profit of hiring an additional driver reduces to zero, and the system enters the second regime. 

{
\begin{remark}
We would like to clarify that the aforementioned result relies on the assumption that the TNC platform has market power that can influence the driver wage\footnote{In a competitive labor market where driver wage is given, the conclusions of this numerical study  no longer hold. }, but does not rely on the assumption that TNC is a monopolistic wage-setter. To validate this, we considered the duopolisitic setting, where two symmetric TNCs compete against each other on both passenger and driver side to maximize their own profits. The numerical study reveals that the number of drivers and number of passengers at the Nash equilibrium demonstrate the same properties as shown in Figure \ref{figure1_trip} and Figure \ref{figure2_trip}, respectively. We believe that this can be further extended to the case of more than two competing TNCs. 
\end{remark}

}

\begin{figure}[ht!]
  \centering
%
%
\definecolor{mycolor1}{rgb}{0.00000,0.44700,0.74100}%
\definecolor{mycolor2}{rgb}{0.85000,0.32500,0.09800}%

\begin{tikzpicture}

\begin{axis}[%
width=2in,
height=1.23in,
at={(1.358in,0.0in)},
scale only axis,
xmin=0,
xmax=3,
xtick={0,1,2,3},
xticklabels={0,1,2,3},
xlabel style={font=\color{white!15!black}},
xlabel={per-trip tax},
ymin=0,
ymax=4,
ylabel style={font=\color{white!15!black}},
ylabel={Elasticity},
axis background/.style={fill=white},
legend style={at={(0.35,0.0)}, anchor=south west, legend cell align=left, align=left, draw=white!15!black}
]
\addplot [color=mycolor1, dashed, line width=1.0pt]
  table[row sep=crcr]{%
0	1.41780320732913\\
0.0303030303030303	1.41780320732913\\
0.0606060606060606	1.41780320732913\\
0.0909090909090909	1.41780320732913\\
0.121212121212121	1.41780320732913\\
0.151515151515152	1.41780320732913\\
0.181818181818182	1.41780320732913\\
0.212121212121212	1.41780320732913\\
0.242424242424242	1.41780320732913\\
0.272727272727273	1.41780320732913\\
0.303030303030303	1.41780320732913\\
0.333333333333333	1.41780320732913\\
0.363636363636364	1.41780320732913\\
0.393939393939394	1.41780320732913\\
0.424242424242424	1.41780320732913\\
0.454545454545455	1.41780320732913\\
0.484848484848485	1.41780320732913\\
0.515151515151515	1.41780320732913\\
0.545454545454545	1.41780320732913\\
0.575757575757576	1.41780320732913\\
0.606060606060606	1.41780320732913\\
0.636363636363636	1.41780320732913\\
0.666666666666667	1.41780320732913\\
0.696969696969697	1.41780320732913\\
0.727272727272727	1.41780320732913\\
0.757575757575758	1.41780320732913\\
0.787878787878788	1.41780320732913\\
0.818181818181818	1.41780320732913\\
0.848484848484849	1.41780320732913\\
0.878787878787879	1.41780320732913\\
0.909090909090909	1.41780320732913\\
0.939393939393939	1.41780320732913\\
0.96969696969697	1.41780320732913\\
1	1.41780320732913\\
1.03030303030303	1.41780320732913\\
1.06060606060606	1.41780320732913\\
1.09090909090909	1.41780320732913\\
1.12121212121212	1.41780320732913\\
1.15151515151515	1.41780320732913\\
1.18181818181818	1.41780320732913\\
1.21212121212121	1.41780320732913\\
1.24242424242424	1.41780320732913\\
1.27272727272727	1.41780320732913\\
1.3030303030303	1.41780320732913\\
1.33333333333333	1.41780320732913\\
1.36363636363636	1.41780320732913\\
1.39393939393939	1.41780320732913\\
1.42424242424242	1.41780320732913\\
1.45454545454545	1.41780320732913\\
1.48484848484848	1.41780320732913\\
1.51515151515152	1.41780320732913\\
1.54545454545455	1.41780320732913\\
1.57575757575758	1.41780320732913\\
1.60606060606061	1.41780320732913\\
1.63636363636364	1.41780320732913\\
1.66666666666667	1.41780320732913\\
1.6969696969697	1.41780320732913\\
1.72727272727273	1.41780320732913\\
1.75757575757576	1.41780320732913\\
1.78787878787879	1.41780320732913\\
1.81818181818182	1.41780320732913\\
1.84848484848485	1.41780320732913\\
1.87878787878788	1.41780320732913\\
1.90909090909091	1.41780320732913\\
1.93939393939394	1.41780320732913\\
1.96969696969697	1.41780320732913\\
2	1.41780320732913\\
2.03030303030303	1.41780320732913\\
2.06060606060606	1.41780320732913\\
2.09090909090909	1.41780320732913\\
2.12121212121212	1.41981185667087\\
2.15151515151515	1.42630470944292\\
2.18181818181818	1.43287482699665\\
2.21212121212121	1.43952352260006\\
2.24242424242424	1.44625262419686\\
2.27272727272727	1.45306353883575\\
2.3030303030303	1.45995808883193\\
2.33333333333333	1.46693784881867\\
2.36363636363636	1.47400466944719\\
2.39393939393939	1.48116026737524\\
2.42424242424242	1.48840650107656\\
2.45454545454545	1.4957453627451\\
2.48484848484848	1.50317891454215\\
2.51515151515152	1.51070899764422\\
2.54545454545455	1.51833789967869\\
2.57575757575758	1.52606775088985\\
2.60606060606061	1.53390066541108\\
2.63636363636364	1.54183921684047\\
2.66666666666667	1.54988559543308\\
2.6969696969697	1.55804251775039\\
2.72727272727273	1.56631230415546\\
2.75757575757576	1.57469768537039\\
2.78787878787879	1.58320142797772\\
2.81818181818182	1.59182645910327\\
2.84848484848485	1.60057561935994\\
2.87878787878788	1.60945198782957\\
2.90909090909091	1.61845870936859\\
2.93939393939394	1.62759928820888\\
2.96969696969697	1.63687678008254\\
3	1.64629499229435\\
};
\addlegendentry{driver}

\addplot [color=black,  line width=1.0pt]
  table[row sep=crcr]{%
0	3.05474826609095\\
0.0303030303030303	3.05780463699217\\
0.0606060606060606	3.06086706961856\\
0.0909090909090909	3.06393578044392\\
0.121212121212121	3.06701033144498\\
0.151515151515152	3.07009074154155\\
0.181818181818182	3.07317735062377\\
0.212121212121212	3.07626994118503\\
0.242424242424242	3.07936836858639\\
0.272727272727273	3.08247324135729\\
0.303030303030303	3.08558423488677\\
0.333333333333333	3.08870130944464\\
0.363636363636364	3.09182444694488\\
0.393939393939394	3.09495364366733\\
0.424242424242424	3.09808923918455\\
0.454545454545455	3.10123117502457\\
0.484848484848485	3.1043791288833\\
0.515151515151515	3.10753340003606\\
0.545454545454545	3.11069396952815\\
0.575757575757576	3.11386063796155\\
0.606060606060606	3.11703393505294\\
0.636363636363636	3.1202134859089\\
0.666666666666667	3.12339930358133\\
0.696969696969697	3.1265916171466\\
0.727272727272727	3.1297906518435\\
0.757575757575758	3.13299583701807\\
0.787878787878788	3.13620717179161\\
0.818181818181818	3.13942549861791\\
0.848484848484849	3.14264993082563\\
0.878787878787879	3.1458811182611\\
0.909090909090909	3.14911891472147\\
0.939393939393939	3.15236308688218\\
0.96969696969697	3.15561396591681\\
1	3.15887137113927\\
1.03030303030303	3.16213578717039\\
1.06060606060606	3.16540649373159\\
1.09090909090909	3.16868406227147\\
1.12121212121212	3.17196799946616\\
1.15151515151515	3.17525905011147\\
1.18181818181818	3.17855665054621\\
1.21212121212121	3.1818611280959\\
1.24242424242424	3.18517230279719\\
1.27272727272727	3.18849048159834\\
1.3030303030303	3.19181565841195\\
1.33333333333333	3.19514744670643\\
1.36363636363636	3.19848601351476\\
1.39393939393939	3.20183164843593\\
1.42424242424242	3.20518420629279\\
1.45454545454545	3.20854395652003\\
1.48484848484848	3.21191025593443\\
1.51515151515152	3.2152840444583\\
1.54545454545455	3.21866461064786\\
1.57575757575758	3.22205230872094\\
1.60606060606061	3.22544711467991\\
1.63636363636364	3.22884898749483\\
1.66666666666667	3.23225824490704\\
1.6969696969697	3.23567417968082\\
1.72727272727273	3.23909780865957\\
1.75757575757576	3.24252840805116\\
1.78787878787879	3.24596631130093\\
1.81818181818182	3.24941145974123\\
1.84848484848485	3.25286391360952\\
1.87878787878788	3.25632386847865\\
1.90909090909091	3.25979068968089\\
1.93939393939394	3.2632653511848\\
1.96969696969697	3.26674733758394\\
2	3.27023660544001\\
2.03030303030303	3.27373329928015\\
2.06060606060606	3.27723773032482\\
2.09090909090909	3.2807492168006\\
2.12121212121212	3.28563803194161\\
2.15151515151515	3.29356455516649\\
2.18181818181818	3.30150975376581\\
2.21212121212121	3.30947329186078\\
2.24242424242424	3.31745564646983\\
2.27272727272727	3.32545669450662\\
2.3030303030303	3.33347649225764\\
2.33333333333333	3.34151502991745\\
2.36363636363636	3.3495725450033\\
2.39393939393939	3.35764883443507\\
2.42424242424242	3.36574403460086\\
2.45454545454545	3.37385834609272\\
2.48484848484848	3.38199175734206\\
2.51515151515152	3.39014419690541\\
2.54545454545455	3.39831586424749\\
2.57575757575758	3.40650679755075\\
2.60606060606061	3.41471691402983\\
2.63636363636364	3.42294642831057\\
2.66666666666667	3.43119540460452\\
2.6969696969697	3.43946389868966\\
2.72727272727273	3.44775185782911\\
2.75757575757576	3.45605944378823\\
2.78787878787879	3.46438664799759\\
2.81818181818182	3.47273361202287\\
2.84848484848485	3.48110036754201\\
2.87878787878788	3.48948692784395\\
2.90909090909091	3.49789338396601\\
2.93939393939394	3.50632003290763\\
2.96969696969697	3.51476660765014\\
3	3.52323338917441\\
};
\addlegendentry{passenger}

\end{axis}

\end{tikzpicture}%
\caption{Absolute value of passenger price elasticity and driver wage elasticity under different trip-based tax}.
\label{elasticity}
\end{figure}
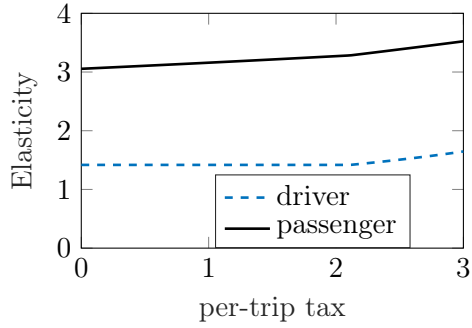

Figures \ref{figure6_trip}-\ref{figure8_trip} show that the tax burden primarily falls on the ride-hailing platform as opposed to passengers and drivers. As the trip-based  charge increases, the passenger cost  increases slightly, the driver wage remains unchanged, while  platform profit reduces significantly. In particular, under a trip-based tax of \$2/trip,  passenger cost increases by 0.6\%,  driver wage remains constant, and  platform profit declines by 59.5\%. This is because  drivers are protected by the minimum wage, and the passenger's price elasticity\footnote{The passenger price elasticity is   $\dfrac{\partial \lambda}{\partial p_f}\dfrac{p_f}{\lambda}$.  We calculate $\dfrac{\partial \lambda}{\partial p_f}$  assuming that the waiting time $t_w$ is fixed under different $p_t$. This is a reasonable approximation since $t_w$ does not change significantly under distinct $p_t$ (Figure \ref{figure5_trip}).} is relatively high (Figure \ref{elasticity}) so that the platform has to refrain from significantly increasing the ride fare.

We  show that the result reported in Figure \ref{figure1_trip}-Figure \ref{figure9_trip} (including number of drivers, number of passengers, platform revenue, and tax revenue) is robust for a large range of model parameters. For notation convenience, let $\tilde w$ be the optimal driver wage set by the platform in the absence of any regulation (i.e., ${p_t} = {w_0} = 0$), and denote by ${N^*_t}({p_t})$  the optimal number of drivers to (\ref{optimalpricing_trip})  under a fixed wage floor, which depends on $p_t$. We then  have the following result. 
\begin{theorem}
Assume that (\ref{optimalpricing_trip}) has a unique solution. For any model parameters $\lambda_0, N_0$ and $\alpha$, any strictly decreasing function ${F_p}(c)$, any strictly increasing function ${F_d}(w)$, any pickup time function ${t_p}$ that satisfies Assumption \ref{assumption1}, and any speed-density relation $v(N)$ that satisfies Assumption \ref{assumption2}, there exists ${w_1} >  \tilde{w},$ such that for any $\tilde{w} < {w_0} < {w_1}$, there exists ${\bar p_t} > 0$, so that $\partial {N^*_t}/\partial{p_t} = 0$ for $ {p_t} \in (0,{\bar p_t})$.
\label{theorem1}
\end{theorem}

The proof of Theorem 1 is can be found in Appendix C. It states that for any wage floor in an appropriate range, there is always a regime in which the congestion charge does not affect the number of TNC vehicles or drivers. In this case, the congestion charge will not directly curb the congestion by reducing the number of TNC vehicle on the city's streets. Instead, it can only indirectly mitigate the traffic congestion by collecting taxes to subsidize public transit to attract passengers. Note that $\tilde{w}$ and $w_1$ can be calculated numerically, and ${\bar p_t}$ depends on the wage floor ${w_0}$. For the case of San Francisco, we calculate that $\tilde{w}=\$21.55$/hour,  $w_1=\$29.20$/hour,  and $\bar{p}_t=\$2.1$/trip when $w_0=\$26.35$/hour.

\section{ Profit maximization under time-based congestion surcharge}
\label{time-based}
This section considers the profit maximization problem under a wage floor and a time-based congestion charge. Under the time-based charge,   each vehicle is penalized based on the total time it stays active on the platform (whether there is a passenger on board or not). Let $p_h$ denote the per-vehicle per-unit-time congestion charge. The total charge (per unit time) is $N p_h$,  and the profit maximization problem  is cast as
(\ref{optimalpricing_time}). For sake of exposition, we will first present a numerical example for San Francisco. The insights derived from the numerical study will be examined by theoretical analysis later to demonstrate its independence on model parameters.

In the numerical study, we will solve the profit maximization problem (\ref{optimalpricing_time}) for different time-based congestion charge $p_h$ under a fixed wage floor $w_0=\$26.35$/hour. The model parameters of (\ref{optimalpricing_time}) are the same as those in Section \ref{parameter_section}.

\begin{figure*}[bt]%
\begin{minipage}[b]{0.32\linewidth}
\centering
%
%
\definecolor{mycolor1}{rgb}{0.00000,0.44700,0.74100}%
\begin{tikzpicture}

\pgfplotsset{every axis y label/.style={
at={(-0.47,0.5)},
xshift=30pt,
rotate=90}}

\begin{axis}[%
width=1.794in,
height=1.03in,
at={(1.358in,0.0in)},
scale only axis,
xmin=0,
xmax=10,
xlabel style={font=\color{white!15!black}},
xlabel={time-based tax},
ymin=3000,
ymax=4200,
ytick={3000,3500,4000},
yticklabels={{3K},{3.5K},{4K}},
ylabel style={font=\color{white!15!black}},
ylabel={Number of drivers},
axis background/.style={fill=white},
legend style={legend cell align=left, align=left, draw=white!15!black}
]
\addplot [color=black, line width=1.0pt]
  table[row sep=crcr]{%
0	3968.14712146414\\
0.101010101010101	3968.14712146414\\
0.202020202020202	3968.14712146414\\
0.303030303030303	3968.14712146414\\
0.404040404040404	3968.14712146414\\
0.505050505050505	3968.14712146414\\
0.606060606060606	3968.14712146414\\
0.707070707070707	3968.14712146414\\
0.808080808080808	3968.14712146414\\
0.909090909090909	3968.14712146414\\
1.01010101010101	3968.14712146414\\
1.11111111111111	3968.14712146414\\
1.21212121212121	3968.14712146414\\
1.31313131313131	3968.14712146414\\
1.41414141414141	3968.14712146414\\
1.51515151515152	3968.14712146414\\
1.61616161616162	3968.14712146414\\
1.71717171717172	3968.14712146414\\
1.81818181818182	3968.14712146414\\
1.91919191919192	3968.14712146414\\
2.02020202020202	3968.14712146414\\
2.12121212121212	3968.14712146414\\
2.22222222222222	3968.14712146414\\
2.32323232323232	3968.14712146414\\
2.42424242424242	3968.14712146414\\
2.52525252525253	3968.14712146414\\
2.62626262626263	3968.14712146414\\
2.72727272727273	3968.14712146414\\
2.82828282828283	3968.14712146414\\
2.92929292929293	3968.14712146414\\
3.03030303030303	3968.14712146414\\
3.13131313131313	3968.14712146414\\
3.23232323232323	3968.14712146414\\
3.33333333333333	3968.14712146414\\
3.43434343434343	3968.14712146414\\
3.53535353535354	3968.14712146414\\
3.63636363636364	3968.14712146414\\
3.73737373737374	3968.14712146414\\
3.83838383838384	3968.14712146414\\
3.93939393939394	3968.14712146414\\
4.04040404040404	3968.14712146414\\
4.14141414141414	3968.14712146414\\
4.24242424242424	3968.14712146414\\
4.34343434343434	3968.14712146414\\
4.44444444444444	3968.14712146414\\
4.54545454545454	3968.14712146414\\
4.64646464646465	3968.14712146414\\
4.74747474747475	3968.14712146414\\
4.84848484848485	3968.14712146414\\
4.94949494949495	3968.14712146414\\
5.05050505050505	3968.14712146414\\
5.15151515151515	3968.14712146414\\
5.25252525252525	3968.14712146414\\
5.35353535353535	3968.14712146414\\
5.45454545454546	3968.14712146414\\
5.55555555555556	3968.14712146414\\
5.65656565656566	3968.14712146414\\
5.75757575757576	3968.14712146414\\
5.85858585858586	3968.14712146414\\
5.95959595959596	3968.14712146414\\
6.06060606060606	3968.14712146414\\
6.16161616161616	3968.14712146414\\
6.26262626262626	3963.43059992327\\
6.36363636363636	3944.4556416108\\
6.46464646464646	3925.46567386878\\
6.56565656565657	3906.45946764197\\
6.66666666666667	3887.43681657938\\
6.76767676767677	3868.3966708167\\
6.86868686868687	3849.33839588616\\
6.96969696969697	3830.26168131001\\
7.07070707070707	3811.16570169624\\
7.17171717171717	3792.04964741393\\
7.27272727272727	3772.91277905263\\
7.37373737373737	3753.75446059231\\
7.47474747474747	3734.5742174448\\
7.57575757575758	3715.37054614458\\
7.67676767676768	3696.14330416115\\
7.77777777777778	3676.89142952682\\
7.87878787878788	3657.61405156825\\
7.97979797979798	3638.31036606345\\
8.08080808080808	3618.97969497472\\
8.18181818181818	3599.62062695937\\
8.28282828282828	3580.23266049606\\
8.38383838383838	3560.81428955609\\
8.48484848484848	3541.36528083539\\
8.58585858585859	3521.88385937158\\
8.68686868686869	3502.36936272887\\
8.78787878787879	3482.82076013936\\
8.88888888888889	3463.23651999442\\
8.98989898989899	3443.61595557184\\
9.09090909090909	3423.95737255749\\
9.19191919191919	3404.25968398083\\
9.29292929292929	3384.52186673961\\
9.39393939393939	3364.74232246738\\
9.49494949494949	3344.91934939533\\
9.5959595959596	3325.05218479773\\
9.6969696969697	3305.13871858587\\
9.7979797979798	3285.17768732858\\
9.8989898989899	3265.16724443007\\
10	3245.10609381859\\
};

\end{axis}
\end{tikzpicture}%
\vspace*{-0.3in}
\caption{Number of drivers under different time-based congestion surcharge. }
\label{figure1_time}
\end{minipage}
\begin{minipage}[b]{0.005\linewidth}
\hfill
\end{minipage}
\begin{minipage}[b]{0.32\linewidth}
\centering
%
%
\definecolor{mycolor1}{rgb}{0.00000,0.44700,0.74100}%
\begin{tikzpicture}

\pgfplotsset{every axis y label/.style={
at={(-0.45,0.5)},
xshift=32pt,
rotate=90}}

\begin{axis}[%
width=1.794in,
height=1.03in,
at={(1.358in,0.0in)},
scale only axis,
xmin=0,
xmax=10,
xlabel style={font=\color{white!15!black}},
xlabel={time-based tax},
ymin=160,
ymax=220,
ylabel style={font=\color{white!15!black}},
ylabel={Passenger arrival},
axis background/.style={fill=white},
legend style={legend cell align=left, align=left, draw=white!15!black}
]
\addplot [color=black, line width=1.0pt]
  table[row sep=crcr]{%
0	208.456289196894\\
0.101010101010101	208.456289196894\\
0.202020202020202	208.456289196894\\
0.303030303030303	208.456289196894\\
0.404040404040404	208.456289196894\\
0.505050505050505	208.456289196894\\
0.606060606060606	208.456289196894\\
0.707070707070707	208.456289196894\\
0.808080808080808	208.456289196894\\
0.909090909090909	208.456289196894\\
1.01010101010101	208.456289196894\\
1.11111111111111	208.456289196894\\
1.21212121212121	208.456289196894\\
1.31313131313131	208.456289196894\\
1.41414141414141	208.456289196894\\
1.51515151515152	208.456289196894\\
1.61616161616162	208.456289196894\\
1.71717171717172	208.456289196894\\
1.81818181818182	208.456289196894\\
1.91919191919192	208.456289196894\\
2.02020202020202	208.456289196894\\
2.12121212121212	208.456289196894\\
2.22222222222222	208.456289196894\\
2.32323232323232	208.456289196894\\
2.42424242424242	208.456289196894\\
2.52525252525253	208.456289196894\\
2.62626262626263	208.456289196894\\
2.72727272727273	208.456289196894\\
2.82828282828283	208.456289196894\\
2.92929292929293	208.456289196894\\
3.03030303030303	208.456289196894\\
3.13131313131313	208.456289196894\\
3.23232323232323	208.456289196894\\
3.33333333333333	208.456289196894\\
3.43434343434343	208.456289196894\\
3.53535353535354	208.456289196894\\
3.63636363636364	208.456289196894\\
3.73737373737374	208.456289196894\\
3.83838383838384	208.456289196894\\
3.93939393939394	208.456289196894\\
4.04040404040404	208.456289196894\\
4.14141414141414	208.456289196894\\
4.24242424242424	208.456289196894\\
4.34343434343434	208.456289196894\\
4.44444444444444	208.456289196894\\
4.54545454545454	208.456289196894\\
4.64646464646465	208.456289196894\\
4.74747474747475	208.456289196894\\
4.84848484848485	208.456289196894\\
4.94949494949495	208.456289196894\\
5.05050505050505	208.456289196894\\
5.15151515151515	208.456289196894\\
5.25252525252525	208.456289196894\\
5.35353535353535	208.456289196894\\
5.45454545454546	208.456289196894\\
5.55555555555556	208.456289196894\\
5.65656565656566	208.456289196894\\
5.75757575757576	208.456289196894\\
5.85858585858586	208.456289196894\\
5.95959595959596	208.456289196894\\
6.06060606060606	208.456289196894\\
6.16161616161616	208.456289196894\\
6.26262626262626	208.216310851355\\
6.36363636363636	207.249885613567\\
6.46464646464646	206.281233856883\\
6.56565656565657	205.310290879597\\
6.66666666666667	204.337053446597\\
6.76767676767677	203.361459136469\\
6.86868686868687	202.383481405233\\
6.96969696969697	201.403102126841\\
7.07070707070707	200.420285157392\\
7.17171717171717	199.434985511578\\
7.27272727272727	198.447168782186\\
7.37373737373737	197.456795667531\\
7.47474747474747	196.463845680477\\
7.57575757575758	195.468244335302\\
7.67676767676768	194.469977957733\\
7.77777777777778	193.468996167103\\
7.87878787878788	192.465256252192\\
7.97979797979798	191.458710407992\\
8.08080808080808	190.449324081149\\
8.18181818181818	189.437025317164\\
8.28282828282828	188.421785377132\\
8.38383838383838	187.40352187648\\
8.48484848484848	186.382227489432\\
8.58585858585859	185.357805340147\\
8.68686868686869	184.330224214341\\
8.78787878787879	183.299422243154\\
8.88888888888889	182.265319994941\\
8.98989898989899	181.22787801933\\
9.09090909090909	180.187007498674\\
9.19191919191919	179.142648886375\\
9.29292929292929	178.094746154089\\
9.39393939393939	177.04321341738\\
9.49494949494949	175.987954076955\\
9.5959595959596	174.928929655892\\
9.6969696969697	173.866024003897\\
9.7979797979798	172.799165547296\\
9.8989898989899	171.728253682768\\
10	170.653219140105\\
};

\end{axis}
\end{tikzpicture}%
\vspace*{-0.3in}
\caption{Passengers arrival rate under different  time-based congestion charge.} 
\label{figure2_time}
\end{minipage}
\begin{minipage}[b]{0.005\linewidth}
\hfill
\end{minipage}
\begin{minipage}[b]{0.32\linewidth}
\centering
%
%
\definecolor{mycolor1}{rgb}{0.00000,0.44700,0.74100}%
\begin{tikzpicture}

\pgfplotsset{every axis y label/.style={
at={(-0.45,0.5)},
xshift=32pt,
rotate=90}}

\begin{axis}[%
width=1.794in,
height=1.03in,
at={(1.358in,0.0in)},
scale only axis,
xmin=0,
xmax=10,
xlabel style={font=\color{white!15!black}},
xlabel={time-based tax},
ymin=0.58,
ymax=0.605,
ylabel style={font=\color{white!15!black}},
ylabel={Occupancy},
axis background/.style={fill=white},
legend style={legend cell align=left, align=left, draw=white!15!black}
]
\addplot [color=black, line width=1.0pt]
  table[row sep=crcr]{%
0	0.597762201599234\\
0.101010101010101	0.597762201599234\\
0.202020202020202	0.597762201599234\\
0.303030303030303	0.597762201599234\\
0.404040404040404	0.597762201599234\\
0.505050505050505	0.597762201599234\\
0.606060606060606	0.597762201599234\\
0.707070707070707	0.597762201599234\\
0.808080808080808	0.597762201599234\\
0.909090909090909	0.597762201599234\\
1.01010101010101	0.597762201599234\\
1.11111111111111	0.597762201599234\\
1.21212121212121	0.597762201599234\\
1.31313131313131	0.597762201599234\\
1.41414141414141	0.597762201599234\\
1.51515151515152	0.597762201599234\\
1.61616161616162	0.597762201599234\\
1.71717171717172	0.597762201599234\\
1.81818181818182	0.597762201599234\\
1.91919191919192	0.597762201599234\\
2.02020202020202	0.597762201599234\\
2.12121212121212	0.597762201599234\\
2.22222222222222	0.597762201599234\\
2.32323232323232	0.597762201599234\\
2.42424242424242	0.597762201599234\\
2.52525252525253	0.597762201599234\\
2.62626262626263	0.597762201599234\\
2.72727272727273	0.597762201599234\\
2.82828282828283	0.597762201599234\\
2.92929292929293	0.597762201599234\\
3.03030303030303	0.597762201599234\\
3.13131313131313	0.597762201599234\\
3.23232323232323	0.597762201599234\\
3.33333333333333	0.597762201599234\\
3.43434343434343	0.597762201599234\\
3.53535353535354	0.597762201599234\\
3.63636363636364	0.597762201599234\\
3.73737373737374	0.597762201599234\\
3.83838383838384	0.597762201599234\\
3.93939393939394	0.597762201599234\\
4.04040404040404	0.597762201599234\\
4.14141414141414	0.597762201599234\\
4.24242424242424	0.597762201599234\\
4.34343434343434	0.597762201599234\\
4.44444444444444	0.597762201599234\\
4.54545454545454	0.597762201599234\\
4.64646464646465	0.597762201599234\\
4.74747474747475	0.597762201599234\\
4.84848484848485	0.597762201599234\\
4.94949494949495	0.597762201599234\\
5.05050505050505	0.597762201599234\\
5.15151515151515	0.597762201599234\\
5.25252525252525	0.597762201599234\\
5.35353535353535	0.597762201599234\\
5.45454545454546	0.597762201599234\\
5.55555555555556	0.597762201599234\\
5.65656565656566	0.597762201599234\\
5.75757575757576	0.597762201599234\\
5.85858585858586	0.597762201599234\\
5.95959595959596	0.597762201599234\\
6.06060606060606	0.597762201599234\\
6.16161616161616	0.597762201599234\\
6.26262626262626	0.597722881144498\\
6.36363636363636	0.597562510203645\\
6.46464646464646	0.597398759802542\\
6.56565656565657	0.597231570584184\\
6.66666666666667	0.597060918269294\\
6.76767676767677	0.596886723489281\\
6.86868686868687	0.596708952430472\\
6.96969696969697	0.596527549393477\\
7.07070707070707	0.596342477840147\\
7.17171717171717	0.596153671946704\\
7.27272727272727	0.59596108632906\\
7.37373737373737	0.595764644960582\\
7.47474747474747	0.595564303763335\\
7.57575757575758	0.59536000482634\\
7.67676767676768	0.595151673986919\\
7.77777777777778	0.594939259455799\\
7.87878787878788	0.594722703602595\\
7.97979797979798	0.594501921874591\\
8.08080808080808	0.594276850299185\\
8.18181818181818	0.5940474178436\\
8.28282828282828	0.593813549223234\\
8.38383838383838	0.593575152084157\\
8.48484848484848	0.593332174424798\\
8.58585858585859	0.593084516179767\\
8.68686868686869	0.592832112566686\\
8.78787878787879	0.592574856127755\\
8.88888888888889	0.592312659243409\\
8.98989898989899	0.592045428672301\\
9.09090909090909	0.591773068082544\\
9.19191919191919	0.591495478009751\\
9.29292929292929	0.591212558019355\\
9.39393939393939	0.590924199320349\\
9.49494949494949	0.590630271641389\\
9.5959595959596	0.590330683136481\\
9.6969696969697	0.590025299327511\\
9.7979797979798	0.589713993577192\\
9.8989898989899	0.589396634285671\\
10	0.589073101845723\\
};

\end{axis}
\end{tikzpicture}%
\vspace*{-0.3in}
\caption{Occupancy rate under different  time-based congestion charge.}
\label{figure3_time}
\end{minipage}
\begin{minipage}[b]{0.32\linewidth}
\centering
%
%
\definecolor{mycolor1}{rgb}{0.00000,0.44700,0.74100}%
\definecolor{mycolor2}{rgb}{0.85000,0.32500,0.09800}%
\begin{tikzpicture}

\pgfplotsset{every axis y label/.style={
at={(-0.42,0.5)},
xshift=32pt,
rotate=90}}

\begin{axis}[%
width=1.794in,
height=1.03in,
at={(1.358in,0.0in)},
scale only axis,
xmin=0,
xmax=10,
xlabel style={font=\color{white!15!black}},
xlabel={time-based tax},
ymin=7,
ymax=13,
ylabel style={font=\color{white!15!black}},
ylabel={price/payment},
axis background/.style={fill=white},
legend style={at={(0.65,0.4)}, anchor=south west, legend cell align=left, align=left, draw=white!15!black}
]
\addplot [color=black, line width=1.0pt]
  table[row sep=crcr]{%
0	11.6282174100135\\
0.101010101010101	11.6282174100135\\
0.202020202020202	11.6282174100135\\
0.303030303030303	11.6282174100135\\
0.404040404040404	11.6282174100135\\
0.505050505050505	11.6282174100135\\
0.606060606060606	11.6282174100135\\
0.707070707070707	11.6282174100135\\
0.808080808080808	11.6282174100135\\
0.909090909090909	11.6282174100135\\
1.01010101010101	11.6282174100135\\
1.11111111111111	11.6282174100135\\
1.21212121212121	11.6282174100135\\
1.31313131313131	11.6282174100135\\
1.41414141414141	11.6282174100135\\
1.51515151515152	11.6282174100135\\
1.61616161616162	11.6282174100135\\
1.71717171717172	11.6282174100135\\
1.81818181818182	11.6282174100135\\
1.91919191919192	11.6282174100135\\
2.02020202020202	11.6282174100135\\
2.12121212121212	11.6282174100135\\
2.22222222222222	11.6282174100135\\
2.32323232323232	11.6282174100135\\
2.42424242424242	11.6282174100135\\
2.52525252525253	11.6282174100135\\
2.62626262626263	11.6282174100135\\
2.72727272727273	11.6282174100135\\
2.82828282828283	11.6282174100135\\
2.92929292929293	11.6282174100135\\
3.03030303030303	11.6282174100135\\
3.13131313131313	11.6282174100135\\
3.23232323232323	11.6282174100135\\
3.33333333333333	11.6282174100135\\
3.43434343434343	11.6282174100135\\
3.53535353535354	11.6282174100135\\
3.63636363636364	11.6282174100135\\
3.73737373737374	11.6282174100135\\
3.83838383838384	11.6282174100135\\
3.93939393939394	11.6282174100135\\
4.04040404040404	11.6282174100135\\
4.14141414141414	11.6282174100135\\
4.24242424242424	11.6282174100135\\
4.34343434343434	11.6282174100135\\
4.44444444444444	11.6282174100135\\
4.54545454545454	11.6282174100135\\
4.64646464646465	11.6282174100135\\
4.74747474747475	11.6282174100135\\
4.84848484848485	11.6282174100135\\
4.94949494949495	11.6282174100135\\
5.05050505050505	11.6282174100135\\
5.15151515151515	11.6282174100135\\
5.25252525252525	11.6282174100135\\
5.35353535353535	11.6282174100135\\
5.45454545454546	11.6282174100135\\
5.55555555555556	11.6282174100135\\
5.65656565656566	11.6282174100135\\
5.75757575757576	11.6282174100135\\
5.85858585858586	11.6282174100135\\
5.95959595959596	11.6282174100135\\
6.06060606060606	11.6282174100135\\
6.16161616161616	11.6282174100135\\
6.26262626262626	11.6293118095835\\
6.36363636363636	11.6336986187072\\
6.46464646464646	11.6380575564872\\
6.56565656565657	11.6423885151698\\
6.66666666666667	11.6466906038034\\
6.76767676767677	11.6509640125124\\
6.86868686868687	11.6552079789661\\
6.96969696969697	11.6594221164922\\
7.07070707070707	11.6636056429744\\
7.17171717171717	11.6677583010673\\
7.27272727272727	11.6718793844454\\
7.37373737373737	11.6759687378347\\
7.47474747474747	11.6800255329073\\
7.57575757575758	11.6840491654277\\
7.67676767676768	11.6880392908935\\
7.77777777777778	11.6919950732999\\
7.87878787878788	11.695915731585\\
7.97979797979798	11.6998009555628\\
8.08080808080808	11.7036499744807\\
8.18181818181818	11.7074620682693\\
8.28282828282828	11.7112365628439\\
8.38383838383838	11.7149730094596\\
8.48484848484848	11.7186701523053\\
8.58585858585859	11.722327525939\\
8.68686868686869	11.7259439504823\\
8.78787878787879	11.7295189883429\\
8.88888888888889	11.7330516815336\\
8.98989898989899	11.7365411623754\\
9.09090909090909	11.7399864404863\\
9.19191919191919	11.7433865417268\\
9.29292929292929	11.7467404204106\\
9.39393939393939	11.7500470243517\\
9.49494949494949	11.7533055934692\\
9.5959595959596	11.7565146084769\\
9.6969696969697	11.7596730921405\\
9.7979797979798	11.7627798951339\\
9.8989898989899	11.7658337410963\\
10	11.7688330584048\\
};
\addlegendentry{$p_f$}

\addplot [color=mycolor1, dashed, line width=1.0pt]
  table[row sep=crcr]{%
0	8.35992020624679\\
0.101010101010101	8.35992020624679\\
0.202020202020202	8.35992020624679\\
0.303030303030303	8.35992020624679\\
0.404040404040404	8.35992020624679\\
0.505050505050505	8.35992020624679\\
0.606060606060606	8.35992020624679\\
0.707070707070707	8.35992020624679\\
0.808080808080808	8.35992020624679\\
0.909090909090909	8.35992020624679\\
1.01010101010101	8.35992020624679\\
1.11111111111111	8.35992020624679\\
1.21212121212121	8.35992020624679\\
1.31313131313131	8.35992020624679\\
1.41414141414141	8.35992020624679\\
1.51515151515152	8.35992020624679\\
1.61616161616162	8.35992020624679\\
1.71717171717172	8.35992020624679\\
1.81818181818182	8.35992020624679\\
1.91919191919192	8.35992020624679\\
2.02020202020202	8.35992020624679\\
2.12121212121212	8.35992020624679\\
2.22222222222222	8.35992020624679\\
2.32323232323232	8.35992020624679\\
2.42424242424242	8.35992020624679\\
2.52525252525253	8.35992020624679\\
2.62626262626263	8.35992020624679\\
2.72727272727273	8.35992020624679\\
2.82828282828283	8.35992020624679\\
2.92929292929293	8.35992020624679\\
3.03030303030303	8.35992020624679\\
3.13131313131313	8.35992020624679\\
3.23232323232323	8.35992020624679\\
3.33333333333333	8.35992020624679\\
3.43434343434343	8.35992020624679\\
3.53535353535354	8.35992020624679\\
3.63636363636364	8.35992020624679\\
3.73737373737374	8.35992020624679\\
3.83838383838384	8.35992020624679\\
3.93939393939394	8.35992020624679\\
4.04040404040404	8.35992020624679\\
4.14141414141414	8.35992020624679\\
4.24242424242424	8.35992020624679\\
4.34343434343434	8.35992020624679\\
4.44444444444444	8.35992020624679\\
4.54545454545454	8.35992020624679\\
4.64646464646465	8.35992020624679\\
4.74747474747475	8.35992020624679\\
4.84848484848485	8.35992020624679\\
4.94949494949495	8.35992020624679\\
5.05050505050505	8.35992020624679\\
5.15151515151515	8.35992020624679\\
5.25252525252525	8.35992020624679\\
5.35353535353535	8.35992020624679\\
5.45454545454546	8.35992020624679\\
5.55555555555556	8.35992020624679\\
5.65656565656566	8.35992020624679\\
5.75757575757576	8.35992020624679\\
5.85858585858586	8.35992020624679\\
5.95959595959596	8.35992020624679\\
6.06060606060606	8.35992020624679\\
6.16161616161616	8.35992020624679\\
6.26262626262626	8.35960736224736\\
6.36363636363636	8.35838066116328\\
6.46464646464646	8.35720071513355\\
6.56565656565657	8.35606815188267\\
6.66666666666667	8.35498329753799\\
6.76767676767677	8.35394709735509\\
6.86868686868687	8.35295993751802\\
6.96969696969697	8.35202257203865\\
7.07070707070707	8.35113539537108\\
7.17171717171717	8.35029921764958\\
7.27272727272727	8.34951457845818\\
7.37373737373737	8.34878246844016\\
7.47474747474747	8.34810346307646\\
7.57575757575758	8.34747814782076\\
7.67676767676768	8.34690758675137\\
7.77777777777778	8.34639236669127\\
7.87878787878788	8.34593319469367\\
7.97979797979798	8.34553117148839\\
8.08080808080808	8.3451871359711\\
8.18181818181818	8.34490189739635\\
8.28282828282828	8.34467649403781\\
8.38383838383838	8.34451202680223\\
8.48484848484848	8.34440926467479\\
8.58585858585859	8.34436937828996\\
8.68686868686869	8.34439324869779\\
8.78787878787879	8.34448229628829\\
8.88888888888889	8.34463757782573\\
8.98989898989899	8.34486038802111\\
9.09090909090909	8.34515188963273\\
9.19191919191919	8.34551340607807\\
9.29292929292929	8.34594629305044\\
9.39393939393939	8.34645192790655\\
9.49494949494949	8.3470319809527\\
9.5959595959596	8.34768775732432\\
9.6969696969697	8.34842093058827\\
9.7979797979798	8.34923322564837\\
9.8989898989899	8.35012634260212\\
10	8.35110192109514\\
};
\addlegendentry{$p_d$}

\end{axis}
\end{tikzpicture}%
\vspace*{-0.3in}
\caption{Per-trip ride price and driver payment under different time-based congestion charge.} 
\label{figure4_time}
\end{minipage}
\begin{minipage}[b]{0.005\linewidth}
\hfill
\end{minipage}
\begin{minipage}[b]{0.32\linewidth}
\centering
%
%
\definecolor{mycolor1}{rgb}{0.00000,0.44700,0.74100}%
\begin{tikzpicture}

\pgfplotsset{every axis y label/.style={
at={(-0.42,0.5)},
xshift=32pt,
rotate=90}}

\begin{axis}[%
width=1.794in,
height=1.03in,
at={(1.358in,0.0in)},
scale only axis,
xmin=0,
xmax=10,
xlabel style={font=\color{white!15!black}},
xlabel={time-based tax},
ymin=4.3,
ymax=5.0,
ylabel style={font=\color{white!15!black}},
ylabel={pickup time},
axis background/.style={fill=white},
legend style={legend cell align=left, align=left, draw=white!15!black}
]
\addplot [color=black, line width=1.0pt]
  table[row sep=crcr]{%
0	4.5113226177619\\
0.101010101010101	4.5113226177619\\
0.202020202020202	4.5113226177619\\
0.303030303030303	4.5113226177619\\
0.404040404040404	4.5113226177619\\
0.505050505050505	4.5113226177619\\
0.606060606060606	4.5113226177619\\
0.707070707070707	4.5113226177619\\
0.808080808080808	4.5113226177619\\
0.909090909090909	4.5113226177619\\
1.01010101010101	4.5113226177619\\
1.11111111111111	4.5113226177619\\
1.21212121212121	4.5113226177619\\
1.31313131313131	4.5113226177619\\
1.41414141414141	4.5113226177619\\
1.51515151515152	4.5113226177619\\
1.61616161616162	4.5113226177619\\
1.71717171717172	4.5113226177619\\
1.81818181818182	4.5113226177619\\
1.91919191919192	4.5113226177619\\
2.02020202020202	4.5113226177619\\
2.12121212121212	4.5113226177619\\
2.22222222222222	4.5113226177619\\
2.32323232323232	4.5113226177619\\
2.42424242424242	4.5113226177619\\
2.52525252525253	4.5113226177619\\
2.62626262626263	4.5113226177619\\
2.72727272727273	4.5113226177619\\
2.82828282828283	4.5113226177619\\
2.92929292929293	4.5113226177619\\
3.03030303030303	4.5113226177619\\
3.13131313131313	4.5113226177619\\
3.23232323232323	4.5113226177619\\
3.33333333333333	4.5113226177619\\
3.43434343434343	4.5113226177619\\
3.53535353535354	4.5113226177619\\
3.63636363636364	4.5113226177619\\
3.73737373737374	4.5113226177619\\
3.83838383838384	4.5113226177619\\
3.93939393939394	4.5113226177619\\
4.04040404040404	4.5113226177619\\
4.14141414141414	4.5113226177619\\
4.24242424242424	4.5113226177619\\
4.34343434343434	4.5113226177619\\
4.44444444444444	4.5113226177619\\
4.54545454545454	4.5113226177619\\
4.64646464646465	4.5113226177619\\
4.74747474747475	4.5113226177619\\
4.84848484848485	4.5113226177619\\
4.94949494949495	4.5113226177619\\
5.05050505050505	4.5113226177619\\
5.15151515151515	4.5113226177619\\
5.25252525252525	4.5113226177619\\
5.35353535353535	4.5113226177619\\
5.45454545454546	4.5113226177619\\
5.55555555555556	4.5113226177619\\
5.65656565656566	4.5113226177619\\
5.75757575757576	4.5113226177619\\
5.85858585858586	4.5113226177619\\
5.95959595959596	4.5113226177619\\
6.06060606060606	4.5113226177619\\
6.16161616161616	4.5113226177619\\
6.26262626262626	4.5133196459385\\
6.36363636363636	4.5213836681822\\
6.46464646464646	4.52950381064092\\
6.56565656565657	4.53768124086525\\
6.66666666666667	4.54591686176354\\
6.76767676767677	4.55421169809308\\
6.86868686868687	4.56256685967705\\
6.96969696969697	4.57098319943113\\
7.07070707070707	4.57946195992755\\
7.17171717171717	4.58800424195261\\
7.27272727272727	4.59661125908114\\
7.37373737373737	4.60528402998741\\
7.47474747474747	4.61402370301903\\
7.57575757575758	4.62283191422324\\
7.67676767676768	4.6317095392832\\
7.77777777777778	4.64065809498341\\
7.87878787878788	4.64967900548615\\
7.97979797979798	4.65877354532955\\
8.08080808080808	4.66794307886816\\
8.18181818181818	4.67718937134925\\
8.28282828282828	4.68651371478698\\
8.38383838383838	4.69591790128553\\
8.48484848484848	4.7054032992546\\
8.58585858585859	4.71497191010915\\
8.68686868686869	4.72462537045712\\
8.78787878787879	4.73436532968108\\
8.88888888888889	4.74419389995714\\
8.98989898989899	4.75411273303143\\
9.09090909090909	4.76412411284259\\
9.19191919191919	4.77423001801787\\
9.29292929292929	4.78443244855959\\
9.39393939393939	4.79473377221198\\
9.49494949494949	4.8051363754991\\
9.5959595959596	4.81564237420759\\
9.6969696969697	4.82625454522727\\
9.7979797979798	4.83697527785022\\
9.8989898989899	4.84780738906848\\
10	4.85875351441087\\
};

\end{axis}
\end{tikzpicture}%
\vspace*{-0.3in}
\caption{Passenger pickup time in minutes under different time-based congestion charge.}
\label{figure5_time}
\end{minipage}
\begin{minipage}[b]{0.005\linewidth}
\hfill
\end{minipage}
\begin{minipage}[b]{0.32\linewidth}
\centering
%
%
\definecolor{mycolor1}{rgb}{0.00000,0.44700,0.74100}%
\begin{tikzpicture}

\pgfplotsset{every axis y label/.style={
at={(-0.45,0.5)},
xshift=32pt,
rotate=90}}

\begin{axis}[%
width=1.794in,
height=1.03in,
at={(1.358in,0.0in)},
scale only axis,
xmin=0,
xmax=10,
xlabel style={font=\color{white!15!black}},
xlabel={time-based tax},
ymin=35,
ymax=37,
ylabel style={font=\color{white!15!black}},
ylabel={passenger cost},
axis background/.style={fill=white},
legend style={legend cell align=left, align=left, draw=white!15!black}
]
\addplot [color=black, line width=1.0pt]
  table[row sep=crcr]{%
0	35.4300486745168\\
0.101010101010101	35.4300486745168\\
0.202020202020202	35.4300486745168\\
0.303030303030303	35.4300486745168\\
0.404040404040404	35.4300486745168\\
0.505050505050505	35.4300486745168\\
0.606060606060606	35.4300486745168\\
0.707070707070707	35.4300486745168\\
0.808080808080808	35.4300486745168\\
0.909090909090909	35.4300486745168\\
1.01010101010101	35.4300486745168\\
1.11111111111111	35.4300486745168\\
1.21212121212121	35.4300486745168\\
1.31313131313131	35.4300486745168\\
1.41414141414141	35.4300486745168\\
1.51515151515152	35.4300486745168\\
1.61616161616162	35.4300486745168\\
1.71717171717172	35.4300486745168\\
1.81818181818182	35.4300486745168\\
1.91919191919192	35.4300486745168\\
2.02020202020202	35.4300486745168\\
2.12121212121212	35.4300486745168\\
2.22222222222222	35.4300486745168\\
2.32323232323232	35.4300486745168\\
2.42424242424242	35.4300486745168\\
2.52525252525253	35.4300486745168\\
2.62626262626263	35.4300486745168\\
2.72727272727273	35.4300486745168\\
2.82828282828283	35.4300486745168\\
2.92929292929293	35.4300486745168\\
3.03030303030303	35.4300486745168\\
3.13131313131313	35.4300486745168\\
3.23232323232323	35.4300486745168\\
3.33333333333333	35.4300486745168\\
3.43434343434343	35.4300486745168\\
3.53535353535354	35.4300486745168\\
3.63636363636364	35.4300486745168\\
3.73737373737374	35.4300486745168\\
3.83838383838384	35.4300486745168\\
3.93939393939394	35.4300486745168\\
4.04040404040404	35.4300486745168\\
4.14141414141414	35.4300486745168\\
4.24242424242424	35.4300486745168\\
4.34343434343434	35.4300486745168\\
4.44444444444444	35.4300486745168\\
4.54545454545454	35.4300486745168\\
4.64646464646465	35.4300486745168\\
4.74747474747475	35.4300486745168\\
4.84848484848485	35.4300486745168\\
4.94949494949495	35.4300486745168\\
5.05050505050505	35.4300486745168\\
5.15151515151515	35.4300486745168\\
5.25252525252525	35.4300486745168\\
5.35353535353535	35.4300486745168\\
5.45454545454546	35.4300486745168\\
5.55555555555556	35.4300486745168\\
5.65656565656566	35.4300486745168\\
5.75757575757576	35.4300486745168\\
5.85858585858586	35.4300486745168\\
5.95959595959596	35.4300486745168\\
6.06060606060606	35.4300486745168\\
6.16161616161616	35.4300486745168\\
6.26262626262626	35.4344327992255\\
6.36363636363636	35.4521268630069\\
6.46464646464646	35.4699242209714\\
6.56565656565657	35.48782713872\\
6.66666666666667	35.5058367783808\\
6.76767676767677	35.5239554236865\\
6.86868686868687	35.5421847254499\\
6.96969696969697	35.560526204845\\
7.07070707070707	35.5789817485279\\
7.17171717171717	35.5975534424185\\
7.27272727272727	35.6162432076225\\
7.37373737373737	35.6350530919839\\
7.47474747474747	35.6539848198834\\
7.57575757575758	35.6730411825951\\
7.67676767676768	35.6922238490633\\
7.77777777777778	35.7115352290333\\
7.87878787878788	35.7309776269074\\
7.97979797979798	35.7505534886938\\
8.08080808080808	35.7702650462693\\
8.18181818181818	35.7901153086201\\
8.28282828282828	35.8101064877239\\
8.38383838383838	35.8302419015701\\
8.48484848484848	35.850523436815\\
8.58585858585859	35.870954808479\\
8.68686868686869	35.8915384843278\\
8.78787878787879	35.9122776000853\\
8.88888888888889	35.9331757095845\\
8.98989898989899	35.9542356239985\\
9.09090909090909	35.9754612217109\\
9.19191919191919	35.9968558587417\\
9.29292929292929	36.0184228885723\\
9.39393939393939	36.0401663544457\\
9.49494949494949	36.0620906078373\\
9.5959595959596	36.0841988762159\\
9.6969696969697	36.1064960921211\\
9.7979797979798	36.1289863524193\\
9.8989898989899	36.1516744618242\\
10	36.1745646631592\\
};

\end{axis}
\end{tikzpicture}%
\vspace*{-0.3in}
\caption{Passenger travel cost in \$ under different time-based congestion charge.}
\label{figure6_time}
\end{minipage}
\begin{minipage}[b]{0.32\linewidth}
\centering
%
%
\definecolor{mycolor1}{rgb}{0.00000,0.44700,0.74100}%
\begin{tikzpicture}

\pgfplotsset{every axis y label/.style={
at={(-0.42,0.5)},
xshift=32pt,
rotate=90}}

\begin{axis}[%
width=1.794in,
height=1.03in,
at={(1.358in,0.0in)},
scale only axis,
xmin=0,
xmax=10,
xlabel style={font=\color{white!15!black}},
xlabel={time-based tax},
ymin=26.349999999997,
ymax=26.350000000003,
ylabel style={font=\color{white!15!black}},
ylabel={driver revenue},
axis background/.style={fill=white},
legend style={legend cell align=left, align=left, draw=white!15!black}
]
\addplot [color=black, line width=1.0pt]
  table[row sep=crcr]{%
0	26.35\\
0.151515151515152	26.35\\
0.303030303030303	26.35\\
0.454545454545455	26.35\\
0.606060606060606	26.35\\
0.757575757575758	26.35\\
0.909090909090909	26.35\\
1.06060606060606	26.35\\
1.21212121212121	26.35\\
1.36363636363636	26.35\\
1.51515151515152	26.35\\
1.66666666666667	26.35\\
1.81818181818182	26.35\\
1.96969696969697	26.35\\
2.12121212121212	26.35\\
2.27272727272727	26.35\\
2.42424242424242	26.35\\
2.57575757575758	26.35\\
2.72727272727273	26.35\\
2.87878787878788	26.35\\
3.03030303030303	26.35\\
3.18181818181818	26.35\\
3.33333333333333	26.35\\
3.48484848484848	26.35\\
3.63636363636364	26.35\\
3.78787878787879	26.35\\
3.93939393939394	26.35\\
4.09090909090909	26.35\\
4.24242424242424	26.35\\
4.39393939393939	26.35\\
4.54545454545455	26.35\\
4.6969696969697	26.35\\
4.84848484848485	26.35\\
5	26.35\\
5.15151515151515	26.35\\
5.3030303030303	26.35\\
5.45454545454546	26.35\\
5.60606060606061	26.35\\
5.75757575757576	26.35\\
5.90909090909091	26.35\\
6.06060606060606	26.35\\
6.21212121212121	26.35\\
6.36363636363636	26.35\\
6.51515151515152	26.35\\
6.66666666666667	26.35\\
6.81818181818182	26.35\\
6.96969696969697	26.35\\
7.12121212121212	26.35\\
7.27272727272727	26.35\\
7.42424242424242	26.35\\
7.57575757575758	26.35\\
7.72727272727273	26.35\\
7.87878787878788	26.35\\
8.03030303030303	26.35\\
8.18181818181818	26.35\\
8.33333333333333	26.35\\
8.48484848484848	26.35\\
8.63636363636364	26.35\\
8.78787878787879	26.35\\
8.93939393939394	26.35\\
9.09090909090909	26.35\\
9.24242424242424	26.35\\
9.39393939393939	26.35\\
9.54545454545454	26.35\\
9.6969696969697	26.35\\
9.84848484848485	26.35\\
10	26.35\\
10.1515151515152	26.35\\
10.3030303030303	26.35\\
10.4545454545455	26.35\\
10.6060606060606	26.35\\
10.7575757575758	26.35\\
10.9090909090909	26.35\\
11.0606060606061	26.35\\
11.2121212121212	26.35\\
11.3636363636364	26.35\\
11.5151515151515	26.35\\
11.6666666666667	26.35\\
11.8181818181818	26.35\\
11.969696969697	26.35\\
12.1212121212121	26.35\\
12.2727272727273	26.35\\
12.4242424242424	26.35\\
12.5757575757576	26.35\\
12.7272727272727	26.35\\
12.8787878787879	26.35\\
13.030303030303	26.35\\
13.1818181818182	26.35\\
13.3333333333333	26.35\\
13.4848484848485	26.35\\
13.6363636363636	26.35\\
13.7878787878788	26.35\\
13.9393939393939	26.35\\
14.0909090909091	26.35\\
14.2424242424242	26.35\\
14.3939393939394	26.35\\
14.5454545454545	26.35\\
14.6969696969697	26.35\\
14.8484848484848	26.35\\
15	26.35\\
};

\end{axis}
\end{tikzpicture}%
\vspace*{-0.3in}
\caption{Per-hour driver wage under different time-based congestion charge.}
\label{figure7_time}
\end{minipage}
\begin{minipage}[b]{0.005\linewidth}
\hfill
\end{minipage}
\begin{minipage}[b]{0.32\linewidth}
\centering
%
%
\definecolor{mycolor1}{rgb}{0.00000,0.44700,0.74100}%
\begin{tikzpicture}

\begin{axis}[%
width=1.794in,
height=1.03in,
at={(1.358in,0.0in)},
scale only axis,
xmin=0,
xmax=10,
xlabel style={font=\color{white!15!black}},
xlabel={time-based tax},
ymin=2000,
ymax=45000,
ylabel style={font=\color{white!15!black}},
ylabel={platform revenue},
axis background/.style={fill=white},
legend style={legend cell align=left, align=left, draw=white!15!black}
]
\addplot [color=black, line width=1.0pt]
  table[row sep=crcr]{%
0	40877.8264253881\\
0.101010101010101	40477.0034838261\\
0.202020202020202	40076.1805422641\\
0.303030303030303	39675.357600702\\
0.404040404040404	39274.53465914\\
0.505050505050505	38873.711717578\\
0.606060606060606	38472.8887760159\\
0.707070707070707	38072.0658344539\\
0.808080808080808	37671.2428928919\\
0.909090909090909	37270.4199513298\\
1.01010101010101	36869.5970097678\\
1.11111111111111	36468.7740682057\\
1.21212121212121	36067.9511266437\\
1.31313131313131	35667.1281850817\\
1.41414141414141	35266.3052435196\\
1.51515151515152	34865.4823019576\\
1.61616161616162	34464.6593603956\\
1.71717171717172	34063.8364188335\\
1.81818181818182	33663.0134772715\\
1.91919191919192	33262.1905357095\\
2.02020202020202	32861.3675941474\\
2.12121212121212	32460.5446525854\\
2.22222222222222	32059.7217110234\\
2.32323232323232	31658.8987694613\\
2.42424242424242	31258.0758278993\\
2.52525252525253	30857.2528863373\\
2.62626262626263	30456.4299447752\\
2.72727272727273	30055.6070032132\\
2.82828282828283	29654.7840616512\\
2.92929292929293	29253.9611200891\\
3.03030303030303	28853.1381785271\\
3.13131313131313	28452.3152369651\\
3.23232323232323	28051.492295403\\
3.33333333333333	27650.669353841\\
3.43434343434343	27249.846412279\\
3.53535353535354	26849.0234707169\\
3.63636363636364	26448.2005291549\\
3.73737373737374	26047.3775875929\\
3.83838383838384	25646.5546460308\\
3.93939393939394	25245.7317044688\\
4.04040404040404	24844.9087629068\\
4.14141414141414	24444.0858213447\\
4.24242424242424	24043.2628797827\\
4.34343434343434	23642.4399382206\\
4.44444444444444	23241.6169966586\\
4.54545454545454	22840.7940550966\\
4.64646464646465	22439.9711135345\\
4.74747474747475	22039.1481719725\\
4.84848484848485	21638.3252304105\\
4.94949494949495	21237.5022888484\\
5.05050505050505	20836.6793472864\\
5.15151515151515	20435.8564057244\\
5.25252525252525	20035.0334641623\\
5.35353535353535	19634.2105226003\\
5.45454545454546	19233.3875810383\\
5.55555555555556	18832.5646394762\\
5.65656565656566	18431.7416979142\\
5.75757575757576	18030.9187563522\\
5.85858585858586	17630.0958147901\\
5.95959595959596	17229.2728732281\\
6.06060606060606	16828.4499316661\\
6.16161616161616	16427.626990104\\
6.26262626262626	16026.8632907397\\
6.36363636363636	15627.4749672371\\
6.46464646464646	15230.0040498666\\
6.56565656565657	14834.4521304064\\
6.66666666666667	14440.8208623925\\
6.76767676767677	14049.111962596\\
6.86868686868687	13659.3272125568\\
6.96969696969697	13271.4684601778\\
7.07070707070707	12885.5376213824\\
7.17171717171717	12501.5366818377\\
7.27272727272727	12119.4676987472\\
7.37373737373737	11739.3328027165\\
7.47474747474747	11361.134199695\\
7.57575757575758	10984.8741729991\\
7.67676767676768	10610.5550854191\\
7.77777777777778	10238.1793814152\\
7.87878787878788	9867.74958940901\\
7.97979797979798	9499.2683241713\\
8.08080808080808	9132.73828931635\\
8.18181818181818	8768.16227990541\\
8.28282828282828	8405.54318516595\\
8.38383838383838	8044.88399133511\\
8.48484848484848	7686.18778463116\\
8.58585858585859	7329.45775436362\\
8.68686868686869	6974.69719618843\\
8.78787878787879	6621.90951551689\\
8.88888888888889	6271.09823108836\\
8.98989898989899	5922.26697871608\\
9.09090909090909	5575.41951521696\\
9.19191919191919	5230.55972253766\\
9.29292929292929	4887.69161208778\\
9.39393939393939	4546.81932929527\\
9.49494949494949	4207.94715839811\\
9.5959595959596	3871.07952748725\\
9.6969696969697	3536.22101381836\\
9.7979797979798	3203.37634941147\\
9.8989898989899	2872.55042695686\\
10	2543.74830605038\\
};

\end{axis}
\end{tikzpicture}%
\vspace*{-0.3in}
\caption{Per-hour TNC profit under different time-based congestion charge.}
\label{figure8_time}
\end{minipage}
\begin{minipage}[b]{0.005\linewidth}
\hfill
\end{minipage}
\begin{minipage}[b]{0.32\linewidth}
\centering
%
%
\definecolor{mycolor1}{rgb}{0.00000,0.44700,0.74100}%
\begin{tikzpicture}

\begin{axis}[%
width=1.794in,
height=1.03in,
at={(1.358in,0.0in)},
scale only axis,
xmin=0,
xmax=10,
xlabel style={font=\color{white!15!black}},
xlabel={time-based tax},
ymin=0,
ymax=35000,
ylabel style={font=\color{white!15!black}},
ylabel={tax revenue},
axis background/.style={fill=white},
legend style={legend cell align=left, align=left, draw=white!15!black}
]
\addplot [color=black, line width=1.0pt]
  table[row sep=crcr]{%
0	0\\
0.101010101010101	400.822941562034\\
0.202020202020202	801.645883124069\\
0.303030303030303	1202.4688246861\\
0.404040404040404	1603.29176624814\\
0.505050505050505	2004.11470781017\\
0.606060606060606	2404.93764937221\\
0.707070707070707	2805.76059093424\\
0.808080808080808	3206.58353249628\\
0.909090909090909	3607.40647405831\\
1.01010101010101	4008.22941562034\\
1.11111111111111	4409.05235718238\\
1.21212121212121	4809.87529874441\\
1.31313131313131	5210.69824030645\\
1.41414141414141	5611.52118186848\\
1.51515151515152	6012.34412343052\\
1.61616161616162	6413.16706499255\\
1.71717171717172	6813.99000655459\\
1.81818181818182	7214.81294811662\\
1.91919191919192	7615.63588967865\\
2.02020202020202	8016.45883124069\\
2.12121212121212	8417.28177280272\\
2.22222222222222	8818.10471436476\\
2.32323232323232	9218.92765592679\\
2.42424242424242	9619.75059748883\\
2.52525252525253	10020.5735390509\\
2.62626262626263	10421.3964806129\\
2.72727272727273	10822.2194221749\\
2.82828282828283	11223.042363737\\
2.92929292929293	11623.865305299\\
3.03030303030303	12024.688246861\\
3.13131313131313	12425.5111884231\\
3.23232323232323	12826.3341299851\\
3.33333333333333	13227.1570715471\\
3.43434343434343	13627.9800131092\\
3.53535353535354	14028.8029546712\\
3.63636363636364	14429.6258962332\\
3.73737373737374	14830.4488377953\\
3.83838383838384	15231.2717793573\\
3.93939393939394	15632.0947209193\\
4.04040404040404	16032.9176624814\\
4.14141414141414	16433.7406040434\\
4.24242424242424	16834.5635456054\\
4.34343434343434	17235.3864871675\\
4.44444444444444	17636.2094287295\\
4.54545454545454	18037.0323702915\\
4.64646464646465	18437.8553118536\\
4.74747474747475	18838.6782534156\\
4.84848484848485	19239.5011949777\\
4.94949494949495	19640.3241365397\\
5.05050505050505	20041.1470781017\\
5.15151515151515	20441.9700196638\\
5.25252525252525	20842.7929612258\\
5.35353535353535	21243.6159027878\\
5.45454545454546	21644.4388443499\\
5.55555555555556	22045.2617859119\\
5.65656565656566	22446.0847274739\\
5.75757575757576	22846.907669036\\
5.85858585858586	23247.730610598\\
5.95959595959596	23648.55355216\\
6.06060606060606	24049.3764937221\\
6.16161616161616	24450.1994352841\\
6.26262626262626	24821.484565176\\
6.36363636363636	25101.0813557051\\
6.46464646464646	25376.7477906669\\
6.56565656565657	25648.4712521948\\
6.66666666666667	25916.2454438625\\
6.76767676767677	26180.0582772444\\
6.86868686868687	26439.9000929554\\
6.96969696969697	26695.7632333728\\
7.07070707070707	26947.6362746199\\
7.17171717171717	27195.5075723625\\
7.27272727272727	27439.3656658373\\
7.37373737373737	27679.1995579029\\
7.47474747474747	27914.9992011026\\
7.57575757575758	28146.7465617013\\
7.67676767676768	28374.4334460856\\
7.77777777777778	28598.0444518752\\
7.87878787878788	28817.5652547802\\
7.97979797979798	29032.9817089911\\
8.08080808080808	29244.2803634321\\
8.18181818181818	29451.441493304\\
8.28282828282828	29654.4523394623\\
8.38383838383838	29853.2915185005\\
8.48484848484848	30047.9478373912\\
8.58585858585859	30238.3967723823\\
8.68686868686869	30424.6227469376\\
8.78787878787879	30606.6066800126\\
8.88888888888889	30784.3246221727\\
8.98989898989899	30957.7596005953\\
9.09090909090909	31126.8852050681\\
9.19191919191919	31291.6799234602\\
9.29292929292929	31452.1223979843\\
9.39393939393939	31608.1854534814\\
9.49494949494949	31759.840287188\\
9.5959595959596	31907.0664197762\\
9.6969696969697	32049.8299984084\\
9.7979797979798	32188.1046132194\\
9.8989898989899	32321.8575711259\\
10	32451.0609381859\\
};

\end{axis}
\end{tikzpicture}%
\vspace*{-0.3in}
\caption{Per-hour tax revenue  under different time-based congestion charge.}
\label{figure9_time}
\end{minipage}
\end{figure*}


Figure \ref{figure1_time} - Figure \ref{figure3_time} display the number of drivers, passenger arrival rates, and  vehicle occupancy as a function of the time-based congestion charge. Figure \ref{figure4_time} shows the ride fare and per-trip driver payment. Figure \ref{figure5_time} and Figure \ref{figure6_time} show the passenger pickup time and total travel cost. Figure \ref{figure7_time} shows the driver wage. Figure \ref{figure8_time} and Figure \ref{figure9_time} present the platform profit and tax revenue, respectively. Clearly, the plots in Figure \ref{figure1_time}-\ref{figure9_time} have two distinct  regimes:
\begin{itemize}
    \item when $p_h\leq \$6.2$/hour the number of TNC drivers and the passenger arrival rate remain constant. So do the occupancy rate, ride fare, per-trip driver payment, pickup time, passenger travel cost and driver wage. The platform revenue reduces linearly, and the tax revenue also increases  linearly.
    \item when $p_h> \$6.2$/hour the numbers of drivers and passengers decline. Vehicle occupancy, ride fare ($p_f$) and per-trip driver payment ($p_d$) also decline. The pickup time and passenger travel cost increase. The driver wage is constant and equals the minimum wage. The platform profit reduces  and the tax revenue increases. 
\end{itemize}

Simulation results suggest that the time-based congestion charge does not affect the number of TNC vehicles unless the charge is greater than $\$6.2$/hour. In that case the effect of the congestion charge on congestion relief is mitigated by the minimum wage on TNC drivers. This observation is consistent with the results in Section \ref{trip_analysis_sec} and for the same reason. However, in contrast with the trip-based  charge, the time-based charge does not affect  passenger arrivals (Figure \ref{figure2_time}). This indicates that the time-based charge leads to a direct money transfer from the platform to the city in the first regime without affecting the passengers or drivers. This is evidenced by the linear curves in the first regime of Figure \ref{figure8_time}-\ref{figure9_time}.

The quantitative results in Figure \ref{figure1_time}-Figure \ref{figure9_time} are robust with respect  to the variation of model parameters. Formally,   denote by ${N^*_h}({p_h})$ and $\lambda^*_h(p_h)$  the optimal number of drivers and passenger arrival rates to (\ref{optimalpricing_time})  under a fixed wage floor. We have the following result. 
\begin{theorem}
Assume that (\ref{optimalpricing_time}) has a unique solution. For any model parameters $\lambda_0, N_0$, and $\alpha$, any strictly decreasing function ${F_p}(c)$, any strictly increasing function ${F_d}(w)$, any pickup time function ${t_p}$ that satisfies Assumption \ref{assumption1}, and any speed-density relation $v(N)$ that satisfies Assumption \ref{assumption2}, there exists ${w_2} >  \tilde{w},$ such that for any $\tilde{w} < {w_0} < {w_2}, $ there exists ${\bar p_h} > 0, $ so that $\partial {N^*_h}/\partial{p_h} = 0$ and $\partial  {\lambda^*_h}/\partial{p_h} = 0$ for $\forall {p_h} \in (0,{\bar p_h}).$
\label{theorem2}
\end{theorem}

The proof of Theorem \ref{theorem2} can be found in Appendix D. Theorem \ref{theorem2}  states that there exists a regime under which both the number of TNC drivers and the passenger arrival rates are unaffected by the congestion charge. This  indicates that the ride fare, driver wage and passenger cost remain constant in this regime and the congestion charge is entirely imposed on the platform through a direct money transfer from the platform to the city. In this scheme, congestion charge will not directly curb the congestion by reducing  traffic in the city. Instead, it can only indirectly mitigate traffic congestion by collecting taxes to subsidize public transit.

\section{Comparison between time-based and trip-based charges}

\begin{figure*}[bt]%
\begin{minipage}[b]{0.32\linewidth}
\centering
%
%
\definecolor{mycolor1}{rgb}{0.00000,0.44700,0.74100}%
\definecolor{mycolor2}{rgb}{0.85000,0.32500,0.09800}%
\begin{tikzpicture}

\pgfplotsset{every axis y label/.style={
at={(-0.47,0.5)},
xshift=30pt,
rotate=90}}

\begin{axis}[%
width=1.794in,
height=1.03in,
at={(1.358in,0.0in)},
scale only axis,
xmin=0,
xmax=35000,
xlabel style={font=\color{white!15!black}},
xlabel={tax revenue/hour},
ymin=3200,
ymax=4200,
ytick={3200,3600,4000},
yticklabels={{3.2K},{3.6K},{4K}},
ylabel style={font=\color{white!15!black}},
ylabel={Number of drivers},
axis background/.style={fill=white},
legend style={at={(0.0,0.0)}, anchor=south west, legend cell align=left, align=left, draw=white!15!black}
]
\addplot [color=black, line width=1.0pt]
  table[row sep=crcr]{%
0	3968.14712146414\\
400.822941562034	3968.14712146414\\
801.645883124069	3968.14712146414\\
1202.4688246861	3968.14712146414\\
1603.29176624814	3968.14712146414\\
2004.11470781017	3968.14712146414\\
2404.93764937221	3968.14712146414\\
2805.76059093424	3968.14712146414\\
3206.58353249628	3968.14712146414\\
3607.40647405831	3968.14712146414\\
4008.22941562034	3968.14712146414\\
4409.05235718238	3968.14712146414\\
4809.87529874441	3968.14712146414\\
5210.69824030645	3968.14712146414\\
5611.52118186848	3968.14712146414\\
6012.34412343052	3968.14712146414\\
6413.16706499255	3968.14712146414\\
6813.99000655459	3968.14712146414\\
7214.81294811662	3968.14712146414\\
7615.63588967865	3968.14712146414\\
8016.45883124069	3968.14712146414\\
8417.28177280272	3968.14712146414\\
8818.10471436476	3968.14712146414\\
9218.92765592679	3968.14712146414\\
9619.75059748883	3968.14712146414\\
10020.5735390509	3968.14712146414\\
10421.3964806129	3968.14712146414\\
10822.2194221749	3968.14712146414\\
11223.042363737	3968.14712146414\\
11623.865305299	3968.14712146414\\
12024.688246861	3968.14712146414\\
12425.5111884231	3968.14712146414\\
12826.3341299851	3968.14712146414\\
13227.1570715471	3968.14712146414\\
13627.9800131092	3968.14712146414\\
14028.8029546712	3968.14712146414\\
14429.6258962332	3968.14712146414\\
14830.4488377953	3968.14712146414\\
15231.2717793573	3968.14712146414\\
15632.0947209193	3968.14712146414\\
16032.9176624814	3968.14712146414\\
16433.7406040434	3968.14712146414\\
16834.5635456054	3968.14712146414\\
17235.3864871675	3968.14712146414\\
17636.2094287295	3968.14712146414\\
18037.0323702915	3968.14712146414\\
18437.8553118536	3968.14712146414\\
18838.6782534156	3968.14712146414\\
19239.5011949777	3968.14712146414\\
19640.3241365397	3968.14712146414\\
20041.1470781017	3968.14712146414\\
20441.9700196638	3968.14712146414\\
20842.7929612258	3968.14712146414\\
21243.6159027878	3968.14712146414\\
21644.4388443499	3968.14712146414\\
22045.2617859119	3968.14712146414\\
22446.0847274739	3968.14712146414\\
22846.907669036	3968.14712146414\\
23247.730610598	3968.14712146414\\
23648.55355216	3968.14712146414\\
24049.3764937221	3968.14712146414\\
24450.1994352841	3968.14712146414\\
24821.484565176	3963.43059992327\\
25101.0813557051	3944.4556416108\\
25376.7477906669	3925.46567386878\\
25648.4712521948	3906.45946764197\\
25916.2454438625	3887.43681657938\\
26180.0582772444	3868.3966708167\\
26439.9000929554	3849.33839588616\\
26695.7632333728	3830.26168131001\\
26947.6362746199	3811.16570169624\\
27195.5075723625	3792.04964741393\\
27439.3656658373	3772.91277905263\\
27679.1995579029	3753.75446059231\\
27914.9992011026	3734.5742174448\\
28146.7465617013	3715.37054614458\\
28374.4334460856	3696.14330416115\\
28598.0444518752	3676.89142952682\\
28817.5652547802	3657.61405156825\\
29032.9817089911	3638.31036606345\\
29244.2803634321	3618.97969497472\\
29451.441493304	3599.62062695937\\
29654.4523394623	3580.23266049606\\
29853.2915185005	3560.81428955609\\
30047.9478373912	3541.36528083539\\
30238.3967723823	3521.88385937158\\
30424.6227469376	3502.36936272887\\
30606.6066800126	3482.82076013936\\
30784.3246221727	3463.23651999442\\
30957.7596005953	3443.61595557184\\
31126.8852050681	3423.95737255749\\
31291.6799234602	3404.25968398083\\
31452.1223979843	3384.52186673961\\
31608.1854534814	3364.74232246738\\
31759.840287188	3344.91934939533\\
31907.0664197762	3325.05218479773\\
32049.8299984084	3305.13871858587\\
32188.1046132194	3285.17768732858\\
32321.8575711259	3265.16724443007\\
32451.0609381859	3245.10609381859\\
};
\addlegendentry{time-based tax}

\addplot [color=mycolor1, dashed, line width=1.0pt]
  table[row sep=crcr]{%
0	3968.14712146414\\
378.707477583736	3968.14712146414\\
756.805539388852	3968.14712146414\\
1134.291868209	3968.14712146414\\
1511.16432016694	3968.14712146414\\
1887.42069745547	3968.14712146414\\
2263.05860177174	3968.14712146414\\
2638.07588716184	3968.14712146414\\
3012.47043557456	3968.14712146414\\
3386.23950918242	3968.14712146414\\
3759.38105741824	3968.14712146414\\
4131.89284581028	3968.14712146414\\
4503.7726290217	3968.14712146414\\
4875.01814949863	3968.14712146414\\
5245.62666936206	3968.14712146414\\
5615.59591099217	3968.14712146414\\
5984.92404239672	3968.14712146414\\
6353.60830180276	3968.14712146414\\
6721.646380254	3968.14712146414\\
7089.0363196865	3968.14712146414\\
7455.77477845368	3968.14712146414\\
7821.8601072554	3968.14712146414\\
8187.28995065928	3968.14712146414\\
8552.06144798892	3968.14712146414\\
8916.17165420025	3968.14712146414\\
9279.61953370031	3968.14712146414\\
9642.4027915463	3968.14712146414\\
10004.5168400589	3968.14712146414\\
10365.9616248833	3968.14712146414\\
10726.7329621335	3968.14712146414\\
11086.8288101225	3968.14712146414\\
11446.2474588592	3968.14712146414\\
11804.9854704934	3968.14712146414\\
12163.040976672	3968.14712146414\\
12520.4098979644	3968.14712146414\\
12877.0922173005	3968.14712146414\\
13233.0835118354	3968.14712146414\\
13588.3831083545	3968.14712146414\\
13942.9857758188	3968.14712146414\\
14296.8911421645	3968.14712146414\\
14650.0954580335	3968.14712146414\\
15002.5969463215	3968.14712146414\\
15354.3918313298	3968.14712146414\\
15705.4775755397	3968.14712146414\\
16055.8533419442	3968.14712146414\\
16405.5159253404	3968.14712146414\\
16754.4614792134	3968.14712146414\\
17102.6881127994	3968.14712146414\\
17450.1919733931	3968.14712146414\\
17796.9736424881	3968.14712146414\\
18143.0259189637	3968.14712146414\\
18488.3497800187	3968.14712146414\\
18832.9408586547	3968.14712146414\\
19176.7966820294	3968.14712146414\\
19519.9148813274	3968.14712146414\\
19862.2910944653	3968.14712146414\\
20203.9266788049	3968.14712146414\\
20544.8132648811	3968.14712146414\\
20884.952359649	3968.14712146414\\
21224.3394062267	3968.14712146414\\
21562.972111313	3968.14712146414\\
21900.8474659068	3968.14712146414\\
22237.9615791096	3968.14712146414\\
22574.3158257629	3968.14712146414\\
22909.9013189879	3968.14712146414\\
23244.7186685271	3968.14712146414\\
23578.7655131032	3968.14712146414\\
23912.03822128	3968.14712146414\\
24244.5319409983	3968.14712146414\\
24576.2487105883	3968.14712146414\\
24873.2654724643	3962.5332677229\\
25092.4435329626	3944.49494467638\\
25307.4914513819	3926.40837145426\\
25518.3993497632	3908.27355554703\\
25725.1454133268	3890.08920145607\\
25927.7176563731	3871.8552668891\\
26126.0979754106	3853.57070110621\\
26320.272163713	3835.23522860657\\
26510.2210242812	3816.84796024135\\
26695.9307796062	3798.4084773862\\
26877.3836569126	3779.91611290089\\
27054.5600445229	3761.36998722856\\
27227.4414880799	3742.76918172402\\
27396.0133222173	3724.11346244639\\
27560.2534302226	3705.40162183681\\
27720.1434845531	3686.63299036701\\
27875.6668946439	3667.80707697129\\
28026.7991690679	3648.92243919869\\
28173.5226237634	3629.97871103748\\
28315.8133051028	3610.97444509346\\
28453.6540396952	3591.90928976276\\
28587.021095821	3572.78210810502\\
28715.8923798213	3553.59186553545\\
28840.2431105799	3534.33735429618\\
28960.0513685829	3515.01775231061\\
29075.2930182	3495.63190359766\\
29185.9432295207	3476.17871459978\\
29291.971418415	3456.65653501054\\
29393.3589281575	3437.06489359696\\
29490.0731683358	3417.40194940702\\
};
\addlegendentry{trip-based tax}

\end{axis}
\end{tikzpicture}%
\vspace*{-0.3in}
\caption{Number of drivers under different schemes of congestion surcharge. }
\label{figure1_compare}
\end{minipage}
\begin{minipage}[b]{0.005\linewidth}
\hfill
\end{minipage}
\begin{minipage}[b]{0.32\linewidth}
\centering
%
%
\definecolor{mycolor1}{rgb}{0.00000,0.44700,0.74100}%
\definecolor{mycolor2}{rgb}{0.85000,0.32500,0.09800}%
\begin{tikzpicture}

\pgfplotsset{every axis y label/.style={
at={(-0.47,0.5)},
xshift=30pt,
rotate=90}}

\begin{axis}[%
width=1.794in,
height=1.03in,
at={(1.358in,0.0in)},
scale only axis,
xmin=0,
xmax=35000,
xlabel style={font=\color{white!15!black}},
xlabel={tax revenue/hour},
ymin=150,
ymax=220,
ylabel style={font=\color{white!15!black}},
ylabel={Passenger arrival},
axis background/.style={fill=white},
legend style={at={(0.0,0.0)}, anchor=south west, legend cell align=left, align=left, draw=white!15!black}
]
\addplot [color=black, line width=1.0pt]
  table[row sep=crcr]{%
0	208.456289196894\\
400.822941562034	208.456289196894\\
801.645883124069	208.456289196894\\
1202.4688246861	208.456289196894\\
1603.29176624814	208.456289196894\\
2004.11470781017	208.456289196894\\
2404.93764937221	208.456289196894\\
2805.76059093424	208.456289196894\\
3206.58353249628	208.456289196894\\
3607.40647405831	208.456289196894\\
4008.22941562034	208.456289196894\\
4409.05235718238	208.456289196894\\
4809.87529874441	208.456289196894\\
5210.69824030645	208.456289196894\\
5611.52118186848	208.456289196894\\
6012.34412343052	208.456289196894\\
6413.16706499255	208.456289196894\\
6813.99000655459	208.456289196894\\
7214.81294811662	208.456289196894\\
7615.63588967865	208.456289196894\\
8016.45883124069	208.456289196894\\
8417.28177280272	208.456289196894\\
8818.10471436476	208.456289196894\\
9218.92765592679	208.456289196894\\
9619.75059748883	208.456289196894\\
10020.5735390509	208.456289196894\\
10421.3964806129	208.456289196894\\
10822.2194221749	208.456289196894\\
11223.042363737	208.456289196894\\
11623.865305299	208.456289196894\\
12024.688246861	208.456289196894\\
12425.5111884231	208.456289196894\\
12826.3341299851	208.456289196894\\
13227.1570715471	208.456289196894\\
13627.9800131092	208.456289196894\\
14028.8029546712	208.456289196894\\
14429.6258962332	208.456289196894\\
14830.4488377953	208.456289196894\\
15231.2717793573	208.456289196894\\
15632.0947209193	208.456289196894\\
16032.9176624814	208.456289196894\\
16433.7406040434	208.456289196894\\
16834.5635456054	208.456289196894\\
17235.3864871675	208.456289196894\\
17636.2094287295	208.456289196894\\
18037.0323702915	208.456289196894\\
18437.8553118536	208.456289196894\\
18838.6782534156	208.456289196894\\
19239.5011949777	208.456289196894\\
19640.3241365397	208.456289196894\\
20041.1470781017	208.456289196894\\
20441.9700196638	208.456289196894\\
20842.7929612258	208.456289196894\\
21243.6159027878	208.456289196894\\
21644.4388443499	208.456289196894\\
22045.2617859119	208.456289196894\\
22446.0847274739	208.456289196894\\
22846.907669036	208.456289196894\\
23247.730610598	208.456289196894\\
23648.55355216	208.456289196894\\
24049.3764937221	208.456289196894\\
24450.1994352841	208.456289196894\\
24821.484565176	208.216310851355\\
25101.0813557051	207.249885613567\\
25376.7477906669	206.281233856883\\
25648.4712521948	205.310290879597\\
25916.2454438625	204.337053446597\\
26180.0582772444	203.361459136469\\
26439.9000929554	202.383481405233\\
26695.7632333728	201.403102126841\\
26947.6362746199	200.420285157392\\
27195.5075723625	199.434985511578\\
27439.3656658373	198.447168782186\\
27679.1995579029	197.456795667531\\
27914.9992011026	196.463845680477\\
28146.7465617013	195.468244335302\\
28374.4334460856	194.469977957733\\
28598.0444518752	193.468996167103\\
28817.5652547802	192.465256252192\\
29032.9817089911	191.458710407992\\
29244.2803634321	190.449324081149\\
29451.441493304	189.437025317164\\
29654.4523394623	188.421785377132\\
29853.2915185005	187.40352187648\\
30047.9478373912	186.382227489432\\
30238.3967723823	185.357805340147\\
30424.6227469376	184.330224214341\\
30606.6066800126	183.299422243154\\
30784.3246221727	182.265319994941\\
30957.7596005953	181.22787801933\\
31126.8852050681	180.187007498674\\
31291.6799234602	179.142648886375\\
31452.1223979843	178.094746154089\\
31608.1854534814	177.04321341738\\
31759.840287188	175.987954076955\\
31907.0664197762	174.928929655892\\
32049.8299984084	173.866024003897\\
32188.1046132194	172.799165547296\\
32321.8575711259	171.728253682768\\
32451.0609381859	170.653219140105\\
};
\addlegendentry{time-based tax}

\addplot [color=mycolor1, dashed, line width=1.0pt]
  table[row sep=crcr]{%
0	208.456289196894\\
378.707477583736	208.289112671055\\
756.805539388852	208.121523331934\\
1134.291868209	207.953509171651\\
1511.16432016694	207.785094022954\\
1887.42069745547	207.616276720102\\
2263.05860177174	207.447038495743\\
2638.07588716184	207.277391134145\\
3012.47043557456	207.107342445751\\
3386.23950918242	206.936858894481\\
3759.38105741824	206.765958158003\\
4131.89284581028	206.594642290514\\
4503.7726290217	206.422912163495\\
4875.01814949863	206.250767863404\\
5245.62666936206	206.078190582081\\
5615.59591099217	205.905183403046\\
5984.92404239672	205.731763957387\\
6353.60830180276	205.55791564656\\
6721.646380254	205.38363939665\\
7089.0363196865	205.208946096188\\
7455.77477845368	205.033806407476\\
7821.8601072554	204.858240904308\\
8187.28995065928	204.682248766482\\
8552.06144798892	204.505817234518\\
8916.17165420025	204.328933742089\\
9279.61953370031	204.151629741407\\
9642.4027915463	203.973905205787\\
10004.5168400589	203.795713408607\\
10365.9616248833	203.617103345923\\
10726.7329621335	203.438038937015\\
11086.8288101225	203.25852818558\\
11446.2474588592	203.078583947501\\
11804.9854704934	202.898187774106\\
12163.040976672	202.717349611199\\
12520.4098979644	202.53604246707\\
12877.0922173005	202.354306271866\\
13233.0835118354	202.172109208596\\
13588.3831083545	201.989478637702\\
13942.9857758188	201.806373071061\\
14296.8911421645	201.622823799756\\
14650.0954580335	201.438812547961\\
15002.5969463215	201.254349279923\\
15354.3918313298	201.069416838842\\
15705.4775755397	200.884015501089\\
16055.8533419442	200.698166774303\\
16405.5159253404	200.511861309716\\
16754.4614792134	200.325082903638\\
17102.6881127994	200.137839617865\\
17450.1919733931	199.950116361796\\
17796.9736424881	199.761949048336\\
18143.0259189637	199.573285108601\\
18488.3497800187	199.384164294319\\
18832.9408586547	199.194566774232\\
19176.7966820294	199.004493870116\\
19519.9148813274	198.813947865372\\
19862.2910944653	198.622910944653\\
20203.9266788049	198.431422738262\\
20544.8132648811	198.239426240081\\
20884.952359649	198.046962031154\\
21224.3394062267	197.854011413978\\
21562.972111313	197.660577687036\\
21900.8474659068	197.466657479488\\
22237.9615791096	197.272239814682\\
22574.3158257629	197.077360383644\\
22909.9013189879	196.881964460052\\
23244.7186685271	196.686081041383\\
23578.7655131032	196.489712609193\\
23912.03822128	196.292851070209\\
24244.5319409983	196.095478934545\\
24576.2487105883	195.897634649617\\
24873.2654724643	195.432800140791\\
25092.4435329626	194.378083706048\\
25307.4914513819	193.321115253612\\
25518.3993497632	192.261912909174\\
25725.1454133268	191.20040509905\\
25927.7176563731	190.136596146736\\
26126.0979754106	189.070445874682\\
26320.272163713	188.001944026521\\
26510.2210242812	186.931045684034\\
26695.9307796062	185.857745933967\\
26877.3836569126	184.782012641274\\
27054.5600445229	183.703802771451\\
27227.4414880799	182.623083151755\\
27396.0133222173	181.539847315898\\
27560.2534302226	180.454040316934\\
27720.1434845531	179.365634311814\\
27875.6668946439	178.274613861095\\
28026.7991690679	177.180914287211\\
28173.5226237634	176.084516398521\\
28315.8133051028	174.985363121421\\
28453.6540396952	173.883441353693\\
28587.021095821	172.778698930786\\
28715.8923798213	171.671095748932\\
28840.2431105799	170.560577535687\\
28960.0513685829	169.447109071496\\
29075.2930182	168.330643789579\\
29185.9432295207	167.211133085796\\
29291.971418415	166.088497733281\\
29393.3589281575	164.962728678435\\
29490.0731683358	163.833739824088\\
};
\addlegendentry{trip-based tax}

\end{axis}
\end{tikzpicture}%
\vspace*{-0.3in}
\caption{Comparison of passenger arrival rate (per minute).} 
\label{figure2_compare}
\end{minipage}
\begin{minipage}[b]{0.005\linewidth}
\hfill
\end{minipage}
\begin{minipage}[b]{0.32\linewidth}
\centering
%
%
\definecolor{mycolor1}{rgb}{0.00000,0.44700,0.74100}%
\definecolor{mycolor2}{rgb}{0.85000,0.32500,0.09800}%
\begin{tikzpicture}

\pgfplotsset{every axis y label/.style={
at={(-0.45,0.5)},
xshift=32pt,
rotate=90}}

\begin{axis}[%
width=1.794in,
height=1.03in,
at={(1.358in,0.0in)},
scale only axis,
xmin=0,
xmax=35000,
xlabel style={font=\color{white!15!black}},
xlabel={tax revenue/hour},
ymin=0.50,
ymax=0.62,
ylabel style={font=\color{white!15!black}},
ylabel={Occupancy},
axis background/.style={fill=white},
legend style={at={(0.0,0.0)}, anchor=south west, legend cell align=left, align=left, draw=white!15!black}
]
\addplot [color=black, line width=1.0pt]
  table[row sep=crcr]{%
0	0.597762201599234\\
400.822941562034	0.597762201599234\\
801.645883124069	0.597762201599234\\
1202.4688246861	0.597762201599234\\
1603.29176624814	0.597762201599234\\
2004.11470781017	0.597762201599234\\
2404.93764937221	0.597762201599234\\
2805.76059093424	0.597762201599234\\
3206.58353249628	0.597762201599234\\
3607.40647405831	0.597762201599234\\
4008.22941562034	0.597762201599234\\
4409.05235718238	0.597762201599234\\
4809.87529874441	0.597762201599234\\
5210.69824030645	0.597762201599234\\
5611.52118186848	0.597762201599234\\
6012.34412343052	0.597762201599234\\
6413.16706499255	0.597762201599234\\
6813.99000655459	0.597762201599234\\
7214.81294811662	0.597762201599234\\
7615.63588967865	0.597762201599234\\
8016.45883124069	0.597762201599234\\
8417.28177280272	0.597762201599234\\
8818.10471436476	0.597762201599234\\
9218.92765592679	0.597762201599234\\
9619.75059748883	0.597762201599234\\
10020.5735390509	0.597762201599234\\
10421.3964806129	0.597762201599234\\
10822.2194221749	0.597762201599234\\
11223.042363737	0.597762201599234\\
11623.865305299	0.597762201599234\\
12024.688246861	0.597762201599234\\
12425.5111884231	0.597762201599234\\
12826.3341299851	0.597762201599234\\
13227.1570715471	0.597762201599234\\
13627.9800131092	0.597762201599234\\
14028.8029546712	0.597762201599234\\
14429.6258962332	0.597762201599234\\
14830.4488377953	0.597762201599234\\
15231.2717793573	0.597762201599234\\
15632.0947209193	0.597762201599234\\
16032.9176624814	0.597762201599234\\
16433.7406040434	0.597762201599234\\
16834.5635456054	0.597762201599234\\
17235.3864871675	0.597762201599234\\
17636.2094287295	0.597762201599234\\
18037.0323702915	0.597762201599234\\
18437.8553118536	0.597762201599234\\
18838.6782534156	0.597762201599234\\
19239.5011949777	0.597762201599234\\
19640.3241365397	0.597762201599234\\
20041.1470781017	0.597762201599234\\
20441.9700196638	0.597762201599234\\
20842.7929612258	0.597762201599234\\
21243.6159027878	0.597762201599234\\
21644.4388443499	0.597762201599234\\
22045.2617859119	0.597762201599234\\
22446.0847274739	0.597762201599234\\
22846.907669036	0.597762201599234\\
23247.730610598	0.597762201599234\\
23648.55355216	0.597762201599234\\
24049.3764937221	0.597762201599234\\
24450.1994352841	0.597762201599234\\
24821.484565176	0.597722881144498\\
25101.0813557051	0.597562510203645\\
25376.7477906669	0.597398759802542\\
25648.4712521948	0.597231570584184\\
25916.2454438625	0.597060918269294\\
26180.0582772444	0.596886723489281\\
26439.9000929554	0.596708952430472\\
26695.7632333728	0.596527549393477\\
26947.6362746199	0.596342477840147\\
27195.5075723625	0.596153671946704\\
27439.3656658373	0.59596108632906\\
27679.1995579029	0.595764644960582\\
27914.9992011026	0.595564303763335\\
28146.7465617013	0.59536000482634\\
28374.4334460856	0.595151673986919\\
28598.0444518752	0.594939259455799\\
28817.5652547802	0.594722703602595\\
29032.9817089911	0.594501921874591\\
29244.2803634321	0.594276850299185\\
29451.441493304	0.5940474178436\\
29654.4523394623	0.593813549223234\\
29853.2915185005	0.593575152084157\\
30047.9478373912	0.593332174424798\\
30238.3967723823	0.593084516179767\\
30424.6227469376	0.592832112566686\\
30606.6066800126	0.592574856127755\\
30784.3246221727	0.592312659243409\\
30957.7596005953	0.592045428672301\\
31126.8852050681	0.591773068082544\\
31291.6799234602	0.591495478009751\\
31452.1223979843	0.591212558019355\\
31608.1854534814	0.590924199320349\\
31759.840287188	0.590630271641389\\
31907.0664197762	0.590330683136481\\
32049.8299984084	0.590025299327511\\
32188.1046132194	0.589713993577192\\
32321.8575711259	0.589396634285671\\
32451.0609381859	0.589073101845723\\
};
\addlegendentry{time-based tax}

\addplot [color=mycolor1, dashed, line width=1.0pt]
  table[row sep=crcr]{%
0	0.597762201599234\\
378.707477583736	0.597282811850303\\
756.805539388852	0.596802238331974\\
1134.291868209	0.596320446610851\\
1511.16432016694	0.595837505028923\\
1887.42069745547	0.595353410243346\\
2263.05860177174	0.594868108437499\\
2638.07588716184	0.594381633403547\\
3012.47043557456	0.593894007538456\\
3386.23950918242	0.593405134675296\\
3759.38105741824	0.592915065505949\\
4131.89284581028	0.592423805920959\\
4503.7726290217	0.591931358419358\\
4875.01814949863	0.591437723249071\\
5245.62666936206	0.590942846476454\\
5615.59591099217	0.590446736943704\\
5984.92404239672	0.589949445209132\\
6353.60830180276	0.589450923675313\\
6721.646380254	0.588951174997862\\
7089.0363196865	0.588450230400359\\
7455.77477845368	0.587948005755014\\
7821.8601072554	0.587444560058542\\
8187.28995065928	0.586939890958961\\
8552.06144798892	0.586433961867616\\
8916.17165420025	0.585926736749736\\
9279.61953370031	0.585418305796617\\
9642.4027915463	0.584908668931739\\
10004.5168400589	0.584397692163421\\
10365.9616248833	0.583885515991097\\
10726.7329621335	0.583372036950915\\
11086.8288101225	0.582857277994003\\
11446.2474588592	0.58234127598546\\
11804.9854704934	0.581823978022496\\
12163.040976672	0.581305412625363\\
12520.4098979644	0.580785502393545\\
12877.0922173005	0.580264361829379\\
13233.0835118354	0.579741899695545\\
13588.3831083545	0.579218194449915\\
13942.9857758188	0.578693127122554\\
14296.8911421645	0.578166787442711\\
14650.0954580335	0.577639123003246\\
15002.5969463215	0.577110162377297\\
15354.3918313298	0.576579856366604\\
15705.4775755397	0.576048205763682\\
16055.8533419442	0.575515272243103\\
16405.5159253404	0.574981028996664\\
16754.4614792134	0.574445429559207\\
17102.6881127994	0.573908497048464\\
17450.1919733931	0.573370188191141\\
17796.9736424881	0.572830605969872\\
18143.0259189637	0.572289599639898\\
18488.3497800187	0.571747283191933\\
18832.9408586547	0.57120359975853\\
19176.7966820294	0.57065855312997\\
19519.9148813274	0.570112149854056\\
19862.2910944653	0.569564338844107\\
20203.9266788049	0.569015233742631\\
20544.8132648811	0.568464671080817\\
20884.952359649	0.567912767227494\\
21224.3394062267	0.567359468566333\\
21562.972111313	0.566804784555954\\
21900.8474659068	0.56624870553025\\
22237.9615791096	0.565691200012944\\
22574.3158257629	0.565132370350419\\
22909.9013189879	0.564572059611319\\
23244.7186685271	0.564010350947831\\
23578.7655131032	0.563447251475981\\
23912.03822128	0.562882737987773\\
24244.5319409983	0.562316760329803\\
24576.2487105883	0.561749428752579\\
24873.2654724643	0.561141512204177\\
25092.4435329626	0.560444208576862\\
25307.4914513819	0.559742851350545\\
25518.3993497632	0.559037425819922\\
25725.1454133268	0.558327825813069\\
25927.7176563731	0.557613999999983\\
26126.0979754106	0.556895898121656\\
26320.272163713	0.556173453866022\\
26510.2210242812	0.555446578107108\\
26695.9307796062	0.554715235833543\\
26877.3836569126	0.553979342632672\\
27054.5600445229	0.553238807765524\\
27227.4414880799	0.552493573240331\\
27396.0133222173	0.551743567180067\\
27560.2534302226	0.550988700841582\\
27720.1434845531	0.550228892216719\\
27875.6668946439	0.549464070606931\\
28026.7991690679	0.548694142462944\\
28173.5226237634	0.547919003270441\\
28315.8133051028	0.547138578291814\\
28453.6540396952	0.546352774700609\\
28587.021095821	0.545561484318982\\
28715.8923798213	0.544764619531068\\
28840.2431105799	0.543962069304597\\
28960.0513685829	0.543153725199225\\
29075.2930182	0.542339486795899\\
29185.9432295207	0.541519234776503\\
29291.971418415	0.540692826962566\\
29393.3589281575	0.539860174902659\\
29490.0731683358	0.539021133748207\\
};
\addlegendentry{trip-based tax}

\end{axis}
\end{tikzpicture}%
\vspace*{-0.3in}
\caption{Occupancy rate under different congestion surcharges.}
\label{figure3_compare}
\end{minipage}
\begin{minipage}[b]{0.32\linewidth}
\centering
%
%
\definecolor{mycolor1}{rgb}{0.00000,0.44700,0.74100}%
\definecolor{mycolor2}{rgb}{0.85000,0.32500,0.09800}%
\begin{tikzpicture}

\begin{axis}[%
width=1.794in,
height=1.03in,
at={(1.358in,0.0in)},
scale only axis,
xmin=0,
xmax=35000,
xlabel style={font=\color{white!15!black}},
xlabel={tax revenue/hour},
ymin=11.5,
ymax=13,
ylabel style={font=\color{white!15!black}},
ylabel={Ride fare},
axis background/.style={fill=white},
legend style={at={(0.0,0.59)}, anchor=south west, legend cell align=left, align=left, draw=white!15!black}
]
\addplot [color=black, line width=1.0pt]
  table[row sep=crcr]{%
0	11.6282174100135\\
400.822941562034	11.6282174100135\\
801.645883124069	11.6282174100135\\
1202.4688246861	11.6282174100135\\
1603.29176624814	11.6282174100135\\
2004.11470781017	11.6282174100135\\
2404.93764937221	11.6282174100135\\
2805.76059093424	11.6282174100135\\
3206.58353249628	11.6282174100135\\
3607.40647405831	11.6282174100135\\
4008.22941562034	11.6282174100135\\
4409.05235718238	11.6282174100135\\
4809.87529874441	11.6282174100135\\
5210.69824030645	11.6282174100135\\
5611.52118186848	11.6282174100135\\
6012.34412343052	11.6282174100135\\
6413.16706499255	11.6282174100135\\
6813.99000655459	11.6282174100135\\
7214.81294811662	11.6282174100135\\
7615.63588967865	11.6282174100135\\
8016.45883124069	11.6282174100135\\
8417.28177280272	11.6282174100135\\
8818.10471436476	11.6282174100135\\
9218.92765592679	11.6282174100135\\
9619.75059748883	11.6282174100135\\
10020.5735390509	11.6282174100135\\
10421.3964806129	11.6282174100135\\
10822.2194221749	11.6282174100135\\
11223.042363737	11.6282174100135\\
11623.865305299	11.6282174100135\\
12024.688246861	11.6282174100135\\
12425.5111884231	11.6282174100135\\
12826.3341299851	11.6282174100135\\
13227.1570715471	11.6282174100135\\
13627.9800131092	11.6282174100135\\
14028.8029546712	11.6282174100135\\
14429.6258962332	11.6282174100135\\
14830.4488377953	11.6282174100135\\
15231.2717793573	11.6282174100135\\
15632.0947209193	11.6282174100135\\
16032.9176624814	11.6282174100135\\
16433.7406040434	11.6282174100135\\
16834.5635456054	11.6282174100135\\
17235.3864871675	11.6282174100135\\
17636.2094287295	11.6282174100135\\
18037.0323702915	11.6282174100135\\
18437.8553118536	11.6282174100135\\
18838.6782534156	11.6282174100135\\
19239.5011949777	11.6282174100135\\
19640.3241365397	11.6282174100135\\
20041.1470781017	11.6282174100135\\
20441.9700196638	11.6282174100135\\
20842.7929612258	11.6282174100135\\
21243.6159027878	11.6282174100135\\
21644.4388443499	11.6282174100135\\
22045.2617859119	11.6282174100135\\
22446.0847274739	11.6282174100135\\
22846.907669036	11.6282174100135\\
23247.730610598	11.6282174100135\\
23648.55355216	11.6282174100135\\
24049.3764937221	11.6282174100135\\
24450.1994352841	11.6282174100135\\
24821.484565176	11.6293118095835\\
25101.0813557051	11.6336986187072\\
25376.7477906669	11.6380575564872\\
25648.4712521948	11.6423885151698\\
25916.2454438625	11.6466906038034\\
26180.0582772444	11.6509640125124\\
26439.9000929554	11.6552079789661\\
26695.7632333728	11.6594221164922\\
26947.6362746199	11.6636056429744\\
27195.5075723625	11.6677583010673\\
27439.3656658373	11.6718793844454\\
27679.1995579029	11.6759687378347\\
27914.9992011026	11.6800255329073\\
28146.7465617013	11.6840491654277\\
28374.4334460856	11.6880392908935\\
28598.0444518752	11.6919950732999\\
28817.5652547802	11.695915731585\\
29032.9817089911	11.6998009555628\\
29244.2803634321	11.7036499744807\\
29451.441493304	11.7074620682693\\
29654.4523394623	11.7112365628439\\
29853.2915185005	11.7149730094596\\
30047.9478373912	11.7186701523053\\
30238.3967723823	11.722327525939\\
30424.6227469376	11.7259439504823\\
30606.6066800126	11.7295189883429\\
30784.3246221727	11.7330516815336\\
30957.7596005953	11.7365411623754\\
31126.8852050681	11.7399864404863\\
31291.6799234602	11.7433865417268\\
31452.1223979843	11.7467404204106\\
31608.1854534814	11.7500470243517\\
31759.840287188	11.7533055934692\\
31907.0664197762	11.7565146084769\\
32049.8299984084	11.7596730921405\\
32188.1046132194	11.7627798951339\\
32321.8575711259	11.7658337410963\\
32451.0609381859	11.7688330584048\\
};
\addlegendentry{time-based tax}

\addplot [color=mycolor1, dashed, line width=1.0pt]
  table[row sep=crcr]{%
0	11.6282174100135\\
378.707477583736	11.6375382566739\\
756.805539388852	11.646872749036\\
1134.291868209	11.6562215224015\\
1511.16432016694	11.66558321836\\
1887.42069745547	11.6749578715563\\
2263.05860177174	11.6843464925866\\
2638.07588716184	11.6937483957887\\
3012.47043557456	11.7031631182581\\
3386.23950918242	11.7125924847753\\
3759.38105741824	11.7220354845881\\
4131.89284581028	11.7314919730709\\
4503.7726290217	11.7409618714867\\
4875.01814949863	11.7504451447636\\
5245.62666936206	11.7599427974303\\
5615.59591099217	11.7694546271539\\
5984.92404239672	11.7789796339134\\
6353.60830180276	11.7885186990463\\
6721.646380254	11.7980717403788\\
7089.0363196865	11.8076381309175\\
7455.77477845368	11.8172194458493\\
7821.8601072554	11.8268145278226\\
8187.28995065928	11.8364233917109\\
8552.06144798892	11.8460467033699\\
8916.17165420025	11.855685115647\\
9279.61953370031	11.8653368839353\\
9642.4027915463	11.8750019817269\\
10004.5168400589	11.8846829186697\\
10365.9616248833	11.8943770018654\\
10726.7329621335	11.9040861599154\\
11086.8288101225	11.9138099271861\\
11446.2474588592	11.923547577592\\
11804.9854704934	11.9333000796086\\
12163.040976672	11.9430668652828\\
12520.4098979644	11.9528493610592\\
12877.0922173005	11.9626453824736\\
13233.0835118354	11.9724566154568\\
13588.3831083545	11.9822815575125\\
13942.9857758188	11.9921224105985\\
14296.8911421645	12.0019774625206\\
14650.0954580335	12.0118476653818\\
15002.5969463215	12.0217324551972\\
15354.3918313298	12.0316327215709\\
15705.4775755397	12.0415484189174\\
16055.8533419442	12.0514783681784\\
16405.5159253404	12.061423041336\\
16754.4614792134	12.0713832741061\\
17102.6881127994	12.0813586070503\\
17450.1919733931	12.0913498143566\\
17796.9736424881	12.1013549567967\\
18143.0259189637	12.1113768007371\\
18488.3497800187	12.1214132050906\\
18832.9408586547	12.1314651946632\\
19176.7966820294	12.1415326706636\\
19519.9148813274	12.1516154840371\\
19862.2910944653	12.1617145485204\\
20203.9266788049	12.1718277421028\\
20544.8132648811	12.1819580475214\\
20884.952359649	12.1921032947185\\
21224.3394062267	12.2022644424042\\
21562.972111313	12.2124412891648\\
21900.8474659068	12.2226339849732\\
22237.9615791096	12.2328430786684\\
22574.3158257629	12.2430666701677\\
22909.9013189879	12.2533076034905\\
23244.7186685271	12.2635643302835\\
23578.7655131032	12.2738366947568\\
23912.03822128	12.2841250943647\\
24244.5319409983	12.2944304157039\\
24576.2487105883	12.3047506257847\\
24873.2654724643	12.3163786923096\\
25092.4435329626	12.3308617723836\\
25307.4914513819	12.3453465208373\\
25518.3993497632	12.359831824301\\
25725.1454133268	12.3743183148737\\
25927.7176563731	12.3888054753346\\
26126.0979754106	12.4032928171334\\
26320.272163713	12.4177800371664\\
26510.2210242812	12.4322672514539\\
26695.9307796062	12.4467535157665\\
26877.3836569126	12.4612387547428\\
27054.5600445229	12.4757229715953\\
27227.4414880799	12.4902055252307\\
27396.0133222173	12.5046859395198\\
27560.2534302226	12.5191640398296\\
27720.1434845531	12.5336394458354\\
27875.6668946439	12.5481115114573\\
28026.7991690679	12.562579975219\\
28173.5226237634	12.5770446764671\\
28315.8133051028	12.5915048777697\\
28453.6540396952	12.6059600780723\\
28587.021095821	12.6204099989822\\
28715.8923798213	12.6348539125674\\
28840.2431105799	12.6492914349662\\
28960.0513685829	12.6637220552638\\
29075.2930182	12.6781450328597\\
29185.9432295207	12.6925598808141\\
29291.971418415	12.7069664175224\\
29393.3589281575	12.7213634252656\\
29490.0731683358	12.7357505650028\\
};
\addlegendentry{trip-based tax}

\end{axis}
\end{tikzpicture}%
\vspace*{-0.3in}
\caption{Per-trip ride price under different congestion surcharges.} 
\label{figure4_compare}
\end{minipage}
\begin{minipage}[b]{0.005\linewidth}
\hfill
\end{minipage}
\begin{minipage}[b]{0.32\linewidth}
\centering
%
%
\definecolor{mycolor1}{rgb}{0.00000,0.44700,0.74100}%
\definecolor{mycolor2}{rgb}{0.85000,0.32500,0.09800}%
\begin{tikzpicture}

\begin{axis}[%
width=1.794in,
height=1.03in,
at={(1.358in,0.0in)},
scale only axis,
xmin=0,
xmax=35000,
xlabel style={font=\color{white!15!black}},
xlabel={tax revenue/hour},
ymin=4.3,
ymax=5.0,
ylabel style={font=\color{white!15!black}},
ylabel={Pickup time},
axis background/.style={fill=white},
legend style={at={(0.0,0.59)}, anchor=south west, legend cell align=left, align=left, draw=white!15!black}
]
\addplot [color=black, line width=1.0pt]
  table[row sep=crcr]{%
0	4.5113226177619\\
400.822941562034	4.5113226177619\\
801.645883124069	4.5113226177619\\
1202.4688246861	4.5113226177619\\
1603.29176624814	4.5113226177619\\
2004.11470781017	4.5113226177619\\
2404.93764937221	4.5113226177619\\
2805.76059093424	4.5113226177619\\
3206.58353249628	4.5113226177619\\
3607.40647405831	4.5113226177619\\
4008.22941562034	4.5113226177619\\
4409.05235718238	4.5113226177619\\
4809.87529874441	4.5113226177619\\
5210.69824030645	4.5113226177619\\
5611.52118186848	4.5113226177619\\
6012.34412343052	4.5113226177619\\
6413.16706499255	4.5113226177619\\
6813.99000655459	4.5113226177619\\
7214.81294811662	4.5113226177619\\
7615.63588967865	4.5113226177619\\
8016.45883124069	4.5113226177619\\
8417.28177280272	4.5113226177619\\
8818.10471436476	4.5113226177619\\
9218.92765592679	4.5113226177619\\
9619.75059748883	4.5113226177619\\
10020.5735390509	4.5113226177619\\
10421.3964806129	4.5113226177619\\
10822.2194221749	4.5113226177619\\
11223.042363737	4.5113226177619\\
11623.865305299	4.5113226177619\\
12024.688246861	4.5113226177619\\
12425.5111884231	4.5113226177619\\
12826.3341299851	4.5113226177619\\
13227.1570715471	4.5113226177619\\
13627.9800131092	4.5113226177619\\
14028.8029546712	4.5113226177619\\
14429.6258962332	4.5113226177619\\
14830.4488377953	4.5113226177619\\
15231.2717793573	4.5113226177619\\
15632.0947209193	4.5113226177619\\
16032.9176624814	4.5113226177619\\
16433.7406040434	4.5113226177619\\
16834.5635456054	4.5113226177619\\
17235.3864871675	4.5113226177619\\
17636.2094287295	4.5113226177619\\
18037.0323702915	4.5113226177619\\
18437.8553118536	4.5113226177619\\
18838.6782534156	4.5113226177619\\
19239.5011949777	4.5113226177619\\
19640.3241365397	4.5113226177619\\
20041.1470781017	4.5113226177619\\
20441.9700196638	4.5113226177619\\
20842.7929612258	4.5113226177619\\
21243.6159027878	4.5113226177619\\
21644.4388443499	4.5113226177619\\
22045.2617859119	4.5113226177619\\
22446.0847274739	4.5113226177619\\
22846.907669036	4.5113226177619\\
23247.730610598	4.5113226177619\\
23648.55355216	4.5113226177619\\
24049.3764937221	4.5113226177619\\
24450.1994352841	4.5113226177619\\
24821.484565176	4.5133196459385\\
25101.0813557051	4.5213836681822\\
25376.7477906669	4.52950381064092\\
25648.4712521948	4.53768124086525\\
25916.2454438625	4.54591686176354\\
26180.0582772444	4.55421169809308\\
26439.9000929554	4.56256685967705\\
26695.7632333728	4.57098319943113\\
26947.6362746199	4.57946195992755\\
27195.5075723625	4.58800424195261\\
27439.3656658373	4.59661125908114\\
27679.1995579029	4.60528402998741\\
27914.9992011026	4.61402370301903\\
28146.7465617013	4.62283191422324\\
28374.4334460856	4.6317095392832\\
28598.0444518752	4.64065809498341\\
28817.5652547802	4.64967900548615\\
29032.9817089911	4.65877354532955\\
29244.2803634321	4.66794307886816\\
29451.441493304	4.67718937134925\\
29654.4523394623	4.68651371478698\\
29853.2915185005	4.69591790128553\\
30047.9478373912	4.7054032992546\\
30238.3967723823	4.71497191010915\\
30424.6227469376	4.72462537045712\\
30606.6066800126	4.73436532968108\\
30784.3246221727	4.74419389995714\\
30957.7596005953	4.75411273303143\\
31126.8852050681	4.76412411284259\\
31291.6799234602	4.77423001801787\\
31452.1223979843	4.78443244855959\\
31608.1854534814	4.79473377221198\\
31759.840287188	4.8051363754991\\
31907.0664197762	4.81564237420759\\
32049.8299984084	4.82625454522727\\
32188.1046132194	4.83697527785022\\
32321.8575711259	4.84780738906848\\
32451.0609381859	4.85875351441087\\
};
\addlegendentry{time-based tax}

\addplot [color=mycolor1, dashed, line width=1.0pt]
  table[row sep=crcr]{%
0	4.5113226177619\\
378.707477583736	4.508636705822\\
756.805539388852	4.50594897088109\\
1134.291868209	4.50325924207303\\
1511.16432016694	4.50056792207342\\
1887.42069745547	4.49787501235266\\
2263.05860177174	4.49518023443258\\
2638.07588716184	4.49248379708062\\
3012.47043557456	4.48978584465945\\
3386.23950918242	4.48708586635222\\
3759.38105741824	4.48438416399627\\
4131.89284581028	4.48168079102637\\
4503.7726290217	4.47897578196405\\
4875.01814949863	4.47626915883364\\
5245.62666936206	4.47356064731129\\
5615.59591099217	4.470850317793\\
5984.92404239672	4.46813846753639\\
6353.60830180276	4.46542485791879\\
6721.646380254	4.46270952493447\\
7089.0363196865	4.45999265920068\\
7455.77477845368	4.45727382613659\\
7821.8601072554	4.45455336711559\\
8187.28995065928	4.45183129050797\\
8552.06144798892	4.44910742055095\\
8916.17165420025	4.44638158583286\\
9279.61953370031	4.44365429276609\\
9642.4027915463	4.44092556125877\\
10004.5168400589	4.43819469611709\\
10365.9616248833	4.43546247029369\\
10726.7329621335	4.43272835330399\\
11086.8288101225	4.42999248992262\\
11446.2474588592	4.42725509769379\\
11804.9854704934	4.42451591774503\\
12163.040976672	4.4217751231611\\
12520.4098979644	4.4190323275667\\
12877.0922173005	4.41628815711498\\
13233.0835118354	4.41354215305559\\
13588.3831083545	4.41079474954899\\
13942.9857758188	4.40804534478471\\
14296.8911421645	4.405294431284\\
14650.0954580335	4.4025417571912\\
15002.5969463215	4.3997874938905\\
15354.3918313298	4.39703140776381\\
15705.4775755397	4.39427352611842\\
16055.8533419442	4.39151419108294\\
16405.5159253404	4.38875328546715\\
16754.4614792134	4.38599059175441\\
17102.6881127994	4.38322625206437\\
17450.1919733931	4.38046006670197\\
17796.9736424881	4.37769258700496\\
18143.0259189637	4.37492306257858\\
18488.3497800187	4.3721521001894\\
18832.9408586547	4.36937943158733\\
19176.7966820294	4.36660509911158\\
19519.9148813274	4.36382915887214\\
19862.2910944653	4.36105137467907\\
20203.9266788049	4.35827234514252\\
20544.8132648811	4.35549126595164\\
20884.952359649	4.35270874853756\\
21224.3394062267	4.34992454520666\\
21562.972111313	4.34713872673787\\
21900.8474659068	4.34435126769059\\
22237.9615791096	4.34156203390439\\
22574.3158257629	4.33877155994032\\
22909.9013189879	4.33597908539685\\
23244.7186685271	4.33318504889571\\
23578.7655131032	4.33038950865653\\
23912.03822128	4.32759237243591\\
24244.5319409983	4.32479341567535\\
24576.2487105883	4.3219932066038\\
24873.2654724643	4.32152618124085\\
25092.4435329626	4.32625170386345\\
25307.4914513819	4.33103590818659\\
25518.3993497632	4.3358797719104\\
25725.1454133268	4.34078444677008\\
25927.7176563731	4.34575080761424\\
26126.0979754106	4.35078023050977\\
26320.272163713	4.35587368246876\\
26510.2210242812	4.36103233794517\\
26695.9307796062	4.36625744743878\\
26877.3836569126	4.37155016987785\\
27054.5600445229	4.37691177178285\\
27227.4414880799	4.3823437384861\\
27396.0133222173	4.38784719015231\\
27560.2534302226	4.39342368069486\\
27720.1434845531	4.39907457501159\\
27875.6668946439	4.40480124412952\\
28026.7991690679	4.41060547055077\\
28173.5226237634	4.41648848707385\\
28315.8133051028	4.42245227059409\\
28453.6540396952	4.42849819815224\\
28587.021095821	4.43462802658879\\
28715.8923798213	4.4408436161793\\
28840.2431105799	4.44714686253146\\
28960.0513685829	4.45353952901893\\
29075.2930182	4.46002365646573\\
29185.9432295207	4.46660122490694\\
29291.971418415	4.47327447595706\\
29393.3589281575	4.48004533772495\\
29490.0731683358	4.48691628984236\\
};
\addlegendentry{trip-based tax}

\end{axis}
\end{tikzpicture}%
\vspace*{-0.3in}
\caption{Comparison of passenger pickup time (minute).}
\label{figure5_compare}
\end{minipage}
\begin{minipage}[b]{0.005\linewidth}
\hfill
\end{minipage}
\begin{minipage}[b]{0.32\linewidth}
\centering
%
%
\definecolor{mycolor1}{rgb}{0.00000,0.44700,0.74100}%
\definecolor{mycolor2}{rgb}{0.85000,0.32500,0.09800}%
\begin{tikzpicture}

\begin{axis}[%
width=1.794in,
height=1.03in,
at={(1.358in,0.0in)},
scale only axis,
xmin=0,
xmax=35000,
xlabel style={font=\color{white!15!black}},
xlabel={tax revenue/hour},
ymin=2000,
ymax=50000,
ylabel style={font=\color{white!15!black}},
ylabel={Platform revenue},
axis background/.style={fill=white},
legend style={at={(0.24,0.59)}, anchor=south west, legend cell align=left, align=left, draw=white!15!black}
]
\addplot [color=black, line width=1.0pt]
  table[row sep=crcr]{%
0	40877.8264253881\\
378.707477583736	40498.9669013202\\
756.805539388852	40120.4117249855\\
1134.291868209	39742.1616371414\\
1511.16432016694	39364.2173808241\\
1887.42069745547	38986.5797013587\\
2263.05860177174	38609.2493463613\\
2638.07588716184	38232.2270657481\\
3012.47043557456	37855.5136117395\\
3386.23950918242	37479.1097388677\\
3759.38105741824	37103.0162039799\\
4131.89284581028	36727.2337662471\\
4503.7726290217	36351.7631871691\\
4875.01814949863	35976.605230579\\
5245.62666936206	35601.760662651\\
5615.59591099217	35227.2302519045\\
5984.92404239672	34853.0147692124\\
6353.60830180276	34479.1149878038\\
6721.646380254	34105.5316832716\\
7089.0363196865	33732.2656335783\\
7455.77477845368	33359.3176190616\\
7821.8601072554	32986.6884224391\\
8187.28995065928	32614.3788288149\\
8552.06144798892	32242.3896256859\\
8916.17165420025	31870.7216029458\\
9279.61953370031	31499.3755528921\\
9642.4027915463	31128.3522702312\\
10004.5168400589	30757.6525520831\\
10365.9616248833	30387.2771979884\\
10726.7329621335	30017.227009913\\
11086.8288101225	29647.5027922537\\
11446.2474588592	29278.1053518433\\
11804.9854704934	28909.0354979568\\
12163.040976672	28540.2940423157\\
12520.4098979644	28171.8817990948\\
12877.0922173005	27803.7995849261\\
13233.0835118354	27436.0482189035\\
13588.3831083545	27068.6285225915\\
13942.9857758188	26701.5413200257\\
14296.8911421645	26334.7874377214\\
14650.0954580335	25968.3677046773\\
15002.5969463215	25602.2829523808\\
15354.3918313298	25236.5340148137\\
15705.4775755397	24871.1217284552\\
16055.8533419442	24506.0469322899\\
16405.5159253404	24141.3104678097\\
16754.4614792134	23776.9131790202\\
17102.6881127994	23412.8559124457\\
17450.1919733931	23049.1395171336\\
17796.9736424881	22685.7648446584\\
18143.0259189637	22322.7327491286\\
18488.3497800187	21960.0440871877\\
18832.9408586547	21597.699718022\\
19176.7966820294	21235.7005033641\\
19519.9148813274	20874.0473074968\\
19862.2910944653	20512.7409972582\\
20203.9266788049	20151.7824420458\\
20544.8132648811	19791.1725138213\\
20884.952359649	19430.9120871128\\
21224.3394062267	19071.0020390221\\
21562.972111313	18711.4432492266\\
21900.8474659068	18352.2365999841\\
22237.9615791096	17993.3829761365\\
22574.3158257629	17634.8832651149\\
22909.9013189879	17276.7383569417\\
23244.7186685271	16918.9491442347\\
23578.7655131032	16561.5165222127\\
23912.03822128	16204.4413886969\\
24244.5319409983	15847.7246441156\\
24576.2487105883	15491.3671915082\\
24873.2654724643	15135.445448984\\
25092.4435329626	14781.0715804206\\
25307.4914513819	14428.6174168321\\
25518.3993497632	14078.0870481204\\
25725.1454133268	13729.4846060128\\
25927.7176563731	13382.814265349\\
26126.0979754106	13038.0802454206\\
26320.272163713	12695.2868113649\\
26510.2210242812	12354.4382756184\\
26695.9307796062	12015.5389994299\\
26877.3836569126	11678.593394438\\
27054.5600445229	11343.6059243175\\
27227.4414880799	11010.5811064952\\
27396.0133222173	10679.5235139412\\
27560.2534302226	10350.4377770407\\
27720.1434845531	10023.3285855467\\
27875.6668946439	9698.20069062258\\
28026.7991690679	9375.05890697717\\
28173.5226237634	9053.90811509731\\
28315.8133051028	8734.75326358504\\
28453.6540396952	8417.5993716045\\
28587.021095821	8102.45153144492\\
28715.8923798213	7789.31491120695\\
28840.2431105799	7478.19475761962\\
28960.0513685829	7169.09639899628\\
29075.2930182	6862.02524833731\\
29185.9432295207	6556.986806589\\
29291.971418415	6253.98666607005\\
29393.3589281575	5953.03051407354\\
29490.0731683358	5654.12413665825\\
};
\addlegendentry{time-based tax}

\addplot [color=mycolor1, dashed, line width=1.0pt]
  table[row sep=crcr]{%
0	40877.8264253881\\
400.822941562034	40477.0034838261\\
801.645883124069	40076.1805422641\\
1202.4688246861	39675.357600702\\
1603.29176624814	39274.53465914\\
2004.11470781017	38873.711717578\\
2404.93764937221	38472.8887760159\\
2805.76059093424	38072.0658344539\\
3206.58353249628	37671.2428928919\\
3607.40647405831	37270.4199513298\\
4008.22941562034	36869.5970097678\\
4409.05235718238	36468.7740682057\\
4809.87529874441	36067.9511266437\\
5210.69824030645	35667.1281850817\\
5611.52118186848	35266.3052435196\\
6012.34412343052	34865.4823019576\\
6413.16706499255	34464.6593603956\\
6813.99000655459	34063.8364188335\\
7214.81294811662	33663.0134772715\\
7615.63588967865	33262.1905357095\\
8016.45883124069	32861.3675941474\\
8417.28177280272	32460.5446525854\\
8818.10471436476	32059.7217110234\\
9218.92765592679	31658.8987694613\\
9619.75059748883	31258.0758278993\\
10020.5735390509	30857.2528863373\\
10421.3964806129	30456.4299447752\\
10822.2194221749	30055.6070032132\\
11223.042363737	29654.7840616512\\
11623.865305299	29253.9611200891\\
12024.688246861	28853.1381785271\\
12425.5111884231	28452.3152369651\\
12826.3341299851	28051.492295403\\
13227.1570715471	27650.669353841\\
13627.9800131092	27249.846412279\\
14028.8029546712	26849.0234707169\\
14429.6258962332	26448.2005291549\\
14830.4488377953	26047.3775875929\\
15231.2717793573	25646.5546460308\\
15632.0947209193	25245.7317044688\\
16032.9176624814	24844.9087629068\\
16433.7406040434	24444.0858213447\\
16834.5635456054	24043.2628797827\\
17235.3864871675	23642.4399382206\\
17636.2094287295	23241.6169966586\\
18037.0323702915	22840.7940550966\\
18437.8553118536	22439.9711135345\\
18838.6782534156	22039.1481719725\\
19239.5011949777	21638.3252304105\\
19640.3241365397	21237.5022888484\\
20041.1470781017	20836.6793472864\\
20441.9700196638	20435.8564057244\\
20842.7929612258	20035.0334641623\\
21243.6159027878	19634.2105226003\\
21644.4388443499	19233.3875810383\\
22045.2617859119	18832.5646394762\\
22446.0847274739	18431.7416979142\\
22846.907669036	18030.9187563522\\
23247.730610598	17630.0958147901\\
23648.55355216	17229.2728732281\\
24049.3764937221	16828.4499316661\\
24450.1994352841	16427.626990104\\
24821.484565176	16026.8632907397\\
25101.0813557051	15627.4749672371\\
25376.7477906669	15230.0040498666\\
25648.4712521948	14834.4521304064\\
25916.2454438625	14440.8208623925\\
26180.0582772444	14049.111962596\\
26439.9000929554	13659.3272125568\\
26695.7632333728	13271.4684601778\\
26947.6362746199	12885.5376213824\\
27195.5075723625	12501.5366818377\\
27439.3656658373	12119.4676987472\\
27679.1995579029	11739.3328027165\\
27914.9992011026	11361.134199695\\
28146.7465617013	10984.8741729991\\
28374.4334460856	10610.5550854191\\
28598.0444518752	10238.1793814152\\
28817.5652547802	9867.74958940901\\
29032.9817089911	9499.2683241713\\
29244.2803634321	9132.73828931635\\
29451.441493304	8768.16227990541\\
29654.4523394623	8405.54318516595\\
29853.2915185005	8044.88399133511\\
30047.9478373912	7686.18778463116\\
30238.3967723823	7329.45775436362\\
30424.6227469376	6974.69719618843\\
30606.6066800126	6621.90951551689\\
30784.3246221727	6271.09823108836\\
30957.7596005953	5922.26697871608\\
31126.8852050681	5575.41951521696\\
31291.6799234602	5230.55972253766\\
31452.1223979843	4887.69161208778\\
31608.1854534814	4546.81932929527\\
31759.840287188	4207.94715839811\\
31907.0664197762	3871.07952748725\\
32049.8299984084	3536.22101381836\\
32188.1046132194	3203.37634941147\\
32321.8575711259	2872.55042695686\\
32451.0609381859	2543.74830605038\\
};
\addlegendentry{trip-based tax}

\end{axis}
\end{tikzpicture}%
\vspace*{-0.3in}
\caption{Platform revenue under different   congestion surcharges.}
\label{figure6_compare}
\end{minipage}
\end{figure*}

This section provides a comparison of the trip-based   and  time-based congestion charges. To ensure a meaningful comparison, we first set a target for the city's tax revenue. This target can be achieved by setting the appropriate charges.  For each  scheme, we find the  charge that exactly attains the targeted tax revenue  and we compare the two schemes for the same target. The model parameters are consistent with previous case studies in Section \ref{parameter_section} and Section \ref{time-based}. 

Figures \ref{figure1_compare}-\ref{figure3_compare} compare the number of drivers, passenger arrival rate and the vehicle occupancy of the two  schemes for different targets for the city's tax revenue. Figure \ref{figure4_compare} and Figure \ref{figure5_compare} compare the ride fare and the pickup time for the two  schemes. Figure \ref{figure6_compare} compares the platform profit under the trip-based and time-based charges. These results reveal that for the same realized tax revenue, the time-based charge is Pareto superior to the trip-based charge (as currently implemented in NYC). 
Under the time-based  charge, the TNC platform earns a higher profit. For drivers, the time-based congestion charge does not affect their surplus  in the first regime as the same number of drivers are hired at the same wage.  For passengers, the time-based  charge leads to 
 a lower ride fare but a longer waiting time. However, the time-based congestion charge also has higher passenger arrival rate (Figure \ref{figure2_compare}). 
Since the demand function $F_p(c)$ is monotonic, this implies that the total travel cost $c$ is lower and the passenger surplus is higher under the time-based congestion charge.  

In summary, the time-based congestion charge leads to higher passenger surplus and higher platform profit (Figure \ref{figure6_compare}), which benefits all participants of the transportation system. This is because the time-based congestion surcharge penalizes idle vehicle hours and motivates the TNC to increase the occupancy rate of the vehicles (see Figure \ref{figure3_compare}). Based on the  data for San Francisco, the surplus resulting from increased vehicle occupancy will be distributed to all market participants, including  the passengers, the TNC platform, and the city. 

While the aforementioned results do not necessarily hold for all levels of  targeted tax revenues,  the conclusion is indeed applicable for a large range of model parameters in the regime of practical interest. To formally present this claim, we define $N_t^*, w_t^*, \lambda_t^*, c_t^*, P_t^*, Tr_t^*$ as the optimal solution to (\ref{optimalpricing_trip}) and denote $N_h^*, w_h^*, \lambda_h^*, c_h^*, P_h^*, Tr_h^*$ as the optimal solution to (\ref{optimalpricing_time}). They are respectively the optimal number of drivers, driver wage, passenger arrival rate, total travel cost, platform profit, and city tax revenue. Note that all variables with subscript $t$ depend on $p_t$ and $w_0$, and all variables with subscript $h$ depend on $p_h$ and $w_0$. We suppress this dependence to simplify the notation whenever it is clear from the context. 
\begin{theorem}
\label{theorem3}
Assume that the profit optimization problems (\ref{optimalpricing_trip}) and (\ref{optimalpricing_time}) both have unique solutions. Assume that  ${F_p}(c)$ and ${F_d}(w)$ satisfy the logit model as specified in (\ref{logit_demand}) and (\ref{logit_supply}), respectively. For any pickup time function ${t_p}$ that satisfies Assumption \ref{assumption1}, any speed-density relation $v(N)$ that satisfies Assumption \ref{assumption2},  and any model parameters ${\Theta}=\{\lambda_0, N_0,  M, L, v_f, \kappa, \alpha,  \epsilon, c_0, \sigma, w_0\}$, there exists $w_3>\tilde{w}$, such that for any $\tilde{w}\leq {w_0}\leq  w_3$,  there exists $\bar{p}_t$ so that for any trip-based congestion surcharge ${p_t}\in [0,\bar{p}_t]$, there exists a time-based congestion surcharge ${p_h}$ that offers a Pareto improvement, i.e.
	\[N_h^* = N_t^*,w_h^* = w_t^* = {w_0},\lambda _h^* > \lambda _t^*,c_h^* < c_t^*,P_h^* > P_t^*,Tr_h^* > Tr_t^*\]	
\end{theorem}
The proof of Theorem \ref{theorem3} can be found in Appendix E. It shows that there exists a regime where a time-based charge offers a Pareto improvement over a trip-based one. In this regime, for any trip-based charge, one can  find an appropriate time-based charge for which the same number of drivers is hired,  more passengers take TNC rides at a lower cost,  the platform earns more profit, and the city collects more tax revenues to subsidize  public transit.   For the case of San Francisco, we calculate $w_3=\$29.20$/hour,  and $\bar{p}_t=\$2.1$/trip when $w_0=\$26.35$/hour.

\begin{remark}
Theorem \ref{theorem3} identified a regime under which a time-based congestion charge offers a Pareto improvement. The caveat is that this regime only applies to the wage floor and congestion charge levels within a certain range, i.e., $w_0\in [\tilde{w},w_3]$, ${p_t}\in [0,\bar{p}_t]$. Outside of this range, the comparison between the two congestion charge schemes may depend on the model parameters. However, we emphasize that it is unlikely for cities to impose very stringent policies that substantially raise the driver payment level (or surcharge level), since this may drive the TNCs out of business.  In practice, regulatory policies are likely to reside in or stay close to the regime identified by this paper. 
\end{remark}

\begin{figure*}[bt]%
\begin{minipage}[b]{0.32\linewidth}
\centering
\include{s1} 
\vspace*{-0.3in}
\caption{Number of drivers as a function of $p_h$ under distinct $\lambda_0$. }
\label{figures1}
\end{minipage}
\begin{minipage}[b]{0.005\linewidth}
\hfill
\end{minipage}
\begin{minipage}[b]{0.32\linewidth}
\centering
\include{s2}
\vspace*{-0.3in}
\caption{Passenger arrival rate (/min) as a function of $p_h$ under distinct $\lambda_0$. }
\label{figures2}
\end{minipage}
\begin{minipage}[b]{0.005\linewidth}
\hfill
\end{minipage}
\begin{minipage}[b]{0.32\linewidth}
\centering
\include{s3}
\vspace*{-0.3in}
\caption{Platform profit (per hour) as a function of $p_h$ under distinct $\lambda_0$. }
\label{figures3}
\end{minipage}
\begin{minipage}[b]{0.32\linewidth}
\centering
\include{s4}
\vspace*{-0.3in}
\caption{Number of drivers as a function of $p_h$ under distinct $N_0$. }
\label{figures4}
\end{minipage}
\begin{minipage}[b]{0.005\linewidth}
\hfill
\end{minipage}
\begin{minipage}[b]{0.32\linewidth}
\centering
\include{s5} 
\vspace*{-0.3in}
\caption{Passenger arrival rate (/min) as a function of $p_h$ under distinct $N_0$. }
\label{figures5}
\end{minipage}
\begin{minipage}[b]{0.005\linewidth}
\hfill
\end{minipage}
\begin{minipage}[b]{0.32\linewidth}
\centering
\include{s6}
\vspace*{-0.3in}
\caption{Platform profit (per hour) as a function of $p_h$ under distinct $N_0$. }
\label{figures6}
\end{minipage}
\begin{minipage}[b]{0.32\linewidth}
\centering
\include{s7}
\vspace*{-0.3in}
\caption{Number of drivers as a function of $p_h$ under distinct $\alpha$. }
\label{figures7}
\end{minipage}
\begin{minipage}[b]{0.005\linewidth}
\hfill
\end{minipage}
\begin{minipage}[b]{0.32\linewidth}
\centering
\include{s8} 
\vspace*{-0.3in}
\caption{Passenger arrival rate (/min) as a function of $p_h$ under distinct $\alpha$. }
\label{figures8}
\end{minipage}
\begin{minipage}[b]{0.005\linewidth}
\hfill
\end{minipage}
\begin{minipage}[b]{0.32\linewidth}
\centering
\include{s9}
\vspace*{-0.3in}
\caption{Platform profit (per hour) as a function of $p_h$ under distinct $\alpha$. }
\label{figures9}
\end{minipage}
\end{figure*}

\section{Sensitivity Analysis}
This section reports a sensitivity analysis to test the robustness of our results with respect to the model parameters. We  vary the model parameters of (\ref{optimalpricing_time}) and evaluate the impact of the time-based congestion charge under distinct parameter values. The nominal values of the  parameters are set to be the same as in Section \ref{parameter_section}. We perturb $\lambda_0, N_0$ and $\alpha$ by $5\%$ and investigate how these perturbations affect passengers, drivers, and the TNC platform under the time-based charge. 

Figure \ref{figures1}-\ref{figures3} show the number of drivers, passenger arrival rate,  and the platform profit as functions of the time-based congestion charge under different $\lambda_0$ (the nominal value is 1049). Clearly, there are two regimes. When $\lambda_0$ increases, the TNC platform has more passengers, and therefore enjoys a higher profit. However, we note that in the first regime, the number of drivers is not affected by  $\lambda_0$. This is because in the first regime, both (\ref{supply_constraint_time}) and (\ref{min_wage_const_time}) are active, which determines  $N$ as $N=N_0F_d(w_0)$.

Figure \ref{figures4}-\ref{figures6} show the number of drivers, passenger arrival rate,  and the platform profit as functions of the time-based charge for different $N_0$ (the nominal value is 10K). There are clearly two regimes for the three values of $N_0$. When $N_0$ increases, the platform hires  more drivers, attracts more passengers and collects a higher  profit. Platform profit is insensitive to the number of potential drivers.

Figure \ref{figures7}-\ref{figures9} show the number of drivers, passenger arrival rate,  and the platform profit as functions of the time-based charge for different $\alpha$ (the nominal value is 2.33). When $\alpha$ increases, both passenger arrival rate and platform profit drop. We note that the platform profit is much more sensitive to $\alpha$ than it is to $\lambda_0$ and $N_0$.

\section{Conclusion}
This paper describes the impact of two proposed congestion charges on TNC: (a) a charge based on vehicle trips, and (b) a charge based on vehicle hours. We used a market equilibrium model to assess the joint effect of minimum wage with either of these two charges. Surprisingly, we find that neither  charging scheme significantly affects the number of TNC vehicles since their effect is mitigated by the wage floor on TNC drivers. Furthermore,  we find that the time-based charge is Pareto superior  compared with the trip-based charge that is currently imposed in New York City. Under the time-based charge,  more passengers take TNC rides at a cheaper overall travel cost, drivers remain unaffected,  the platform earns a higher profit, and the city  collects more tax revenue from the TNC system to subsidize public transit. 

The policy implication of these results are profound. First of all, our results imply that the  TNC driver minimum wage mitigates the effectiveness of the congestion charge (either time-based or trip-based) in reducing the TNC traffic. Therefore, when a driver minimum wage is imposed, the city can not merely count on the congestion charge  to reduce the number of TNC vehicles on the city's street, unless the charge is significant and exceeds certain threshold.  Second, { the TNC profit is rather sensitive to regulations such as minimum wage and congestion charges. Based on calibrated model parameters, we showed that the tax burden mainly falls on the ride-hailing platform as opposed to passengers and drivers. We argue that this effect should be taken into account in policy formulation, and an interesting research direction is to synthesize more effective policies that achieve the regulatory objective without jeopardizing the TNC business model, e.g., \cite{li2020off}.  } Third,   
our result suggests that the time-based congestion charge is superior to the trip-based congestion charge. While most city selects the trip-based congestion charge as a natural candidate of its charge scheme (e.g., NYC, Chicago, Seattle), a shift to the time-based congestion charge is not difficult to implement: the city only needs to periodically audit the operations data of the TNC  and collect the charge based on the accumulated vehicle hours on the platform. 

Future research directions include determining the optimal level of congestion charge that maximizes social welfare, extending the model to capture temporal and spatial aspect of the TNC market, and characterizing the impact of regulatory policies on TNC competition.

\section*{Acknowledgments}
This research was supported by the Hong Kong Research Grant Council project HKUST26200420 and National Science Foundation EAGER
award 1839843.

\bibliographystyle{unsrt}
\bibliography{resourceprocurement}


\section*{Appendix}

\subsection*{\bf{A: Proof of Proposition \ref{prop_concave}}}
{
Given $N$, the first-order derivative of (\ref{optimalpricing_transformed}) with respect to $\lambda$, denoted as $\dfrac{\partial P}{\partial \lambda}$, can be derived as follows:\begin{equation}
\label{1st_order_condition}
\dfrac{\partial P}{\partial \lambda}= F_p^{-1} \left(\dfrac{\lambda}{\lambda_0}\right)- \alpha t_p(N-\lambda/\mu, v)-\beta p_0 + \lambda \left(  \dfrac{\partial F_p^{-1}}{\partial \lambda}+\dfrac{\alpha}{\mu} \dfrac{\partial t_p}{\partial N_I}    
    \right)
\end{equation}
We note that under the assumptions of Proposition \ref{prop_concave}
, each term in the right-hand side of (\ref{1st_order_condition}) is decreasing with respect to $\lambda$. In particular, the first term $F_p^{-1}$ is a decreasing function of $\lambda$ by the definition of $F_p$. The second term $-t_p(N-\lambda\mu, v)$ is a decreasing function of $\lambda$ due to the monotone property of $t_p$. Since $F_p^{-1}$ is the logit model, we have 
\begin{equation}
\lambda \dfrac{\partial F_p^{-1}}{\partial \lambda}=-\dfrac{1}{\epsilon} \left(1+\dfrac{\lambda}{\lambda_0-\lambda} \right)
\end{equation}
which is  decreasing function of $\lambda$. In addition, since $t_p(N_I, v)$ is convex with respect to $N_I$, the last term $\lambda \dfrac{\alpha}{\mu}\dfrac{\partial t_p}{\partial {N_I}}$ a=is a decreasing function\footnote{Note that this term is negative} of $\lambda$. Therefore, the objective function (\ref{optimalpricing_transformed}) is a strictly concave function of $\lambda$ when $N$ is fixed.

Next, to show that there exists a unique $\lambda$ that maximizes the profit, it suffices to show that there exists a unique solution to $\dfrac{\partial P}{\partial \lambda}=0$. Apparently, we have $\left.\dfrac{\partial P}{\partial \lambda}\right|_{\lambda=0}=\infty$ and $\left.\dfrac{\partial P}{\partial \lambda}\right|_{\lambda=N\mu}=-\infty$. Therefore, by continuity, there exists at least one $\lambda\in (0, N\mu)$ that satisfies $\dfrac{\partial P}{\partial \lambda}=0$. The uniqueness of the solution follows from the monotonicity of (\ref{1st_order_condition}). This completes the proof. 

}
\subsection*{\bf{B: Calibration of Model Parameters}}

We describe how  the model parameters for San Francisco are obtained. Eleven parameters are to be estimated, ${\Theta}=\{\lambda_0, N_0,  M, L, v_f, \kappa, \alpha,  \epsilon, c_0, \sigma, w_0\}.$
Some of these parameters are taken directly from published sources, the remaining parameters are based on `reverse engineering': values are selected so that the results of the optimization match source data. 

The number of TNC rides and TNC drivers are from the SFCTA report \cite{castiglione2016tncs}. It gives hourly passenger arrival rates and the number of TNC vehicles over an entire day. We take the rates for an`average'  Wednesday and calculate  the average passenger arrival rate\footnote{The total number of TNC trip is 170000/day (Table 2 of \cite{castiglione2016tncs}). Since there are virtually no rides between 12AM-6AM, we divide 170,000/day by 18 hours to derive 157.4 trips/min.} is $\lambda^*=157.4/\text{min}$ and the average number of TNC vehicles\footnote{To be consistent with passenger arrival rate, we also take the average number of vehicle hours between 6AM to 12AM as the market volume between 12AM-6AM is trivial. } is $N^*=3000$ (Figure 1 of \cite{castiglione2016tncs}). According to \cite{castiglione2016tncs}, TNC vehicles provide  approximately $15\%$ intra-SF vehicles trips. This implies that the arrival rate of potential passengers is $\lambda_0=\lambda^*/0.15=1049/\text{min}$. Assuming that $30\%$ of the for-hire vehicle drivers work for TNC (the rest work for food delivery, package delivery, etc),  the total number of potential drivers is $N_0=N^*/0.3=10,000$. 

The average TNC trip distance in San Francisco is 2.6 miles (Table 4 of \cite{castiglione2016tncs}), and the average arterial traffic speed of San Francisco is about 14 mph \cite{castiglione2018tncs}. This gives the average trip time  $t_0=11.14$ min.  From Lyft's pricing formula   for San Francisco \cite{lyftprice}, a 2.6 mile, $11.14$ min Lyft ride costs $\$11.8$, i.e., $p_f^*=11.8$. This total price is comprised of a service fee of $\$2.70$, a base fare of $\$2.24$, a per-mile fare of $\$0.93$ and a per-minute fare of $\$0.40$ \cite{lyftprice}. The TNC platform keeps $100\%$ of the service fee and $25\%$ of the rest, leading to an average of $42\%$ commission rate. This implies that the driver wage is $w^*=\lambda^*p_f^*\times 0.58/N^*=\$21.55$/hour, since $58\%$ of the total passenger fare is shared by all   $N^*$ drivers. 

Taking the average pickup time as  5 minutes (page 2 of \cite{Uber_Sec}), we  obtain $M$ based on (\ref{pickuptime_func}),
\[
M=v t_p \sqrt{(N-\lambda/\mu)}=41.18.
\]
where $\mu=1/t_0$ is the service rate. To estimate  $v_f$ and $\kappa$ in  Greenshield's model, we need two data points to fix the linear relationship (\ref{greenshiledmodel}). One data point is already available: we know that on average there are $3000$ TNC vehicles and the traffic speed is $v^*=14$ mph. Another data point can be obtained from the fact that arterial traffic speed declined from $18$ mph to $14$ mph between  2010 and 2016. Since TNC vehicles only contributed to part of the decrease of traffic speed. Without TNC vehicles, the traffic speed in 2016 would range between 14 and 18 mph. We take $v_f=15$ mph, indicating that TNCs contribute to 25\% of the traffic speed reduction between 2010-2016 (TNCs may contribute more in the increase of VMT as there is a nonlinear relationship between VMT and traffic speed), which provides another data point (when $N=0$, traffic speed is $v=15$).  We plug in these two points in (\ref{greenshiledmodel}) and obtain $v_f=15$ mph  and $\kappa=0.0003$. 

Empirical study suggests that the value of travel time (VOT) for TNC customers ranges between \$40 and \$100 per hour \cite{schwieterman2018uber}, and  the the value of time while waiting for vehicle pickup is about 1.5 to 2.5 times VOT \cite{wardman2004public, abrantes2011meta}. If we take VOT to be $\$70$/hour and assume that the value of time while waiting is 2 times the value of travel time, then we obtain $\alpha=2.33$.

The parameters of the logit model ($\epsilon, c_0, \sigma$ and $w_0$) can be uniquely determined by the aforementioned parameter values based on reverse engineering. We select $\epsilon, c_0, \sigma$ and $w_0$ so that the optimal solution to the following unregulated profit maximization problem  is consistent with the  values of ${\lambda_0, N_0,  M, L, v_f, \kappa, \alpha}$ which we have just determined:
\begin{equation}
\label{unregulated}
 \hspace{-2cm} \mathop {\max }\limits_{{p_f},{p_d}} \quad \lambda ({p_f} - {p_d})
\end{equation}
\begin{subnumcases}{\label{constraint_unregualted}}
\lambda  = {\lambda _0}{F_p}\left(\alpha {t_p}+ \beta t_0 +  {p_f} \right) \\
N = {N_0}{F_d}\left(\frac{{\lambda {p_d}}}{N}\right) 
\end{subnumcases}
Reverse engineering yields $\epsilon=0.33$, $c_0=31.18$, $\sigma=0.089$, and $w_0=\$31.04$/hour. 

\subsection*{\bf{C: Proof of Theorem \ref{theorem1}}}
First, we note that the unregulated problem (\ref{unregulated}) is equivalent to the following:
\begin{equation}
\label{unregulated_equi}
 \hspace{-2cm} \mathop {\max }\limits_{{p_f},N} \quad \lambda {p_f} - Nw
\end{equation}
\begin{subnumcases}{\label{constraint_unregualted}}
\lambda  = {\lambda _0}{F_p}\left(\alpha {t_p}+ \beta t_0 + {p_f} \right) \\
N = {N_0}{F_d}\left(w\right)  \label{supply_equi}
\end{subnumcases}
where we used the definition of driver wage (\ref{driver_wage_def}) to obtain an optimization problem over ride fare  $p_f$ and driver wage $w$. 
Let  $\tilde{p}_f$ and $\tilde{N}$ be the optimal solution to this unregulated problem (\ref{unregulated}), and let $\tilde{\lambda}$ and $\tilde{w}$ be the corresponding  passenger arrival rate and driver wage. 

When the minimum wage is greater than the optimal wage $\tilde{w}$, i.e., $w_0>\tilde{w}$, the minimum wage constraint (\ref{min_wage_const}) is active. In this case, the regulated profit maximization problem (\ref{optimalpricing_trip}) can be reformulated as:
\begin{equation}
\label{optimalpricing_trip_equivalentform}
 \hspace{-2.5cm} \mathop {\max }\limits_{{p_f},N} \quad \lambda {p_f} - Nw_0
\end{equation}
\begin{subnumcases}{\label{constraint_optimapricing_equi}}
\lambda  = {\lambda _0}{F_p}\left(\alpha {t_p}+ \beta t_0  + {p_f} + {p_t}\right) \label{demand_constraint_trip_equi}\\
N \le {N_0}{F_d}\left(w_0\right) \label{supply_constraint_trip_eqio},
\end{subnumcases}
where $w_0$ is given exogenously\footnote{Compared with (\ref{supply_equi}), the supply function  (\ref{supply_constraint_trip_eqio}) is an inequality  since $w$ is endogenous in (\ref{unregulated_equi}) while $w_0$ is exogenous in (\ref{optimalpricing_trip_equivalentform}).}. Note that (\ref{optimalpricing_trip_equivalentform}) can be  be equivalently viewed as nested maximization, where in the outer loop the platform chooses the number of drivers $N$ and in the inner loop the platform chooses the  ride fare $p_f$ to maximize the profit. To prove Theorem \ref{theorem1}, we first consider the inner problem where $N$ and $w_0$ are given and the platform solves
\begin{align}
\label{inner_problem}
& \mathop {\max }\limits_{{p_f}} \quad \lambda {p_f} \\
& s.t. \quad \lambda  = {\lambda _0}{F_p}\big(\alpha {t_p}+ \beta t_0   + {p_f} + {p_t}\big). \label{demand_constraint_trip_equi}
\end{align}
Denote the optimal value of (\ref{inner_problem}) by $\Gamma(N,p_t)$, which depends on $N$ and $p_t$. Since (\ref{optimalpricing_trip}) has a unique solution and $F_d$ and $F_p$ are continuously differentiable, $\Gamma(N,p_t)$ is a continuous function with respect to both $N$ and $p_t$.  The regulated profit maximization problem (\ref{optimalpricing_trip_equivalentform}) can be written as
\begin{align}
\label{nested_problem_regulated}
& \mathop {\max }\limits_{{N}} \, \Gamma(N,p_t)-Nw_0 \\
& s.t. \, N \le {N_0}{F_d}\left(w_0\right). \label{supply_nested}
\end{align}

Similarly, the unregulated problem (\ref{unregulated_equi}) can be written as
\begin{align}
\label{nested_problem_unregulated_initial}
& \mathop {\max }\limits_{{N}} \, \Gamma(N, 0)-Nw \\
& s.t. \, N = {N_0}{F_d}\left(w\right). \label{supply_nested_initial}
\end{align}
Since the solution to (\ref{nested_problem_unregulated_initial}) is $\tilde{w}$, when the minimum wage $w_0=\tilde{w}$, the optimal number of drivers for  (\ref{nested_problem_regulated}) and (\ref{nested_problem_unregulated_initial}) are the same. Furthermore, based on (\ref{supply_nested_initial}), we have $w=F_d^{-1}\left(\dfrac{N}{N_0}\right)$. Therefore (\ref{nested_problem_unregulated_initial}) can be written as
\begin{equation}
\label{nested_problem_unregulated}
\mathop {\max }\limits_{{N}} \, \Gamma(N,0)-NF_d^{-1}\left(\dfrac{N}{N_0}\right) .
\end{equation}
The first order optimality condition for (\ref{nested_problem_unregulated}) indicates that
\begin{equation}
\label{1st_order_unregulated}
\dfrac{\partial^+\Gamma(\tilde{N},0)}{\partial N}-\tilde{w}-\dfrac{\tilde{N}}{f_d(\tilde{w})}=0,
\end{equation}
where $f_d(w)=\dfrac{\partial F_d(w)}{\partial w}$.  Since $F_d(w)$ is strictly increasing, we have $f_d(w)>0$, and so 
\begin{equation}
\label{1st_order_unregulated2}
\dfrac{\partial^+\Gamma(\tilde{N},0)}{\partial N}-\tilde{w}=\dfrac{\tilde{N}}{f_d(\tilde{w})}>0.
\end{equation}
We apply (\ref{1st_order_unregulated2}) to the regulated case (\ref{nested_problem_regulated}) and conclude that when the minimum wage satisfies $w_0=\tilde{w}$ and the  charge is zero, i.e., $p_f=0$, the right derivative of the objective function (\ref{nested_problem_regulated}) with respect to $N$ is strictly positive (see equation (\ref{1st_order_unregulated2})). Since the first order conditions are continuous with respect to $w_0$, there exists $w_1>\tilde{w}$ such that for all $ w_0\in [\tilde{w},w_1)$ we have 
\begin{equation}
\label{1st_order_unregulated3}
\dfrac{\partial^+\Gamma(N^*_t,0)}{\partial N}-w_0>0,
\end{equation}
where $N^*$ is the corresponding optimal solution for the regulated problem. Furthermore, due to continuity, for each $w_0\in [\tilde{w},w_1)$, there exists $\bar{p}_t>0$ so that for all $ p_t\in [0,\bar{p}_t>0)$
\begin{equation}
\label{1st_order_unregulated4}
\dfrac{\partial^+\Gamma(N^*_t,p_t)}{\partial N}-w_0>0.
\end{equation}
This indicates that  there exists $w_1>\tilde{w}$ such that for $ w_0\in [\tilde{w},w_1)$, the derivative of the profit with respect to the number of drivers is strictly positive. This indicates that the platform will earn extra profit if it hires more drivers. Therefore, it is optimal for the platform to hire all drivers available in the market, which gives $N^*_t=N_0F_d(w_0)$. Clearly, the optimal number of drivers does not depend on $p_t$ when $w_0$ is fixed. This completes the proof. 

\subsection*{\bf{D: Proof of Theorem \ref{theorem2}}}
We can first show that there exists ${w_2} >  \tilde{w},$ such that for any $\tilde{w} < {w_0} < {w_2}, $ there exists ${\bar p_h} > 0, $ so that $\partial {N^*_h}/ \partial {p_h} = 0$ for ${p_h} \in (0,{\bar p_h})$.\footnote{The proof is similar to that of Theorem \ref{theorem1} and is therefore omitted.}  Therefore, we have ${N^*_h}({p_h})=N_0F_d(w_0)$ for  $\forall {p_h} \in (0,{\bar p_h})$.  Hence the optimal number of passengers will not be affected by $p_h$ when $p_h\in (0,{\bar p_h})$. This is because when we know that the optimal number of drivers satisfies $N^*=N_0F_d(w_0)$, the profit maximization problem (\ref{optimalpricing_time}) can be written as
\begin{align}
\label{inner_problem_time}
& \mathop {\max }\limits_{{p_f}} \quad \lambda {p_f} \\
& s.t. \quad \lambda  = {\lambda _0}{F_p}\big(\alpha {t_p}+ \beta t_0   + {p_f}\big). \label{demand_constraint_trip_equi_time}
\end{align}
It is clear that $p_h$ does not affect the optimal solution to (\ref{inner_problem_time}). This completes the proof.

\subsection*{\bf{E: Proof of Theorem \ref{theorem3}}}
To prove Theorem \ref{theorem3}, we first show that there exists $w_3>\tilde{w}$, such that for any $\tilde{w}\leq {w_0}\leq  w_3$,  there is $\bar{p}_t$ so that for any trip-based  charge ${p_t}\in [0,\bar{p}_t]$, there exists a time-based  charge ${p_h}$ that yields the same tax revenue, but leads to higher passenger surplus and higher platform profit, i.e., 
	\[N_h^* = N_t^*,w_h^* = w_t^* = {w_0},\lambda _h^* > \lambda _t^*,c_h^* < c_t^*,P_h^* > P_t^*,Tr_h^* = Tr_t^* .\]	
The result of Theorem \ref{theorem3} then follows by slightly increasing $p_h$ to make sure  $Tr_h^* > Tr_t^*$ without changing the sign of other inequalities. Such $p_h$ exists due to continuity.

Based on Theorem \ref{theorem1} and Theorem \ref{theorem2}, there exists $w_3>\tilde{w}$ such that for  all $w\in [\tilde{w},w_3)$, there is $\bar{p_t}>0$ and $\bar{p_h}>0$ so that for  $ p_t\in [0,\bar{p}_t)$ and $ p_h\in [0,\bar{p}_h)$ we have $N_h^*(p_t)=N_h^*(p_h)=N_0F_d(w_0)$. For notational convenience,  define $\hat{N}(w_0)=N_0F_d(w_0)$ and suppress the dependence on $w_0$ to simplify the notation whenever the context is clear. Since $w_0\geq \tilde{w}$, the minimum wage constraint is active in both (\ref{optimalpricing_trip}) and (\ref{optimalpricing_time}), thus $w_t^*=w_h^*=w_0$. 

To show that $\lambda_h^*>\lambda_t^*$, we first note that since $N_t^*=\hat{N}$, it is clear that $\lambda^*_t$ is the optimal solution to:
\begin{align}
\label{inner_problem_compare1}
& \mathop {\max }\limits_{{p_f}} \quad \lambda {p_f} \\
& s.t. \quad \lambda  = {\lambda _0}{F_p}\big(\alpha {t_p}+ \beta t_0   + {p_f} + {p_t}\big). \label{demand_constraint_trip_compare1}
\end{align}
We can regard this as an optimization problem over $\lambda$. After writing the objective function as solely a function of $\lambda$, the first order optimality condition indicates that:
\begin{equation}
    \label{1st_order_compare1}
    F_p^{-1}\left(\dfrac{\lambda_t^*}{\lambda_0}\right)-\alpha t_p\bigg(\hat{N}-\lambda^*_t/\mu, v \bigg)- \beta t_0+ \dfrac{\lambda_t^*}{\lambda_0 f_p\left(F_p^{-1}\left(\dfrac{\lambda_t^*}{\lambda_0}\right)\right)}+\dfrac{\lambda_t^*\alpha }{\mu} \dfrac{\partial t_p(\hat{N}-\lambda^*_t/\mu, v)}{\partial N_I}-p_t=0. 
\end{equation}
Similarly, since $N_h^*=\hat{N}$, we know that $\lambda^*_h$ is the optimal solution to
\begin{align}
\label{inner_problem_compare2}
& \mathop {\max }\limits_{{p_f}} \quad \lambda {p_f} \\
& s.t. \quad \lambda  = {\lambda _0}{F_p}\big(\alpha {t_p}+ \beta t_0   + {p_f} \big) \label{demand_constraint_trip_compare2}
\end{align}
The first order condition with respect to $\lambda$ gives
\begin{equation}
    \label{1st_order_compare2}
    F_p^{-1}\left(\dfrac{\lambda_h^*}{\lambda_0}\right)-\alpha t_p\bigg(\hat{N}-\lambda^*_h/\mu, v \bigg)- \beta t_0+ \dfrac{\lambda_h^*}{\lambda_0 f_p\left(F_p^{-1}\left(\dfrac{\lambda_h^*}{\lambda_0}\right)\right)}+\dfrac{\lambda_h^*\alpha }{\mu} \dfrac{\partial t_p(\hat{N}-\lambda^*_h/\mu, v)}{\partial N_I}=0. 
\end{equation}
It can be verified that when  ${F_p}(c)$ and ${F_d}(w)$ satisfy the logit model as specified in (\ref{logit_demand}) and (\ref{logit_supply}) and when  Assumption \ref{assumption1} and Assumption \ref{assumption2} hold,  every term of (\ref{1st_order_compare1}) and (\ref{1st_order_compare2}) is an increasing function of $\lambda$. By comparing (\ref{1st_order_compare1}) and (\ref{1st_order_compare2}) \footnote{$\mu$ and $v$ depend on the number of TNC vehicles $N$. Since $N^*_h$ and $N^*_t$ are the same, we do not need to distinguish $\mu$ and $v$ for these two cases. }, we have $\lambda_t^*<\lambda_h^*$. By stricta
 monotonicity, this indicates that $c_t^*>c_h^*$.

Last, we show that there exists $w_3>\tilde{w}$, such that for any $\tilde{w}\leq {w_0}\leq  w_3$,  there is $\bar{p}_t$ so that for any trip-based  charge ${p_t}\in [0,\bar{p}_t]$, there exists a time-based  charge ${p_h}$ which ensures $Tr_h^*=Tr_t^*$ and $P_h^*>P_t^*$. Note that  $Tr_t^*=\lambda_t^*p_t^*$ and $Tr_h^*=N_h^*p_h^*$, and that there exists $\bar{p}_t>0$ so that  for any $p_t^*\in [0,\bar{p}_t]$, we can find $p_h^*$ such that $\lambda_t^*p_t^*=N_h^*p_h^*$. After setting $p_h^*$ and $p_t^*$ so that $Tr_t^*=Tr_h^*$, we apply a change of variable  (i.e., $p_f=p_f+p_t$) to transform (\ref{optimalpricing_trip}) into:
\begin{equation}
\label{optimalpricing_trip_2}
 \hspace{-0.5cm} \mathop {\max }\limits_{{p_f},{p_d}, N} \quad \lambda ({p_f} - {p_d})- \lambda{p_t}
\end{equation}
\begin{subnumcases}{\label{constraint_optimapricing2_2}}
\lambda  = {\lambda _0}{F_p}\left(\alpha {t_p}+ \beta t_0  +  {p_f} \right) \label{demand_constraint2_2}\\
N \le {N_0}{F_d}\left(\frac{{\lambda {p_d}}}{N}\right) \label{supply_constraint2_2} \\
\frac{{\lambda {p_d}}}{N} \ge {w_0},
\end{subnumcases}
Then the two charging schemes (\ref{optimalpricing_trip_2}) and (\ref{optimalpricing_time}) have the same constraints, and the optimal value to (\ref{optimalpricing_trip_2}) satisfies:
\begin{align}
P_t^*&=\lambda_t^*p_{f,t}^*-w_0N_t^*-\lambda_t^*p_t \\&= \lambda_t^*p_{f,t}^*-w_0N_t^*-Tr_t^*,
\end{align}
where $p_{f,t}^*$ is the corresponding optimal ride fare for the trip-based congestion charge. 
The optimal value to (\ref{optimalpricing_time}) satisfies
\begin{align}
P_h^*&=\lambda_h^*p_{f,h}^*-w_0N_h^*-N_h^*p_h \\
&=\lambda_h^*p_{f,h}^*-w_0N_h^*-Tr_h^*,
\end{align}
where $p_{f,h}^*$ is the corresponding optimal ride fare for the time-based  charge. 

Based on Theorem \ref{theorem2}, we have  $\lambda_h^*(p_h)=\lambda_h^*(0)$ and $N_h^*(p_h)=N_h^*(0)=$ for  $ p_h\in [0,\bar{p}_h)$. Therefore, $\lambda_h^*(p_h), N_h^*(p_h)$ and $p_{f,h}^*(p_h)$ are the optimal solutions to the profit maximization problem with minimum wage only  (i.e., $w_0>\tilde{w}$ and $p_h=0$):
\begin{equation}
\label{optimalpricing_minimumwage_only}
 \hspace{-2.5cm} \mathop {\max }\limits_{{p_f},N} \quad \lambda {p_f} - Nw_0
\end{equation}
\begin{subnumcases}{}
\lambda  = {\lambda _0}{F_p}\left(\alpha {t_p}+ \beta t_0   + {p_f} + {p_t}\right)\\
N \le {N_0}{F_d}\left(w_0\right).
\end{subnumcases}
This implies that $\lambda_h^*p_{f,h}^*-w_0N_h^*>\lambda_t^*p_{f,t}^*-w_0N_t^*$
 \footnote{Otherwise, this contradicts with the fact that ${\lambda_h^*, N_h^*, p_{f_h}^*}$ is the optimal solution to (\ref{optimalpricing_minimumwage_only}).}. This indicates that $P_h^*>P_t^*$, which completes the proof.  

\end{document}